\documentclass[12pt]{article}
\usepackage[colorlinks=true,
linkcolor=Green, 
citecolor=Green,
filecolor=Green,
urlcolor=Green,
linktoc=page, 
pdfstartview=FitV,
bookmarksopen=true]{hyperref}
\usepackage{float,tabularx,braket}
\usepackage[dvipsnames]{xcolor}
\usepackage[a4paper, left=2.5cm, right=2.5cm, top= 3cm, bottom=4cm]{geometry}

\usepackage{pgfplots}
\usetikzlibrary{arrows}
\pgfdeclarelayer{background}
\pgfsetlayers{background,main}
\usepackage{graphicx}
\usepackage{graphbox}
\usepackage{cite}
\usepackage{caption,multirow}
\usetikzlibrary{arrows.meta}

\usepackage{amsmath,amssymb,amsfonts,pifont}

\newcommand\B[1]{$\mathcal{B}^{#1}$}
\newcommand\BIII[1]{$\mathcal{B}_{III}^{#1}$}
\newcommand\BIIa[1]{$\mathcal{B}_{IIa}^{#1}$}
\newcommand\BIIb[1]{$\mathcal{B}_{IIb}^{#1}$}
\newcommand\BI[1]{$\mathcal{B}_{I}^{#1}$}

\newcommand\Z{\mathbb{Z}}

\newcommand{\cmark}{\textcolor{Blue}{\ding{51}}}%
\newcommand{\xmark}{\textcolor{Red}{\ding{55}}}%

\numberwithin{equation}{section}

\begin{document}

\begin{titlepage}

\vspace*{-2cm} 
\begin{flushright}
	{\tt \phantom{xxx}   MPP-2025-79} 
	
	{\tt \phantom{xxx} IFT-UAM/CSIC-25-111}
\end{flushright}

\vspace*{0.8cm} 
\begin{center}
{\LARGE Symmetries and dualities in \\ non-supersymmetric CHL strings
}\\

 \vspace*{1.5cm}
Bernardo Fraiman$^{1,2}$ and H\'ector Parra De Freitas$^3$ \\

 \vspace*{1.0cm} 
$^1$ {\it Max-Planck-Institut f\"ur Physik,	85748 Garching bei M\"unchen, Germany}\\[2mm]
$^2$ {\it Instituto de Física Teórica UAM/CSIC C/ Nicolás Cabrera 13-15 Universidad Autónoma de Madrid Cantoblanco, Madrid 28049, Spain}\\

$^3$ {\it Department of Physics, Harvard University, Cambridge, MA 02138, USA}\\

\vspace{1cm}
\small{\bf Abstract} \\[3mm]\end{center}
We chart the classical moduli space of heterotic strings with broken supersymmetry a la Scherk-Schwarz and gauge group rank reduced by 8 in eight dimensions. This space consists of four connected components, each with its own characteristic spectrum and \mbox{T-duality} group. Three of these components uplift to nine dimensions and can be described as Coxeter polyhedra, allowing an exact characterization of their maximal symmetry enhancements and decompactification limits. We determine the maximal enhancements in the eight dimensional theories using lattice based algorithms in the bosonic formulation, and perform an indepth analysis of their massless spectra. Finally we argue that one component has a supersymmetric $\mathcal{N} = 1$  sector described by BPS objects at strong coupling in a non-supersymmetric version of the type IIB string on $T^2/\mathbb{Z}_2$ with one \mbox{$O7^+$-plane}.

\end{titlepage}

\newpage

\tableofcontents	

\section{Introduction}
\label{s:intro}
String compactifications in $D \geq 7$ have either 32, 16 or 0 supercharges. It is natural to ask what is the structure of moduli space in this regime. In the supersymmetric case, we have good reasons to believe that these moduli spaces are completely known \cite{Bedroya:2021fbu,ParraDeFreitas:2022wnz,Montero:2022vva}. In the $N = 0$ case, various different theories in $D \geq 7$ have been constructed in the literature, see e.g. \cite{Dixon:1986iz,Alvarez-Gaume:1986ghj,Kawai:1986vd,Seiberg:1986by, Sagnotti:1995ga,Sagnotti:1996qj,Sugimoto:1999tx,Baykara:2024tjr,Bossard:2024mls,Acharya:2022shu,Blum:1997cs,Blum:1997gw,Antoniadis:1998ki,Coudarchet:2021qwc, Angelantonj:2004cm}.\footnote{We will not dwell on the well known issues that arise in these theories. See \cite{Leone:2025mwo,Mourad:2017rrl} for reviews. See also \cite{ValeixoBento:2025yhz,Chen:2025rkb} for newly proposed scenarios without running directions for the potential function, and \cite{Montero:2025ayi} for a recent study M-theory with implications for duality without supersymmetry.} In particular, string orbifolds without open string sectors have exact moduli spaces at tree level, yielding a laboratory for studying certain aspects of the $N = 0$ phase of string theory with some degree of control. The ``standard component" of this moduli space is given by the Scherk-Schwarz reduction of the heterotic string, or equivalently any circle compactification of the 6 full rank non-supersymmetric heterotic strings \cite{Ginsparg:1986wr,Fraiman:2023cpa}. 

The goal of this paper is to start an in-depth analysis of the remaining connected components of this landscape. Many of these components are obtained through asymmetric orbifolds of the heterotic string acting non-trivially on the gauge bundle, i.e. by topologically non-trivial flat connections. The simplest such operation reduces the rank of the gauge group by 8, and in the supersymmetric setup defines the CHL string \cite{Chaudhuri:1995fk} as constructed in \cite{Chaudhuri:1995bf}. This non-trivial operation on the gauge bundle gives rise to non-simple-connected gauge symmetry groups as well as gauge groups with associated Kac-Moody algebra level $k = 2$. At strong coupling, this operation is seen geometrically by turning on flat fluxes which define so-called frozen singularities \cite{deBoer:2001wca}. Non-supersymmetric analogs of this theory can then be constructed by involving the operator $(-1)^F$ in the orbifold group, where $F$ is the spacetime fermion number. 

What makes these theories particularly interesting is that they are the closest non-supersymmetric analogs of the heterotic strings with 16 supercharges. Since the automorphism on the lattice is purely left-moving, the classical moduli space of these theories is purely of Coulomb branch type, locally of the exact same form as for the CHL string.\footnote{Mixing with right-moving operations on the lattice we also get Higgs branches such as in \cite{Acharya:2022shu} (although in some cases these are trivial, e.g. as in \cite{Baykara:2024tjr}).}  They can be understood as analogs of the CHL string in the same way as the non-supersymmetric standard component is analog to the torus compactifications of the supersymmetric heterotic strings. An important difference, however, is that there are four different such non-supersymmetric theories (up to T-duality) instead of just one \cite{DeFreitas:2024ztt}, each with different features. This makes them a natural laboratory for probing the non-supersymmetric branch of the string landscape beyond the standard component. 

The most salient interaction between supersymmetry breaking and rank reduction is perhaps the possibility of having supersymmetric subsectors in the spectrum. As a hint, the $E_8$ string in $D = 10$ has two fermions of opposite chirality transforming in the adjoint of $E_8$. We may compactify this theory on $S^1$ and turn on a holonomy which projects out half of these fermions, resulting in a theory \cite{Nakajima:2023zsh,Saxena:2024eil} with an exact Bose-Fermi degeneracy in one of its subsectors, as we will show. We will see that this same theory is amenable to the construction of a strong coupling dual in $D = 8$ using the adiabatic argument \cite{Vafa:1995gm}, which coincides with a type IIB orientifold studied in \cite{Coudarchet:2021qwc}, allowing us to interpret the above Bose-Fermi degeneracy in terms of open strings stretching between mutually BPS $D7$-branes.\footnote{This feature is unique to this particular model --- the $E_8$ string has too many adjoint fermions and the other two models, like the $T^2$ compactification of the $SO(16)\times SO(16)$ string \cite{Fraiman:2023cpa}, do not admit adjoints. This model is also special in that it plays a role in the description of a non-BPS 7-branes in the $E_8\times E_8$ heterotic string \cite{Kaidi:2023tqo,Kaidi:2024cbx}.} This orientifold involves an $O7^+$-plane, i.e. a ``frozen singularity" \cite{Bianchi:1991eu,Witten:1997bs,Bhardwaj:2018jgp,Cvetic:2022uuu}, and our observations suggest that singularity freezing may be a good ingredient for controlling non-supersymmetric compactifications.  
 
In this work we carry out a general analysis of the four non-supersymmetric rank reduced theories at the level of their 1-loop partition functions, massless and tachyonic spectrum, and symmetry enhancements accross the tree-level moduli space. In $D = 8$ we use an adapted version of the exploration algorithm of \cite{Font:2020rsk,Font:2021uyw} to find the points of maximal symmetry enhancement as well as the rest of the massless spectrum (see \cite{Hamada:2024cdd,Hamada:2025cpe} for prior results for the $E_8$ string). For the three theories that admit an uplift to $D = 9$ we determine the Coxeter diagrams describing the global geometry of moduli space. These diagrams encode the T-duality group of the respective theory, and gives its symmetry enhancements as well as decompactification limits, corroborating the T-duality relations argued for in \cite{DeFreitas:2024ztt}. We do not focus on the 1-loop potentials beyond reporting on their values at tachyon-free points; a refined analysis is left for future work.

This paper is structured as follows. In Section \ref{s:generalities} we review/introduce various concepts that we will use in the rest of the paper, and introduce the general formula for the 1-loop partition function as well as the massless and tachyonic states that can appear in the spectrum. In Section \ref{s:max} we develop some lattice theoretical concepts applied to the problem of gauge symmetry enhancement as well as to explain various features of the observed spectra, and carry out the exploration algorithm. In Section \ref{s:CD} we focus on the 9D theories and construct their Coxeter diagrams, and determine the T-duality groups from this perspective. In Section \ref{s:duality} we use the adiabatic argument to make the S-duality proposal. We end with some discussion in Section \ref{s:disc}. We leave to appendices \ref{app:Z} and \ref{app:max} the technical details on the computation of the 1-loop partition functions and the results of the exploration algorithm, respectively.

\section{Generalities}
\label{s:generalities}

\subsection{Tree-level moduli spaces}

It was argued in \cite{DeFreitas:2024ztt} that there are four non-supersymmetric analogs of the CHL string of \cite{Chaudhuri:1995bf} up to T-duality. Importantly, for each one there is a T-dual frame described by an asymmetric orbifold of the $E_8\times E_8$ heterotic string on $T^d$, involving a combination of the symmetries 
\begin{equation}\label{operations}
    \theta_L\,, ~~~~~ (-1)^F\,, ~~~~~ \delta_i.
\end{equation}
$\theta_L$ is the outer automorphism of the gauge group exchanging the two $E_8$ factors, $F$ is the spacetime fermion number and $\delta_i$ is a half-period shift along a circle direction $x^i$ acting as $x^i \to x^i + \pi R_i$, with $R_i$ the circle radius. Including the supersymmetric CHL string, the precise constructions are:
\begin{itemize}
	\item \underline{\B{} string:} Orbifold of HE on $S^1$ by $g = \theta \delta$. This is the supersymmetric CHL string as constructed in \cite{Chaudhuri:1995bf}.
	
	\item \underline{\BIII{} string:} Orbifold of HE in 10D by $g = \theta (-1)^{F}$. This is the $E_8$ string \cite{Kawai:1986vd}.
	
	\item \underline{\BIIb{} string:} Orbifold of HE on $S^1$ by $g = \theta (-1)^{F}\delta$. This is the ``non-supersymmetric version of the CHL string" \cite{Nakajima:2023zsh}. It is T-dual to the orbifold of the $E_8$ string on $S^1$ by $g = (-1)^{F_R}\delta$ \cite{Nakajima:2023zsh,Saxena:2024eil} as well as the orbifold of the $SO(16)\times SO(16)$ string\footnote{We have tried to be as accurate as possible when talking in terms of groups rather than algebras concerning their topology, but in many cases such as this one (i.e. when naming string theories) the notation becomes very cumbersome. From context it should be clear that we are refering to the gauge algebra.} on $S^1$ by $g = \theta \delta$ where $\theta$ exchanges the two $SO(16)$ factors \cite{DeFreitas:2024ztt}. 
	
	\item \underline{\BIIa{} string:} Orbifold of HE on $S^1$ by $g_1 = \theta(-1)^{F}$ and $g_2 = (-1)^F \delta$. This is the Scherk-Schwarz reduction of the $E_8$ string. It is T-dual to the orbifold of the $E_7\times SU(2)\times E_7 \times SU(2)$ non-supersymmetric heterotic string by $g = \theta\delta$ with $\theta$ the exchange of the two $E_7\times SU(2)$ factors \cite{DeFreitas:2024ztt}. 
	
	\item \underline{\BI{} string:} Orbifold of HE on $S^1_1\times S^1_2$ by $g_1 = \theta\delta_1$ and $g_2 = (-1)^F \delta_2$. This is the Scherk-Schwarz reduction of the CHL string. We discuss some dual realizations in Section \ref{ss:frames}.
\end{itemize}
Here HE means the ten-dimensional $E_8\times E_8$ heterotic string. The nomenclature follows \cite{DeFreitas:2024ztt,Hohn:2023auw} and is meant to encompass all the T-duality frames, similarly to how in circle compactifications of heterotic and type II strings one usually drops the distinction between the two 10D theories. As we will show in Section \ref{s:CD}, the T-dualities for the \BIIb{} and \BIIa{} string as well as the \BIII{} string on $S^1$ are nicely encoded in their T-duality groups and Coxeter diagrams.

Let us now review some basic aspects of toroidal compactifications of the heterotic string and how gauging the symmetries \eqref{operations} affects the corresponding moduli space.

\subsubsection{Local moduli space}
The $E_8\times E_8$ heterotic string compactified on $T^d$ has a moduli space locally of the form
\begin{equation}\label{Mparent}
    \mathcal{M}_{d,d+16} =   O(d,d+16)/O(d)\times O(d+16)\times \mathbb{R}^+\,,
\end{equation}
where the coset part is parametrized by the metric, B-field and Wilson line moduli $G_{ij}, B_{ij}, A_i^A, A_i^{A'}$, $i = 1,...,d$, $A, A' = 1,...,8$, and $\mathbb{R}^+$ is parametrized by the dilaton modulus. The symmetry $\theta_L$ is present only at the locus 
\begin{eqnarray}\label{WLconst}
    A_i^A = A_i^{A'} \mod Q\,, ~~~~~Q\in \Gamma_{E_8}\,,
\end{eqnarray}
where the Wilson lines act in the same manner on the two $E_8$ factors; $\Gamma_{E_8}$ is the root lattice of the $E_8$ group. The other two symmetries, $(-1)^F$ and $\delta_i$, are instead present at every point in $\mathcal{M}_{d,d+16}$. Therefore, every one of the orbifolds that we consider can be constructed in the subspace defined by \eqref{WLconst}, which is locally of the form
\begin{eqnarray}\label{Mloc8}
    \mathcal{M}_{d,d+8} \simeq O(d,d+8)/O(d)\times O(d+8)\times \mathbb{R}^+\,.
\end{eqnarray}
After orbifolding, the corresponding moduli survive in the untwisted sector as tree level moduli  --- as usual there is a non-vanishing potential at one loop due a lack of Bose-Fermi degeneracy. We work therefore in the weak coupling limit $g_s \to 0$ and drop the $\mathbb{R}^+$. 

As we will see, there are no scalars in the twisted sector with mass vanishing everywhere in $\mathcal{M}_{d,d+8}$, hence  the coset space in \eqref{Mloc8} corresponds precisely to the tree-level moduli space of the non-supersymmetric orbifold. At special loci we do find extra massless scalars, but these are either (1) circle reductions of enhanced gauge bosons or (2) massless states becoming tachyonic in some direction in moduli space. The former are equivalent to the untwisted sector scalars under a change of choice of maximal torus of the enhanced gauge group. The latter signal an instability, hence they do not appear for any candidate to a minimum of the 1-loop potential. For these reasons we do not consider the problem of giving nonzero VEVs to charged scalars, and focus only on the moduli space \ref{Mloc8}.

\subsubsection{Fundamental regions and charge lattices}\label{ss:fund}
States in the HE theory are labeled by integer charges corresponding to the internal canonical momenta 
\begin{align}\label{momenta}
	p_R &= \frac{1}{\sqrt{2}}(n_i - E_{ij}w^j - A_i\cdot \pi)e^{*i}\,,\\
	p_L &= p_R + \sqrt2 G_{ij}w^j e^{*i} + \pi + A_i w^i\,,
\end{align}
where $n_i$ are the Kaluza-Klein momentum numbers, $w^i$ the winding numbers, $\pi\in E_8\oplus E_8$ is the gauge charge vectors, and $e^{*i}$ is the dual basis for the internal $T^d$. We have set $\alpha' = 1$ and conveniently defined the moduli
\begin{eqnarray}
	E_{ij} = G_{ij} + B_{ij} + \frac12 A_i \cdot A_j\,.
\end{eqnarray}

The vectors $(p_L,p_R)$ lie in the even self-dual Narain lattice with inner product
\begin{equation}
	(p_L,p_R) \cdot (p_L', p_R') = p_L \cdot p_L' - p_R \cdot  p_R'\,,
\end{equation}
where the RHS dot products are Euclidean. One shows that 
\begin{equation}\label{equivlat}
	 p_L \cdot p_L' - p_R \cdot  p_R' = \sum_{i = 1}^d  (n_i w'^i + n'_i w^i) + \pi \cdot \pi' \,,
\end{equation}
hence the \textit{charge vectors} 
\begin{equation}
	u \equiv (n_1,...,n_d,w^1,...,w^d; \pi)\,,
\end{equation} 
also form an even self-dual lattice $\Gamma_{d,d+16}$ with inner product given by the RHS of \eqref{equivlat}. Geometrically, the momenta define a moduli-dependent embedding of the abstract lattice $\Gamma_{d,d+16}$ into the ambient space $\mathbb{R}^{d,d+16}$, and locally the moduli space is just the space of corresponding lattice boosts.  

Automorphisms of $\Gamma_{d,d+16}$ preserve the spectrum of the theory yet act non-trivially on the moduli. They form the T-duality symmetry group of the theory, equivalent to the discrete subgroup $O(d,d+16;\mathbb{Z}) \subset O(d,d+16)$, and define the fundamental region of the moduli space
\begin{equation}
	\hat{\mathcal{M}}_{d,d+16} \simeq \text{Aut}(\Gamma_{d,d+16})\backslash O(d,d+16)/O(d)\times O(d+16)\,,
\end{equation}
where again we have dropped the dilaton contribution. 

For the orbifold theories we also define a charge lattice which we denote $\Upsilon_{d,d+8}$. The main difference is that the spectrum is now separated into different classes, and this structure must be respected by the T-duality group. As such, this group is generically some subgroup $\Theta(d,d+8,\mathbb{Z})$  of the automorphism group of $\Upsilon_{d,d+8}$, and we write
\begin{equation}
	\hat{\mathcal{M}}_{d,d+8} \simeq \Theta(d,d+8,\mathbb{Z})\backslash O(d,d+16)/O(d)\times O(d+8)\,.
\end{equation}
The lattice properties of these theories and the T-duality groups are worked out respectively in Sections \ref{s:max} and \ref{s:CD}. The definitions and formulas for the momenta as well as quantum numbers are the same as for the HE string (with a suitable normalization of the moduli fields), differing on the quantization conditions for these numbers and the reduced number of Wilson line moduli.

It is important to note that, just as for the supersymmetric CHL string, we expect that all electric charges are realized perturbatively in the non-supersymmetric theories, i.e. that they correspond to the momentum lattices. In $D \geq 6$ we expect this to be the case since the only non-perturbative objects in the spectrum are NS5-branes and there are not enough compact directions for them to produce particle-like excitations.

\subsection{Partition functions}\label{ss:partition}
We will now present the 1-loop partition function for the rank reduced theories in an unified formulation. The explicit computations are left to Appendix \ref{app:Z}. But first, let us set our notation and give some convenient definitions. 

\subsubsection{Conventions}
The 1-loop partition function of the heterotic strings under study take the generic form
\begin{equation}\label{genericZ}
    Z(\tau,\bar\tau) = \frac{1}{\tau_2^{(D-2)/2}\eta^{24-d}\bar\eta^{8-d}}\left[Z_v(\tau) \bar V_8(\bar \tau) - Z_s(\tau)\bar S_8(\bar\tau) - Z_c(\tau)\bar C_8(\bar\tau) + Z_o(\tau)\bar O_8(\bar\tau)\right]
\end{equation}
where $\tau = \tau_1 + i \tau_2$ is the complex structure of the worldsheet torus, $q = e^{2\pi i\tau}$ is the elliptic nome, $\eta$ is Dedekind's eta function and $d$ and $D$ are the number of compact and non-compact dimensions. In the RHS every unbarred function is holomorphic and every barred function antiholomorphic. We use the $Spin(2n)$ characters 
\begin{equation}
	\begin{split}
		O_{2n} &= \frac{1}{2\eta^{n}}(\vartheta_3^n + \vartheta_4^n)\,, ~~~~~ V_{2n} = \frac{1}{2\eta^n} (\vartheta_3^n - \vartheta_4^n)\,,\\
		S_{2n} &= \frac{1}{2\eta^n} (\vartheta_2^n + \vartheta_1^n)\,, ~~~~~ C_{2n} = \frac{1}{2\eta^n} (\vartheta_2^n - \vartheta_1^n)\,,
	\end{split}
\end{equation}
with $\vartheta_{1,2,3,4}$ the usual Jacobi theta functions evaluated at zero chemical potential. We have suppressed the dependence on $\tau$ and $\bar \tau$, and will do so in the following when convenient. 

To ease notation we define the following holomorphic functions associated to the action of $\theta$ on string oscillator modes:
\begin{equation}\label{fij}
    f_{00} \equiv 1\,, ~~~~~ f_{01} \equiv \left(\frac{2\eta^3}{\vartheta_2}\right)^4\,, ~~~~~ f_{10} \equiv \left(\frac{\eta^3}{\vartheta_4}\right)^4\,, ~~~~~ f_{11} \equiv\left(\frac{\eta^3}{\vartheta_3}\right)^4\,.
\end{equation}
Triality of $Spin(8)$ implies $V_8 = S_8 = C_8$, and we will make use of the $\bar q$-expansions
\begin{equation}\label{expvs}
    \frac{\bar V_8}{\bar \eta^8}(\bar q) = 8 + 128\bar q + ...\,, ~~~~~ \frac{\bar O_8}{\bar \eta^8} (\bar q) = \frac{1}{\bar q^{1/2}} + 36 \bar q^{1/2}+...\,,
\end{equation}
as well as
\begin{equation}\label{expfij}
    \frac{f_{01}}{\eta^{24}}(q) = q^{-1} + 8  +...\,,~~
    \frac{f_{10}}{\eta^{24}}(q) = q^{-1/2} - 8  +...\,,~~
    \frac{f_{11}}{\eta^{24}}(q) = q^{-1/2} + 8  +...\,.
\end{equation}\\

\subsubsection{Unified partition function}
One of the great advantages of working in the T-duality frame defined by orbifolds of the $E_8\times E_8$ heterotic string is that their 1-loop partition functions can be rewritten in such a way that they have the same generic form. As we show in Appendix \ref{app:Z}, the four blocks $Z_{v,s,c,o}$ in \eqref{genericZ} for each of the five rank reduced theories can then be obtained from the following formula
\begin{equation}\label{master}
	\begin{split}
	Z^{(J_i,K_i,M_i)}_{(F,T)} & = 
\prod_{i=1}^{d} 
	\left( \sum_{\begin{smallmatrix}
	  2w_i \in  2^{M_i}\mathbb{Z} +  J_i T \\
 n_i \in 2^{K_i}\mathbb{Z} + 2w_i + F \end{smallmatrix}}\right) \sum_{\pi \in E_8(\tfrac12)} \frac12[f_{10} -(-1)^{(1+F)T+p_L^2 - p_R^2} 
 f_{11}] q^{\tfrac12 p_L^2} \bar{q}^{\tfrac12 p_R^2}\\
 &+ c^{J_i,K_i,M_i}_{F,T}
 \prod_{i=1}^{d} 
 \left( \sum_{\begin{smallmatrix}
 		2w_i\in  2\mathbb{Z} + K_i T\\
 		n_i \in  2^{J_i}\mathbb{Z} + 2w_i + M_i F\end{smallmatrix}}\right)
 \sum_{\pi \in E_8(2)}
 f_{01}q^{\tfrac12 p_L^2}\bar q^{\tfrac12 p_R^2}\,,
\end{split}
\end{equation}
where the two parameters $F, T \in \{0,1\}$ specify the spacetime Lorentz class,\footnote{The nomenclature is motivated by considering a $\mathbb{Z}_2$ orbifold of a supersymmetric heterotic string where the symmetry includes an $(-1)^F$ factor. The untwisted sector contains the classes $Z_v$ and $Z_s$ while the twisted sector contains the classes $Z_c$ and $Z_o$. In this setting, $F$ is the spacetime fermion number and $T = 0$ ($T = 1$) for untwisted (twisted) states.}
\begin{equation}
    Z_{(0,0)} = Z_v\,, ~~~~~ Z_{(0,1)} = Z_0\,, ~~~~~ Z_{(1,0)} = Z_s\,, ~~~~~ Z_{(1,1)} = Z_c\,,
\end{equation}
not to be confused with the $Z_{g^i,g^j}$ of eq. \eqref{std}, and the $3(10-D)$ parameters $J_i,K_i,M_i \in \{0,1\}$ specify the orbifold. There is also an orbifold and class-dependent constant
\begin{equation}
	c^{(J_i,K_i,M_i)}_{(F,T)} = \left(1-T\prod_{i=1}^D M_i  \right) \left(1- F\prod_{i=1}^D  \left(1-J_i+K_i\right)\right) \in \{0,1\}\,,
\end{equation}
specifying if the second line in \eqref{master} is present or not for each class. 

Two compact dimensions are enough to construct all of the orbifolds, and these are specified by the parameter values
\begin{equation}\label{JKM}
\begin{array}{c|ccc|ccc|c}
 \text{Theory}  & J_1 & K_1 & M_1 & J_2 & K_2 & M_2 & c \\ \hline
\mathcal{B} & 1 & 0 & 0 & 0 & 0 & 1 & 1 \\
 \mathcal{B}_{III} & 0 & 0 & 1 & 0 & 0 & 1 & (1-F) (1-T) \\
 \mathcal{B}_{IIb} & 1 & 0 & 1 & 0 & 0 & 1 & 1-T \\
 \mathcal{B}_{IIa} & 1 & 1 & 0 & 0 & 0 & 1 & 1-F \\
 \mathcal{B}_I & 1 & 1 & 1 & 1 & 0 & 0 & 1 \\
\end{array}
\end{equation}
In particular, a triple $(J_i, K_i, M_i)=(0,0,1)$ defines the partition function of a compact boson, i.e. an ordinary circle compactification. We summarize the quantization conditions for $n_i$ and $w_i$ for given $J_i,K_i,M_i$ and $F,T$ in Table \ref{tab:quant}.

\begin{table}[H]
\centering
\begin{tabular}{|c c c|c c|c c|}
\hline
$J_i$ & $K_i$ & $M_i$ & $n_i$ & $w^i$ & $n'_i$ & $w'^i$ \\ \hline\hline
$0$ & $0$ & $1$ & \text{integer} & \text{integer} & \text{integer} & \text{integer} \\ \hline
$1$ & $0$ & $0$ & \text{integer} & \text{half-integer} & \text{even} & \text{integer} \\
    &     &     & \text{integer} & \text{integer} &  &  \\ \hline
$1$ & $0$ & $1$ & \text{integer} & \text{$\frac{T}{2}$ mod 1} & \text{$F$ mod 2} & \text{integer}\\ \hline
$1$ & $1$ & $0$ & \text{$F+1$ mod 2} & \text{half-integer}& \text{$T$ mod 2} & \text{$\frac{T}{2}$ mod 1} \\
    &     &      & \text{$F$ mod 2} & \text{integer}&                &                     \\ \hline
$1$ & $1$ & $1$ & \text{$F+T$ mod 2} & \text{$\frac{T}{2}$ mod 1} & \text{$F+T$ mod 2} & \text{$\frac{T}{2}$ mod 1} \\ \hline
\end{tabular}
\caption{Quantization conditions on the winding and momentum numbers in formula \eqref{master} for given values of $F$ and $T$. The numbers $(n_i, w^i)$ and $(n'_i, w'^i)$ correspond respectively to the sums in the first and second line in the RHS of \eqref{master}. }
\label{tab:quant}
\end{table}

Formula \eqref{master} gives the most natural presentation for each of the rank reduced theories, as it does not involve the data defining the parent theories from which they are constructed as orbifolds. In particular, it is written manifestly in terms of the lattices (more generally sets) of electric charges for each sector in the spacetime spectrum, allowing for example a clean derivation of T-duality groups. 

\subsection{Massless and tachyonic fields}
\label{ss:fields}

We now determine what types of massless and tachyonic fields appear generically in these theories. To this end let us first clarify some aspects of the structure of the spectrum encoded in \eqref{master}. Compared to the usual expression in terms of untwisted and twisted sectors, this formula reflects a rewriting of certain states as excitations of the untwisted vacuum in terms of the twisted vacuum given by the twist field $\sigma$. The fact that this is possible was essentially anticipated in \cite{Mikhailov:1998si}, particularly in the observation that T-duality mixes twisted and untwisted sectors. Concretely, the first and second lines in \eqref{master} are interpreted as counting over states with and without a $\sigma$ insertion. With these facts in mind we analyse first the second and then the first lines of \eqref{master} in the following. We record the specific forms of the quantum states in Table \ref{tab:states}.

\subsubsection{Vector class}
Using \eqref{expvs} and \eqref{expfij} we see that the second line in the RHS of \eqref{master} for $Z_v$ ($F = T = 0$) counts states with 
\begin{equation}\label{masspr1}
   m^2 = m_L^2 + m_R^2\,, ~~~~~ m_L^2 = p_L^2 + 2N_L - 2\,, ~~~~~ m_R^2 = p_R^2 + 2N_R - 1\,,
\end{equation}
where $N_L \in \mathbb{Z}$ is an effective occupation number associated to $f_{01}/\eta^{24}$, counting $\mathbb{Z}$-modded left-moving oscillations in spacetime and $2\mathbb{Z}$-modded oscillations in the internal gauge lattice directions; $N_R \in \mathbb{Z} + \tfrac12$ is associated to $\bar V_8/\bar \eta^8$. Setting $p_L = p_R = 0$, $N_L = 1$ and $N_R = 1/2$ we find the states furnishing the 10D graviton, B-field and dilaton fields $G_{MN},B_{MN},\phi$, suitably reduced on the internal torus. We split the indices $M,N...$ into $i,j...$ and $\mu,\nu,...$ for compact and non-compact directions. Setting instead $p_L^2 - p_R^2 = 2$, $N_L = 0$ and $N_R = 1/2$, we obtain level-matched states with
\begin{equation}\label{masspr2}
    m_L^2 = m_R^2 = p_R^2\,,
\end{equation}
which become massless gauge bosons $A_M^{A'}$ when $p_R = 0$ as a function of the moduli, with the index $A'$ a gauge group index for long roots.  

The first line in the RHS of \eqref{master} counts states with
\begin{equation}
    m_L^2 = p_L^2 + 2N_L' - 1\,, ~~~~~ m_R^2 = p_R^2 + 2N_R - 1\,,
\end{equation}
where $N_L'$ is an effective occupation number with a shifted ground state energy, associated to $(f_{10} \pm f_{11})/\eta^{24}$. It counts $(\mathbb{Z}/2)$-modded oscillations in the internal gauge lattice directions, and is conditioned by the values of the momenta:
\begin{equation}\label{effNL}
    N_L' \in
    \begin{cases}
        \mathbb{Z} & ~~~\text{if}~~~~~~p_L^2 - p_R^2 \in 2\mathbb{Z} + 1\\
        \mathbb{Z} + 1/2 & ~~~\text{if}~~~~~~ p_L^2 - p_R^2 \in 2\mathbb{Z} 
    \end{cases}\,.
\end{equation}
Setting $p_L = p_R = 0$ and $N_L = N_R = 1/2$ we find eight massless gauge bosons $A_M^a$, $a = 1,...,8$, furnishing the Cartan subalgebra of the gauge group $E_8$. Setting $p_L^2 - p_R^2 = 1$, $N_L' = 0$ and $N_R = 1/2$, we find again states with mass given by \eqref{masspr2}, becoming massless gauge bosons $A_M^{A}$ when $p_R = 0$. The indices $a$ and $A$ are respectively gauge indices for the abelian subalgebra and short roots. 

\subsubsection{Spinor class}

To examine $Z_s$ we set $F = 1$ and $T = 0$ in \eqref{master}. From \eqref{JKM} we see that theories \BIIb{} and \BI{} have $c = 1$ in this sector. Since $\bar S_8 = \bar V_8$, the way in which massless fermions appear is just as discussed above for the vector class, i.e. they must have $p_L^2-p_R^2 \in \{0,1,2\}$. They differ from states in $Z_v$ only in the allowed values for their winding and momentum numbers. We denote them by $\psi^A_\alpha$, $\psi^{A'}_\alpha$

On the contrary, the theories \BIII{} and \BIIa{} have $c = 0$, the constraint on momenta reduces to $p_L^2-p_R^2 \in \{0,1\}$. The same considerations apply to $Z_c$ ($F = T = 1$), with the main difference being that $c = 0$ also for the \BIIb{} theory. In the cases where $c =1$, the quantization conditions for fermions in Table \ref{tab:quant} preclude the appearance of fermionic partners to the graviton, B-field and dilaton. In the \BIII{} theory we find fermionic partners to the $E_8$ Cartans both in $Z_s$ and $Z_c$, while in the \BIIb{} theory we find them in $Z_s$.

\subsubsection{Scalar class}
Lastly we set $F = 0$ and $T = 1$ in \eqref{master} to examine the scalar class. Only the theories \BIIa{} and \BI{} have $c = 1$ in this sector, hence their spectra may contain states in the second line of \eqref{master}. These have mass given by \eqref{masspr1} with $N_L,N_R \in \mathbb{Z}$. Setting $N_L = N_R = 0$ and $p_L^2 - p_R^2 = 1$ selects states with
\begin{equation}\label{masstach}
    m_L^2 = m_R^2 = p_R^2 - 1\,,
\end{equation}
which are tachyonic in the moduli space region bounded by the space defined by $p_R^2 = 1$. When $p_R^2 = 1$, these states are massless with $p_L^2 = 2$, and carry a long root gauge index $A'$. The $p_R$ charge might correspond to an ordinary graviphoton $U(1)$ charge, or in the case that there is an enhancement of said $U(1)$ to $SU(2)$ (see below), to a root gauge index $\bar A$. We write these scalars fields generically as $\varphi^{A'}$, suppressing the right-moving gauge group charge/index. These fields also appear in the Scherk-Schwarz reduction of the HE theory (indeed, both the \BIIa{} and \BI{} theories are also Scherk-Schwarz reductions), and similarly give rise to ``knife-edges": regions in moduli space where a variation in the VEV of some modulus makes a massless scalar field tachyonic \cite{Ginsparg:1986wr,Fraiman:2023cpa}. 

Setting instead $N_R = 0$, $N_L = 1$ and $p_L^2 - p_R^2 = -1$ we get states with mass given also by \eqref{masstach}, but now we have a bound $p_R^2 \geq 1$ so that they become massless when $p_R^2 = 1$ or equivalently when $p_L^2 = 0$. Since $N_L = 1$, they transform as spacetime gauge fields $\varphi_M$ reduced on $T^d$. They enhance $U(1)\to SU(2)$ at level 2, and also appear generically in Scherk-Schwarz reductions. 

In the first line in \eqref{master}, the mass formula for states in $Z_o$ differs from $Z_{v,s,c}$ in that $(1+F)T = 1$, hence \eqref{effNL} is modified to
\begin{equation}
    N_L' \in
    \begin{cases}
        \mathbb{Z} & ~~~\text{if}~~~~~~p_L^2 - p_R^2 \in 2\mathbb{Z} \\
        \mathbb{Z} + 1/2 & ~~~\text{if}~~~~~~ p_L^2 - p_R^2 \in 2\mathbb{Z} + 1 
    \end{cases}\,,
\end{equation}
and we also have $N_R \in \mathbb{Z}$. Setting $N_R = N_L' = 0$ and $p_L^2 - p_R^2 = 0$ yields two distinct types of possibly tachyonic states with $m_{L,R}^2 = p_R^2 - 1$. The first has vanishing quantum numbers, hence $p_L = p_R = 0$ and $m^2 = -2$ for all values of the moduli. This state is only present in the \BIII{} and \BIIa{} theories, where $Z_o^\text{short}$ admits null winding and momenta (cf Table \ref{tab:quant}), leading to a generic tachyonic field $\mathcal{T}$. The second type of state generically has $p_R^2 = p_L^2 \neq 0$, becoming extremally tachyonic at infinite distance. An example is given by winding or Kaluza-Klein modes, which have
\begin{equation}
    2p_L^2 = 2p_R^2 = 
    \begin{cases}
         w'^2R^2 \\
        n'^2/R^2
    \end{cases}\,,
\end{equation}
and furnish extremally tachyonic towers as $R \to 0$ or $R \to \infty$, respectively. At $p_L^2 = p_R^2 = 1$ these states are massless, and are the short root counterparts to $\varphi^{A'}$, denoted $\varphi^{A}$.

We finally have states with $N_L' = 1/2$, $N_R = 0$ and $p_L^2 - p_R^2 = -1$. These become massless when $p_L = 0$ and $p_R^2 = 1$, and carry an index $a$. They are the Cartan counterparts $\varphi^{a}$ to $\varphi^{A}$ and $\varphi^{A'}$. 

\begin{table}[H]
\centering
\begin{tabular}{|rl|c|c|c|c|c|c|c|c|}
\hline
States&&Fields& \B{} & \BIII{} & \BIIb{} & \BIIa{} & \BI{} & $\small m^2=0$& \small$m^2=-2$ \\ \hline\hline
$\alpha_{-1}^M\bar\alpha_{-1/2}^N$&$\ket{0,0}$ &$G,~B,~\phi$& \cmark$_{9}$ & \cmark$_{10}$ & \cmark$_{9}$ & \cmark$_{9}$ & \cmark$_{8}$ & $(0,0)$ &  \\ 
$\alpha_{-1/2}^a\bar\alpha_{-1/2}^M$&$\ket{0,\sigma}$&$A_M^a$& \cmark$_{9}$ & \cmark$_{10}$ & \cmark$_{9}$ & \cmark$_{9}$ & \cmark$_{8}$ & $(0,0)$ &  \\
$\bar\alpha_{-1/2}^M$&$\ket{2,0}$&$A_M^{A'}$& \cmark$_{8}$ & \cmark$_{9}$ & \cmark$_{8}$ & \cmark$_{8}$ & \cmark$_{7}$& $(2,0)$ &   \\
$\bar\alpha_{-1/2}^M$&$\ket{1,\sigma}$&$A_M^{A}$& \cmark$_{9}$ & \cmark$_{10}$ & \cmark$_{9}$ & \cmark$_{9}$ & \cmark$_{8}$ & $(1,0)$ &   \\ \hline
$\alpha_{-1}^M $&$\ket{0,\bar S_\alpha}$&$\psi_{\dot \alpha}^M,\lambda_{\dot\alpha}$& \cmark$_{9}$ & \xmark & \xmark & \xmark & \xmark & $(0,0)$ &   \\ 
$\alpha_{-1/2}^a$&$\ket{0,\bar S_\alpha\sigma}$&$\psi_\alpha^a$& \cmark$_{9}$ & \cmark$_{10}$ & \cmark$_{9}$ & \xmark & \xmark & $(0,0)$ &   \\
&$\ket{2,\bar S_\alpha}$&$\psi_\alpha^{A'}$& \cmark$_{8}$ & \xmark & \cmark$_{9}$ & \xmark & \cmark$_{8}$ & $(2,0)$ &   \\
&$\ket{1,\bar S_\alpha\sigma}$&$\psi_\alpha^{A}$& \cmark$_{9}$ & \cmark$_{10}$ & \cmark$_{9}$ & \cmark$_{9}$ & \cmark$_{8}$ & $(1,0)$ &   \\
\hline
$\alpha_{-1}^M $&$\ket{0,\bar C_{\dot\alpha}}$&$\psi_{\alpha}^M,\lambda_{\alpha}$& \xmark & \xmark & \xmark & \xmark & \xmark & $(0,0)$ &   \\ 
$\alpha_{-1/2}^a$&$\ket{0,\bar C_{\dot\alpha}\sigma}$&$\psi_{\dot\alpha}^a$& \xmark & \cmark$_{10}$ & \xmark & \xmark & \xmark & $(0,0)$ &   \\
&$\ket{2,\bar C_{\dot\alpha}}$&$\psi_{\dot\alpha}^{A'}$& \xmark & \xmark & \xmark & \xmark & \cmark$_{8}$ & $(2,0)$ &   \\
&$\ket{1,\bar C_{\dot\alpha}\sigma}$&$\psi_{\dot\alpha}^{A}$& \xmark & \cmark$_{10}$ & \cmark$_{9}$ & \cmark$_{9}$ & \cmark$_{8}$ & $(1,0)$ &   \\
\hline
$\alpha_{-1}^{M}$&$\ket{-1,0}$&$ {\varphi}_{M}$& \xmark & \xmark & \xmark & \cmark$_{8}$ & \cmark$_{8}$ & $(0,1)$ &   \\
$\alpha_{-1/2}^a$&$\ket{-1,\sigma}$&$\varphi^{a}$& \xmark & \cmark$_{9}$ & \cmark$_{9}$ & \cmark$_{9}$ & \cmark$_{8}$ & $(0,1)$ &   \\
&$\ket{1,0}$&$\varphi^{A'}$& \xmark & \xmark & \xmark & \cmark$_{9}$ & \cmark$_{8}$ & $(2,1)$ & $(1,0)$  \\
&$\ket{0',\sigma}$&$\varphi^{A}$& \xmark & \cmark$_{9}$ & \cmark$_{9}$ & \cmark$_{9}$ & \cmark$_{8}$ & $(1,1)$ & $(0,0)$  \\
&$\ket{0,\sigma}$&$\mathcal{T}$& \xmark & \cmark$_{10}$ & \xmark & \cmark$_{9}$ & \xmark &  & $(0,0)$  \\
\hline
\end{tabular}
\caption{States becoming massless and/or tachyonic at special points in moduli space for each of the theories with rank reduction. The kets have the generic form $\ket{p_L^2-p_R^2, \mathcal{O}}$ where $\mathcal{O}$ is a combination of spin and twist fields. The subscript in \cmark denotes the maximal number of spacetime dimensions for which these states appear, as detailed in Section \ref{ss:analysis}. The Lorentz and gauge indices indicate their transformation properties as explained in the text.  We have suppressed the indices in $G_{MN}$ and $B_{MN}$ due to space constraints.}
\label{tab:states}
\end{table}

\section{Maximal symmetry enhancements in $D \geq 8$}
\label{s:max}

In this Section we introduce some lattice theoretical formalism in order to make systematic the determination of symmetry enhancements, their fundamental groups and the rest of the massless spectrum. We then explain how to use an adapted version of the exploration algorithm of \cite{Font:2020rsk} to determine the maximal enhancements, and then carry out an analysis of the spectra in some generality. Finally we make some comments on the tachyonic content and stability of the enhancements. 

\subsection{Charge lattices and gauge symmetries}
The quantization conditions for states in the vector class define two lattices with vectors $v' = (n_i,w^i;\pi)$ and $v = (n'_i, w'^i{},\pi')$, respectively, where $\pi \in E_8(\tfrac12)$ and $\pi' \in E_8(2)$, and $n_i,w^i,n_i',w^i{}'$ as in Table \ref{tab:quant}. These are the lattices of electric charges for the two sectors corresponding to the first and second line in \eqref{master} with $F = T = 0$, and we denote them respectively as $\Gamma_{d,d+8}^v$ and $\Gamma_{d,d+8}^{v'}$. For all theories we find that
\begin{equation}
	\Gamma_{d,d+8}^{v'} \subset \Gamma_{d,d+8}^v\,,
\end{equation}
and so we refer to $\Gamma_{d,d+8}^v$ as the \textit{vector class lattice}. For the four non-supersymmetric theories, this lattice takes the form
\begin{equation}\label{vclass}
	\Gamma^v_{d,d+8} =
	\begin{cases}
		\Gamma_{d,d}\oplus E_8(\tfrac12) & (\mathcal{B}_{III})\\
		\Gamma_{d,d}\oplus E_8(\tfrac12) & (\mathcal{B}_{IIb})\\
		\Gamma_{d-1,d-1}\oplus \mathbb{Z}\oplus \mathbb{Z}(-1)\oplus E_8(\tfrac12) & (\mathcal{B}_{IIa})\\
		\Gamma_{d-2,d-2}\oplus \Gamma_{1,1}(2) \oplus\Gamma_{1,1}(\tfrac12) \oplus E_8(\tfrac12)& (\mathcal{B}_{I})
	\end{cases}\,.
\end{equation}
Any charge vector $v$ in this lattice with $v^2 = 1$ gives rise to a state which for suitable values of the moduli furnishes a massless gauge boson. Elements with $v^2 = 2$ also furnish gauge bosons in the case that they are restricted to the sublattice $\Gamma_{d,d+8}^{v'}$. This restriction can be understood as the condition that $v$ generates a reflection which is a T-duality symmetry (i.e. that $v$ is \textit{reflective}). The union of all charge sets forms the full chage lattice $\Upsilon_{d,d+8}$ (cf. Section \ref{ss:fund}), which takes the form
\begin{equation}\label{chargelat}
	\Upsilon_{d,d+8} =
	\begin{cases}
		\Gamma_{d,d}\oplus E_8(\tfrac12) & (\mathcal{B}_{III})\\
		\Gamma_{d-1,d-1}\oplus \Gamma_{1,1}(\tfrac12)
		\oplus E_8(\tfrac12) & (\mathcal{B}_{IIb})\\
		\Gamma_{d-1,d-1}\oplus \Gamma_{1,1}(\tfrac12)\oplus E_8(\tfrac12) & (\mathcal{B}_{IIa})\\
		\Gamma_{d-2,d-2}\oplus \Gamma_{1,1}(\tfrac12) \oplus\Gamma_{1,1}(\tfrac12) \oplus E_8(\tfrac12)& (\mathcal{B}_{I})	\end{cases}\,.
\end{equation}

The states furnishing gange bosons have mass $m^2 = m_L^2+m_R^2 = 2p_R^2$, and from the definition of $p_R$ in \eqref{momenta}, we see that the masslessness condition $p_R = 0$ defines a constraint on the moduli fields for a given charge vector. A maximal symmetry enhancement is characterized by $d+8$ linearly independent constraints, completely fixing the moduli to some rational values, and each constraint can be associated for example to each of the $d+8$ simple roots furnishing the enhanced gauge algebra $\mathfrak g$. These roots span the \textit{root lattice} $L \subset \Gamma_{d,d+8}^v$, where the sublattice relation is such that there are no more reflective vectors (i.e. roots) in the intersection of the rational span of $L$ and $\Gamma_{d,d+8}^v$ (since these would modify the gauge algebra). 

We are also interested in the topology of the \textit{gauge group} $G$ given by the homotopy groups $\pi_0(G)$ and $\pi_1(G)$. The group $\pi_0(G)$ is given by the T-duality symmetries which become enhanced at a given point in moduli space but do not form part of the Weyl group of $G$, and in general it is a non-trivial task to determine its form; we will not compute these groups in this paper. On the other hand, $\pi_1(G)$ can be computed in a rather straightforward manner from the lattice data using the results of \cite{Cvetic:2021sjm}. Concretely, we have the isomorphism
\begin{equation}\label{top1}
	\pi_1(G) = (P(L,\Upsilon_{d,d+8}))^*/L^\vee\,,
\end{equation}
where $(P(L,\Upsilon_{d,d+8}))^*$ is the dual of the lattice $P(L,\Upsilon_{d,d+8})$ resulting from the projection of $\Upsilon_{d,d+8}$ onto $L \subset \Gamma_{d,d+8} \subset\Upsilon_{d,d+8}$, and $L^\vee$ is the coroot lattice of $\mathfrak g$. As shown in \cite{Cvetic:2021sjm}, \eqref{top1} can be reexpressed as
\begin{equation}\label{top2}
	\pi_1(G) = S(L^\vee, \Upsilon_{d,d+8}^*)/L^\vee\,,
\end{equation}
where $S(L^\vee, \Upsilon_{d,d+8}^*) \equiv L^\vee\otimes_\mathbb{Z}\mathbb{R} \cap \Upsilon_{d,d+8}^*$ is the saturation of $L^\vee \subset \Upsilon_{d,d+8}^*$, i.e. its unique overlattice which is primitively embedded into $\Upsilon_{d,d+8}^*$.  

\subsection{Exploration algorithm}
\label{ss:exploration}

As just explained, a point of maximal symmetry enhancement in the moduli space $\mathcal{M}_{d,d+8}$ is defined by an embedding $L \hookrightarrow \Gamma^v_{d,d+8}$. Such an embedding is completely spacified by the quantum numbers $n_i,w^i,\pi$ for each of the simple roots. Setting $p_R = 0$ (cf. eq. \eqref{momenta}) for each of these vectors imposes a constraint on the moduli fields, which altogether define the point in $\mathcal{M}_{d,d+8}$ with gauge algebra $\mathfrak g$. With these data we also determine $\pi_1(G)$ as well as the rest of the massless and tachyonic spectrum.

Starting from a maximal symmetry enhancement given by some $L \hookrightarrow \Gamma_{d,d+8}^v$, the exploration algorithm of \cite{Font:2020rsk,Font:2021uyw} works as follows:
\begin{enumerate}
	\item Delete one of the simple roots generating $L$. This operation relaxes the constraints on the moduli coming from setting $p_R = 0$ for this root, and so defines a $d$-dimensional subspace $\Sigma_d \subset \mathcal{M}_{d,d+8}$.
	
	\item Generate an arbitrary root that extends the remaining set of eight simple roots to a new set generating a new lattice $\tilde L'$. Compute the saturation $S(\tilde L',\Gamma_{d,d+8}^v)$ and determine its root sublattice $L'$, which generically is an overlattice $L' \supseteq \tilde L'$. If $L'$ is different from $L$, we have found a new point of maximal symmetry enhancement with gauge algebra $\mathfrak g'$.
	
	\item If $L' \simeq L$, we compute (1) the rest of the massless and tachyonic spectrum as well as (2) the fundamental group of the gauge group $\pi_1(G)$. If these data differ, the embedding $L' \hookrightarrow \Gamma_{d,d+8}$ defines a new maximal symmetry enhancement point. Otherwise we classify it as equivalent to the original one.\footnote{It is possible a priori that there are more subtle differences in the massive spectrum (perhaps leading to different $\pi_0(G)$'s), but this problem is well outside the scope of this work.}

	\item Repeat this process for $L \hookrightarrow \Gamma_{d,d+8}$ by deleting and adding roots in different ways to produce its ``neighboring" maximal enhancements, and then iterate it by starting from these new enhancements.  
\end{enumerate}
In practice one may generate a large set of seemingly inequivalent embeddings and then filter them out computing the data beyond $\mathfrak{g}$. 

We note that, given its constructive nature, this algorithm is not a priori exhaustive; as far as we know there is no first principles reason why every maximal symmetry enhancement should be connected along $d$-dimensional spaces corresponding to root deletions. We do know however that this is the case for the supersymmetric CHL string in $D = 9,8$ from the exact results of \cite{Fraiman:2022aik}. We then expect the number of missed maximal enhancements in the non-supersymmetric theories to be very few or none at all. 

We have carried out this algorithm for the four theories in $D = 8$ as well as the three $D = 9$ uplifts. The different maximal enhancements are presented in Appendix \ref{app:max}. In the following we will analyse the spectrum of each theory, explaining the notation used in the Tables in Appendix \ref{app:max}.

\subsection{Analysis of spectrum}
\label{ss:analysis}
\subsubsection{\BIII{}}

The \BIII{} theory is the simplest. As for the supersymmetric CHL string, its vector class lattice is the full charge lattice,
\begin{equation}
	\Gamma_{d,d+8}^v = \Upsilon_{d,d+8}\,.
\end{equation}
Moreover, its charge lattice in $D$ dimensions is exactly the same as that of the CHL string in $D-1$ dimensions up to a $\Gamma_{1,1}$ factor,
\begin{equation}
	\Upsilon_{d,d+8}^\text{\BIII{}}\oplus \Gamma_{1,1} = \Upsilon_{d+1,(d+1)+8}^\text{\B{}}\,, ~~~~ d \geq 1\,.
\end{equation}
This implies that every gauge symmetry group $G$ realized in the $D$-dimensional \BIII{} string is also realized in the $(D-1)$-dimensional CHL string as $G\times U(1)$, or $G\times SU(2)$ at the self-dual radius associated to $\Gamma_{1,1}$. This can indeed be checked by comparing Tables \ref{tab:BIII9D} and \ref{tab:BIII8D} with the results of \cite{Font:2021uyw,Cvetic:2021sjm} for $D-1 = 8$ and those of \cite{Fraiman:2021soq} for $D-1 = 7$.

The two spinor classes are exactly equivalent, and their quantum numbers also form the lattice $\Upsilon_{d,d+8}$. From \eqref{JKM}, however, $c = 1$ only in the vector class, and so only the gauge bosons associated to short roots have massless fermionic pairs. From this it follows that 
every gauge algebra comprising only short roots, which are simply-laced and at level 2, comes paired with two fermionic adjoints. For the remaining algebras, which are at level 1, we have $\mathfrak{sp}(n)$ with with antisymmetric traceless rep $\mathbf{n(n-1)/2 -1}$, $\mathfrak{so}(2n+1)$ with vector rep $\mathbf{2n+1}$, $\mathfrak{f}_4$ with fundamental rep $\mathbf{26}$, and simply laced algebras without massless spinors. 

This theory has in total eight pairs of generic massless fermions, which are in fact required to furnish the above representations as they furnish their 0-weights. Interestingly, this leads to an upper bound 
\begin{equation}
	n \leq 8
\end{equation}
on the rank of level 2 algebras, since the adjoint rep absorbs $n$ such fermions, as well as a bound
\begin{equation}
	n \leq 9
\end{equation}
on the rank of $\mathfrak{sp}(n)$, since the antisymmetric traceless absorbs $n-1$ such fermions. Both bounds are valid for all $D$. Similar bounds can be worked out for combinations of different algebras (the vector representation of $\mathfrak{so}(2n+1)$ has one 0-weight, while the $\mathbf{26}$ of $\mathfrak{f}_4$ has two), easily ruling out many gauge algebras which are indeed not observed as possible enhancements. As an interesting aside, these results necessarily carry over to the $(D-1)$-dimensional CHL string given its relationship explained above.

In the scalar class we find, apart from the tachyonic singlet $\mathcal{T}$, the fields $\varphi^a$ and $\varphi^{A}$ corresponding respectively to 0-weights and 1-weights of $G$, both charged under the right-moving $U(1)$'s. The fields $\varphi^a$ are present in the spectrum whenever there are vectors in $\Gamma_{d,d+8}^o = \Upsilon_{d,d+8}$ with $p_L = 0$ and $p_R^2 = 1$. Adding to such a vector some other vector with $p_L^2 = 1$ and $p_R = 0$ yields another vector furnishing a massless state $\varphi^A$, and these fields join with $\varphi^a$ to furnish a massless field with $U(1)$ charge and which transforms in a representation of $G$ degenerate with that of the massless spinors. Alternatively we may have states $\varphi^A$ whose charge vectors are not combinations of this type, and they furnish minuscule representations, i.e. without 0-weights. Because these representations are not degenerate with others in the spectrum, we refer to them as \textit{accidental}.

In $D$ dimensions, the charge vectors of the $\varphi^a$ form an ADE root system. In $D = 8$ for example we can have either $A_1$, $2\, A_1$ or $A_2$. These vectors correspond to the $U(1)$ charges of the representations which are degenerate with massless spinors, hence they also have this ADE structure. For accidental or mixed representations there may also be an ADE structure on the $U(1)$ charges, and we do observe these patterns. For this reason we find it convenient to specify the representations of massless scalars as
\begin{equation}
	[x_1]\,, ~~~~~ [x_1,x_2]\,, ~~~~~ [x_1,x_2,x_3]\,,
\end{equation}
where each entry $x_i$ corresponds to a simple root in $A_1$, $2A_1$ and $A_2$, respectively. We label the above degenerate representations as $s$ and the accidental representations as $a_i$, the latter of which are tabulated (cf. Table \ref{tab:acc}). For each entry one has two representations given by a pair of charge vectors $\pm(p_L,p_R)$. With these data the whole charge structure of the scalars is specified. There are a few \textit{exceptional} cases which we label $e_i$, in which the $U(1)$ charges do not form an ADE system. 

\subsubsection{\BIIb{}}

The best way to understand the spectrum of this theory is to use its presentation as a shift-orbifold of the CHL string.\footnote{Consider for example the realization of this theory as an orbifold of the $SO(16)\times SO(16)$ string on $S^1$ with $g' = \delta' \theta_L'$. The $SO(16)\times SO(16)$ theory itself is a shift orbifold of the $E_8\times E_8$ string given by $g = \delta (-1)^F $ with $\delta$ breaking each $E_8$ to $SO(16)$. Starting the construction with $g'$ instead of $g$, the \BIIb{} theory is now given as a shift-orbifold of the CHL string. That the full charge lattice is unaltered is empirical.} This orbifold does not alter the full charge lattice $\Upsilon_{d,d+8}$, and in its untwisted sector splits it as
\begin{equation}\label{split}
	\Upsilon_{d,d+8}= \Gamma_{d,d+8}^v \cup \Gamma_{d,d+8}^c
\end{equation}
according to the shift $\delta$.\footnote{The fact that the RHS in \eqref{split} involves $\Gamma_{d,d+8}^c$ rather than $\Gamma_{d,d+8}^s$ is purely due to conventions.} Therefore, every gauge algebra $\mathfrak{g}$ in this theory uplifts to a gauge algebra $\mathfrak{g}'$ in the CHL string by including the charge vectors for the cospinors as roots for gauge bosons. This readily explains the observed patterns for gauge algebras as well as massless cospinors. In particular it explains why these massless fermions arrange into minuscule representations of $G$. 

The spinor sector of this theory behaves essentially the same as for the \BIII{} theory explained above (and the discussion on bounds on gauge group ranks also applies). The important difference is that $c = 1$ in this sector leading to the presence of 2-weights. For example, given an enhanced $\mathfrak{so}(2n)$ gauge symmetry, it is possible a priori to have massless fermions in the adjoint \textit{or} the symmetric traceless representation. Similarly there is a special $\mathfrak{sp}(4)$ enhancement which is paired with the \textbf{42} representation rather than the antisymmetric traceless \textbf{27}. Breaking this algebra to $\mathfrak{su}(2)\oplus \mathfrak{sp}(2)$ and then to $\mathfrak{su}(2)\oplus \mathfrak{su}(2)$ gives also special pairings, namely the \textbf{(3,5)} and the \textbf{(3,3)}. All of these special cases are specified in Tables \ref{tab:BIIb9D} and
\ref{tab:BIII8D} by a prime on the respective root lattice symbol, i.e. $D_n'$, $C_4'$, $(A_1C_2)'$ and $(A_1A_1)'$. In the special case of $D_4$ the symmetric traceless representation has two images under triality, hence instead of using $D_4'$ we use $D_4^{v}$, $D_4^c$ or $D_4^s$.

Here we find the same kinds of scalar class fields as in the \BIII{} theory (except for the generic tachyon $\mathcal{T}$), but they satisfy different quantization conditions. In the case that there are massless fields $\varphi^a$ we find again that there appear fields $\varphi^A$ to furnish a representation of $G$ degenerate with that of the massless spinors. It should be noted that while 2-weights are allowed in the spinor class they are not allowed in the scalar class. We observe that for maximal enhancements the $\varphi^a$ are present only when the representations do not involve 2-weights, but have not proven that this is a generic phenomenon.

The remaining allowed representations are minuscule, but in this case not all of them are accidental. There are situations where they become degenerate with those in which massless cospinors transform. Finally, the structure of the right-moving $U(1)$ charges is just as for the \BIII{} string and we use exactly the same notation. 

Finally we note that one of the important features of this spectrum is that the sectors in class $v$ and $s$ in the first line of \eqref{master} are degenerate, since the quantization conditions are the same (see Table \ref{tab:quant}). We will interpret this degeneracy in terms of open strings stretching between mutually BPS objects in an orientifold dual in Section \ref{s:duality}.

\subsubsection{\BIIa{}}
\label{sss:BIIa}
As with the \BIIb{} theory this one can be understood as a shift orbifold of the supersymmetric CHL string, hence the considerations above for gauge bosons and spinors apply. The difference is that the shift sits in a different class in the charge lattice of the CHL string.\footnote{E.g. a shift breaking $E_8$ to $E_7\times SU(2)$ rather than to $SO(16)$.} Moreover, spinors and cospinors are fully degenerate in this theory, both transforming in minuscule representations of $G$. The interesting behaviors in these theory come from the scalar class, which admits all kinds of states in Table \ref{tab:states}. In particular, the appearance of right-moving $SU(2)$ enhancements as well as the appearance of 2-weights is due to it being a Scherk-Schwarz reduction, namely of the \BIII{} theory.  

The scalar fields $\varphi^a$ in this theory lead also to the appearance of fields $\varphi^A$ and/or $\varphi^{A'}$ in roughly two different ways, depending on if a right-moving $U(1)$ is enhanced to $SU(2)$ or not. This enhancement is due to the presence of states with $p_L = 0$, $p_R^2 = 1$ and $N_L = 1$, and obey the conditions $(n,m,\pi) \in (2\mathbb{Z}+1,\mathbb{Z}+\tfrac12,E_8(2))$. Taking $N_L = 0$ and $N_L' = 1/2$ instead with the same charge vectors we obtain massless states $\varphi^a$, since the quantization conditions for $\varphi^a$ form a superset of those for $\varphi_M$, see Table \ref{tab:quant}. Combining these charge vectors with those of both short and long roots of $G$ we obtain massless fields $\varphi^A$ and $\varphi^{A'}$, and by a suitable inclusion of other massless scalars in the generic spectrum of the compactification we obtain the representation $(\text{Adj},\text{Adj})$ of $G\times SU(2)$. We label these representations with the symbol $v$. Representations of this type are already known to occur for the Scherk-Schwarz reduction of the full rank heterotic string \cite{Fraiman:2023cpa}.

The second way in which the $\varphi^a$ appears is when there is no right-moving $SU(2)$ enhancement. In this case it can be shown that only the short roots of $G$ also lead to the appearance of fields $\varphi^A$, but there could appear fields $\varphi^{A'}$ which are not degenerate with long roots of $G$. There is some variability in the way these 2-weights appear, in a manner analogous to how 2-weights appear for massless spinors in the \BIIb{} string. As such, we use the same method of encoding this information into the root lattice data using primes. We label the associated representations with the symbol $\tilde v$. There are however a few exceptional cases where two representations of this type appear at the same time, for which we use a different notation $[v_i', v_i'']$. 

Apart from these representations we may also have degeneracies with the massless spinors or accidental representations, both of minuscule type. The transformation properties of the right-moving $U(1)$-charges are as before, with the subtlety that now there can be $SU(2)$ enhancements. As we already explained, however, when these enhancements are present we associate to them the representations labeled $v$.

Another interesting feature of the spectrum of this theory is the pairing between certain gauge bosons and extremal tachyons. Perhaps the clearest way to understand this is from the point of view of the heterotic worldsheet fields. Extremal tachyons are associated to pairs of Majorana-Weyl fermions $\lambda_i$, which usually furnish an $SO(2n)$ gauge symmetry with tachyons in the vector representation (as in the 10D heterotic theories). In this case however there is an extra fermion $\lambda'$ associated to the generic tachyon $\mathcal{T}$, hence the gauge symmetry is actually $SO(2n+1)$. When there is such an enhancement, the extremal tachyons transform in the vector representation of this gauge group, with $2n$ of them degenerate with the short roots of its adjoint representation. This degeneration can be seen directly from the quantization conditions in Table \ref{tab:quant}.

\subsubsection{\BI{}}
A special property of this theory is that its charge lattice in eight dimensions is self-dual once scaled by 2, 
\begin{equation}
	\Upsilon_{2,10}(2) = \Gamma_{2,2}\oplus E_8\,.
\end{equation}
As a result, only vectors with norm 1 in the vector class are reflective, and every enhanced symmetry group is of ADE type at level 2. 

It is instructive to compare this theory with the Scherk-Schwarz reduction of the full rank supersymmetric heterotic string, i.e. the $\mathcal{A}_I$. In the latter, gauge bosons have charge vectors with norm 2 while tachyonic states have charge vectors with norm 1. They can appear mixed in symmetry enhancements with $G = SO(2n)$ with tachyons in the vector representation, corresponding to a B-type lattice where long roots furnish gauge bosons and short roots furnish tachyons. 

In the 8D \BI{} theory this situation is not possible, because both types of charge vectors have the same norm. It is in particular inconsistent to have two such charge vectors with a non-zero inner product in the case that they have $p_R = 0$. Thus in configurations involving extremal tachyons, the symmetry enhancement cannot be maximal. As a result, the set of possible maximal enhancements is rather small in comparison to the other three theories. Indeed we have found only twelve such enhancements, see Table \ref{tab:BI8D}. 

Massless fermions in this theory always transform in minuscule representations of $G$, and they become degenerate in the case that there is an extremal tachyon in the spectrum. This is just as in the $\mathcal{A}_I$ theory. From the discussion above, maximal symmetry enhancements cannot exhibit this degeneration, although in principle it could happen purely at the massless level. In any case, we see from Table \ref{tab:BI8D} that there is no such degeneration for the massless fermions. 

As with the \BIIa{} theory we find all kinds of scalar fields, except for the generic tachyon (cf. Table \ref{tab:states}). We use exactly the same notation as above. 

\subsubsection{Comment on the fundamental group $\pi_1$}
Both the \BIIb{} and \BIIa{} theories in $D = 9$ as well as the \BI{} theory in $D = 8$ exhibit the special feature that for any symmetry enhancement $G$, the elements in the fundamental group $\pi_1(G)$ are exactly in correspondence with the minuscule representations in which the rest of the massless spectrum transforms. For the first two theories the correspondence is with cospinors while for the \BI{} theory it involves all spinors and massless states, see Tables \ref{tab:BIIb9D}, \ref{tab:BIIa9D} and \ref{tab:BI8D}. 

For the two $D = 9$ theories this can be understood by first recalling that every symmetry enhancement in the $D = 9$ CHL string is simply-connected. Any $G$ in the non-supersymmetric theory has a root lattice $L$ which is a sublattice of some $L'$ in the CHL string, and the quotient $L'/L$ defines the minuscule representation in which the cospinor transforms. Now use eq. \eqref{top1}, 
\begin{equation}
	\pi_1(G) = (P(L,\Upsilon_{1,9}))^*/L^\vee\,,
\end{equation}
and note that, since $\Upsilon_{1,9}$ is the same in both the parent and orbifold theory, $(P(L,\Upsilon_{d,d+8}))^*$ is the same in both cases. For ADE gauge groups, $L^\vee = L$ and so whenever $L'$ is reduced to $L$, $\pi_1(G') = 1$ is enlarged to $\pi_1(G) = L'/L$. 

The situation for the $D = 8$ \BI{} theory is more involved. Written as a shift-orbifold of the CHL string, the charge lattice is enlarged in such a way as to accomodate two inequivalent sets of minuscule representations for spinors as well as an extra lattice conjugacy class for scalars. These three sets then contribute to the full form of $\pi_1(G)$. We leave this as a curious observation. 

More important is the fact that, as can be checked from our results, all of the fundamental groups $\pi_1(G)$ in $D = 8$ satisfy the 1-form center anomaly cancellation conditions of \cite{Cvetic:2020kuw}, extending the results of the supersymmetric CHL string \cite{Cvetic:2021sjm,Font:2021uyw} to their non-supersymmetric cousins.

\subsubsection{Tachyons and stability}
Maximal enhancements form a subset of the points in moduli space which extremize the 1-loop potential, which can be regular only in the \BIIb{} and the \BI{} theories, as these have tachyon-free regions in moduli space. We have recorded in Tables \ref{tab:BIIb9D}, \ref{tab:BIIb8D} and \ref{tab:BI8D} whether the maximal enhancements are tachyon-free or not. These states generically arrange into representations of the gauge symmetry group, just as the massless scalars. Determining this information is outside the scope of this paper --- we have simply checked if there are tachyons or not. 

In this work we limit ourselves to reporting on the values of the 1-loop potential or cosmological constant (CC) at tachyon-free enhancements, see Table \ref{tab:CC}. We note however that for any such enhancement to be stable in the Narain moduli there cannot be massless scalars, since these always lead to knife-edge instabilities. This in particular rules out every enhancement we have found in the \BI{} theory as a candidate for a point of stable equilibrium. 

\begin{table}[H]
	\centering
	\begin{tabular}{|c|c|c|c|c|}
		\hline
		Theory &\# & $L$& KNF & CC \\ \hline\hline
		\BIIb{} &4&$A_1D_8$  & \xmark  & 312  \\
		\BIIb{} &3&$D_9$  & \cmark  & 308  \\\hline\hline
		\BIIb{} &10&$C_1D_9$ & \cmark & 362  \\ 
		\BIIb{} &11&$C_2D_8$ & \xmark & 260  \\  
		\BIIb{} &13&$A_1C_1D_8$ & \xmark & 366  \\ 
		\BIIb{} &14&$A_1A_1D_8$ & \cmark & 264  \\ 
		\BIIb{} &15&$A_1C_2D_7$ & \cmark & 263  \\ 
		\BIIb{} &16&$C_4D_6$ & \cmark & 264  \\ 
		\BIIb{} &21&$A_1A_1C_2D_6$ & \xmark & 262  \\ 
		\BIIb{} &24&$A_1A_3D_6$ & \cmark & 263  \\ 
		\BIIb{} &25&$A_1A_1A_2D_6$ & \xmark & 244  \\  
		\BIIb{} &27&$D_5D_5$ & \cmark & 264  \\ 
		\BIIb{} &31&$A_1C_4D_5$ & \cmark & 263  \\
		\BIIb{} &35&$A_4C_1D_5$ & \cmark & 244  \\  
		\BIIb{} &37&$A_1A_1D_4D_4$ & \xmark & 262  \\  
		\BIIb{} &59&$A_2C_4C_4$ & \xmark & 244 \\  
		\BIIb{} &60&$A_1A_1C_4C_4$ & \xmark & 262  \\  
		\BIIb{} &100&$A_1A_7C_1C_1$ & \xmark & 257  \\  
		\BI{} &3&$D_5D_5$  & \xmark  & 160  \\ 
		\hline
	\end{tabular}
	\caption{Value of the 1-loop cosmological constant for tachyon-free maximal enhancements in $D = 9$ (first two rows) and $D = 8$ (rest of rows). KNF means free of knife-edges. The CC is written in units of $(4\pi^2 \alpha')^{-9/2}$ and $(4\pi^2 \alpha')^{-4}$ respectively for $D = 9$ and $D = 8$. Entry \# refers to the number in the respective table in Appendix \ref{app:max}. We have computed these values using the same procedure as in \cite{Fraiman:2023cpa}; they are approximate and should be considered as usual as $\mathcal{O}(100)$ numbers.}
	\label{tab:CC}
\end{table}

\section{T-duality and Coxeter diagrams}
\label{s:CD}
\begin{figure}[h]
	\centering
	\begin{tikzpicture}[scale = 1.05]
		\draw(3.5,-0.5)node{$\text{II}_{1,17}$};
		\draw(0,0)--(7,0);
		\draw(0.5,0)--(0.5,1);	
		\draw(6.5,0)--(6.5,1);
		\draw[fill=white](0,0)circle(0.1);
		\draw[fill=white](0.5,0)circle(0.1);
		\draw[fill=white](1,0)circle(0.1);
		\draw[fill=white](1.5,0)circle(0.1);
		\draw[fill=white](2,0)circle(0.1);
		\draw[fill=white](2.5,0)circle(0.1);
		\draw[fill=white](3,0)circle(0.1);
		\draw[fill=BlueGreen](3.5,0)circle(0.1);
		\draw[fill=white](4,0)circle(0.1);
		\draw[fill=white](4.5,0)circle(0.1);
		\draw[fill=white](5,0)circle(0.1);
		\draw[fill=white](5.5,0)circle(0.1);
		\draw[fill=white](6,0)circle(0.1);
		\draw[fill=white](6.5,0)circle(0.1);
		\draw[fill=white](7,0)circle(0.1);
		\draw[fill=white](0.5,0.5)circle(0.1);
		\draw[fill=white](0.5,1)circle(0.1);
		\draw[fill=white](6.5,0.5)circle(0.1);
		\draw[fill=white](6.5,1)circle(0.1);
		\draw(-0.5,1)node[above]{(a)};
		\begin{scope}[shift={(9,0)}]
			\draw(2,-0.5)node{$\text{II}_{1,9}$};
			\draw(-0.5,1)node[above]{(b)};
			\draw(0,0)--(3.5,0);
			\draw(0.5,0)--(0.5,1);	
			\draw[fill=white](0,0)circle(0.1);
			\draw[fill=white](0.5,0)circle(0.1);
			\draw[fill=white](1,0)circle(0.1);
			\draw[fill=white](1.5,0)circle(0.1);
			\draw[fill=white](2,0)circle(0.1);
			\draw[fill=white](2.5,0)circle(0.1);
			\draw[fill=white](3,0)circle(0.1);
			\draw[fill=BlueGreen](3.5,0)circle(0.1);
			\draw[fill=white](0.5,0.5)circle(0.1);
			\draw[fill=white](0.5,1)circle(0.1);
		\end{scope}
		\begin{scope}[shift={(0,-4)}]
			\draw(-0.5,2)node[above]{(c)};
			\draw(1,-0.5)node{$\text{I}_{1,17}$};
			\draw(0,0)--(0.5,0.5);
			\draw(2,0)--(1.5,0.5);
			\draw(0,2)--(0.5,1.5);
			\draw(2,2)--(1.5,1.5);
			
			\draw[double, Red](0.5,0.5)--(1,0.75);
			\draw[double, Red](0.5,1.5)--(1,1.25);
			\draw[double, Red](1,0.75)--(1.5,1.5);
			\draw[double, Red, fill=white](1,1.25)--(1.5,0.5);
			\draw[double, Red](1,0.75)--(1,1.25);
			
			\draw(0,0)--(2,0)--(2,2)--(0,2)--(0,0);
			\draw[fill=white](0,0)circle(0.07);
			\draw[fill=white](0.5,0)circle(0.07);
			\draw[fill=white](1,0)circle(0.07);
			\draw[fill=white](1.5,0)circle(0.07);
			\draw[fill=white](2,0)circle(0.07);
			\draw[fill=white](0,0.5)circle(0.07);
			\draw[fill=white](0,1)circle(0.07);
			\draw[fill=white](0,1.5)circle(0.07);
			\draw[fill=white](2,0.5)circle(0.07);
			\draw[fill=white](2,1)circle(0.07);
			\draw[fill=white](2,1.5)circle(0.07);	
			\draw[fill=white](0,2)circle(0.07);
			\draw[fill=white](0.5,2)circle(0.07);
			\draw[fill=white](1,2)circle(0.07);
			\draw[fill=white](1.5,2)circle(0.07);
			\draw[fill=white](2,2)circle(0.07);
			
			\draw[fill=white](0.5,0.5)circle(0.07);
			\draw[fill=white](0.5,1.5)circle(0.07);
			\draw[fill=white](1.5,0.5)circle(0.07);
			\draw[fill=white](1.5,1.5)circle(0.07);
			\draw[Red, fill=white,double](1,1.25)circle(0.07);
			\draw[Red, fill=white,double](1,0.75)circle(0.07);
			\begin{scope}[shift={(3.5,0)}]
				\draw(1,-0.5)node{$\text{I}_{1,16}$};
				\draw[double, Red](0,0)--(0.75,0.75);
				\draw(2,0)--(1.5,0.5);
				\draw(0,2)--(0.5,1.5);
				\draw[double, Red](2,2)--(1.25,1.25);
				\draw[double, Red](0.75,0.75)--(1.25,1.25);
				
				\draw(0,0)--(2,0)--(2,2)--(0,2)--(0,0);
				\draw[fill=white](0,0)circle(0.07);
				\draw[fill=white](0.5,0)circle(0.07);
				\draw[fill=white](1,0)circle(0.07);
				\draw[fill=white](1.5,0)circle(0.07);
				\draw[fill=white](2,0)circle(0.07);
				\draw[fill=white](0,0.5)circle(0.07);
				\draw[fill=white](0,1)circle(0.07);
				\draw[fill=white](0,1.5)circle(0.07);
				\draw[fill=white](2,0.5)circle(0.07);
				\draw[fill=white](2,1)circle(0.07);
				\draw[fill=white](2,1.5)circle(0.07);	
				\draw[fill=white](0,2)circle(0.07);
				\draw[fill=white](0.5,2)circle(0.07);
				\draw[fill=white](1,2)circle(0.07);
				\draw[fill=white](1.5,2)circle(0.07);
				\draw[fill=white](2,2)circle(0.07);
				
				\draw[fill=white](0.5,1.5)circle(0.07);
				\draw[fill=white](1.5,0.5)circle(0.07);
				\draw[Red, fill=white,double](1.25,1.25)circle(0.07);
				\draw[Red, fill=white,double](0.75,0.75)circle(0.07);
			\end{scope}
			\begin{scope}[shift={(7,0)}]
				\draw(1,-0.5)node{$\text{I}_{1,15}$};
				\draw(2,0)--(1.5,0.5);
				\draw(0,2)--(0.5,1.5);
				
				\draw(0,0)--(2,0)--(2,2)--(0,2)--(0,0);
				
				\draw[Red, fill=white,double](1.5,2)--(2,2);
				\draw[Red, fill=white,double](2,1.5)--(2,2);
				
				\draw[fill=white](0,0)circle(0.07);
				\draw[fill=white](0.5,0)circle(0.07);
				\draw[fill=white](1,0)circle(0.07);
				\draw[fill=white](1.5,0)circle(0.07);
				\draw[fill=white](2,0)circle(0.07);
				\draw[fill=white](0,0.5)circle(0.07);
				\draw[fill=white](0,1)circle(0.07);
				\draw[fill=white](0,1.5)circle(0.07);
				\draw[fill=white](2,0.5)circle(0.07);
				\draw[fill=white](2,1)circle(0.07);
				\draw[fill=white](2,1.5)circle(0.07);	
				\draw[fill=white](0,2)circle(0.07);
				\draw[fill=white](0.5,2)circle(0.07);
				\draw[fill=white](1,2)circle(0.07);
				\draw[fill=white](1.5,2)circle(0.07);
				
				\draw[fill=white](0.5,1.5)circle(0.07);
				\draw[fill=white](1.5,0.5)circle(0.07);
				\draw[Red, fill=white,double](2,2)circle(0.07);
			\end{scope}
			\begin{scope}[shift={(10.5,0)}]
				\draw(1,-0.5)node{$\text{I}_{1,14}$};
				\draw(2,0)--(1.5,0.5);
				\draw(0,2)--(0.5,1.5);
				
				\draw(0,0)--(2,0)--(2,1.5);
				\draw(1.5,2)--(0,2)--(0,0);
				
				\draw[Red, fill=white,double](1.5,2)--(2,1.5);
				\draw[Red, fill=white,double](1,2)--(1.5,2);
				\draw[Red, fill=white,double](2,1)--(2,1.5);
				
				\draw[fill=white](0,0)circle(0.07);
				\draw[fill=white](0.5,0)circle(0.07);
				\draw[fill=white](1,0)circle(0.07);
				\draw[fill=white](1.5,0)circle(0.07);
				\draw[fill=white](2,0)circle(0.07);
				\draw[fill=white](0,0.5)circle(0.07);
				\draw[fill=white](0,1)circle(0.07);
				\draw[fill=white](0,1.5)circle(0.07);
				\draw[fill=white](2,0.5)circle(0.07);
				\draw[fill=white](2,1)circle(0.07);
				\draw[fill=white](2,1.5)circle(0.07);	
				\draw[fill=white](0,2)circle(0.07);
				\draw[fill=white](0.5,2)circle(0.07);
				\draw[fill=white](1,2)circle(0.07);
				\draw[fill=white](1.5,2)circle(0.07);
				
				\draw[fill=white](0.5,1.5)circle(0.07);
				\draw[fill=white](1.5,0.5)circle(0.07);
				\draw[Red, fill=white,double](1.5,2)circle(0.07);
				\draw[Red, fill=white,double](2,1.5)circle(0.07);
			\end{scope}
		\end{scope}
		
	\end{tikzpicture}
	\caption{Coxeter diagrams representing the reflection symmetries of the even self-dual lattices $\text{II}_{1,17}$ and $\text{II}_{1,9}$ as well as the odd self-dual lattices $\text{I}_{1,17},...,\text{I}_{1,14}$. These correspond respectively to (a) the two 10D supersymmetric heterotic strings on $S^1$, (b) the 9D CHL string and (c) the 10D non-supersymmetric heterotic strings with rank 16 gauge group on $S^1$ and three related subcritical strings. In each case the nodes generate the Weyl subgroup of the T-duality group, with outer automorphisms corresponding to symmetries of the diagrams themselves. All of these diagrams were originally constructed by Vinberg \cite{Vinberg:1972}.}
	\label{fig:CoxeterKnown}
\end{figure}
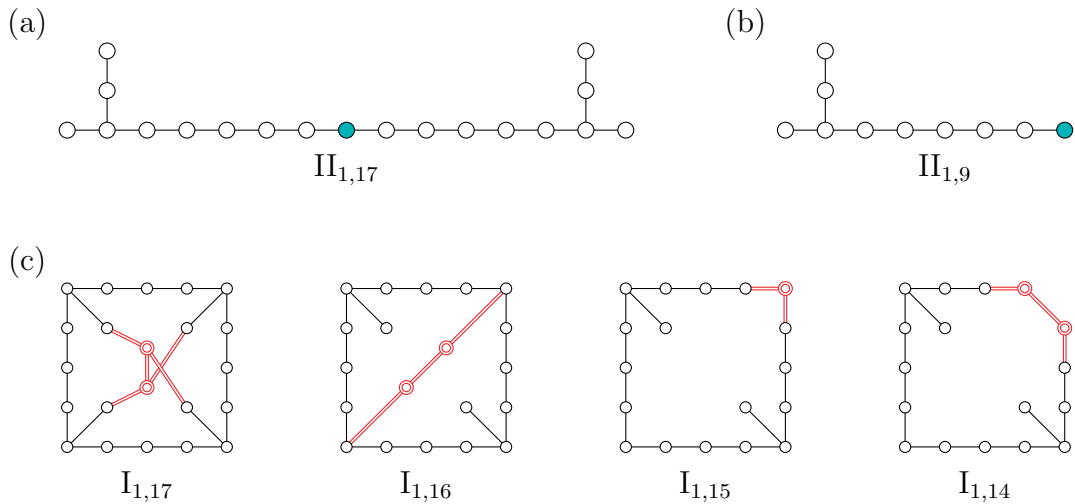

Coxeter diagrams represent the fundamental domain of a hyperbolic space modded by some discrete reflective symmetry group $\Gamma$, i.e. a Coxeter polyhedron. Heterotic strings with 9D Minkowski target space have moduli spaces precisely of this form, where $\Gamma$ is the T-duality symmetry group. Indeed, it has been known for quite some time that the supersymmetric heterotic strings compactified on $S^1$ have a moduli space described by the Coxeter diagram shown in Figure \ref{fig:CoxeterKnown} (a). Similarly, the moduli space of the CHL string is described by the diagram in Figure \ref{fig:CoxeterKnown} (b). In these two cases, the nodes in the diagram represent the codimension 1 boundaries of the fundamental domain, and it is at such loci that the spectrum undergoes a symmetry enhancement $U(1) \to SU(2)$. These diagrams, therefore, encode every possible symmetry enhancement in 9D. We refer to \cite{Cachazo:2000ey,Font:2020rsk} for detailed explanations.\footnote{In the literature, these diagrams are also referred to as extended Dynkin diagrams, generalized Dynkin diagrams or Coxeter-Dynkin diagrams.} 

Likewise, non-supersymmetric heterotic strings of maximal rank compactified on $S^1$ share a tree-level moduli space described by the first Coxeter diagram shown in Figure \ref{fig:CoxeterKnown} (c). In this case there are two special walls at which a pair of tachyons in the spectrum acquire their minimal squared mass $m^2 = -2$. There is a corresponding enhancement of a T-duality symmetry sometimes referred to as thermal T-duality in the context of finite temperature models \cite{Atick:1988si}. These tachyons are in correspondence with worldsheet marginal deformations realizing a process of tachyon condensation aided by a lightlike linear dilaton \cite{Hellerman:2006ff,Kaidi:2020jla}, and flowing to subcritical heterotic strings with moduli spaces encoded in related Coxeter diagrams, three of which are shown in Figure \ref{fig:CoxeterKnown} (c), see \cite{DeFreitas:2024yzr} for details. 

In all of these cases, the nodes in the diagrams correspond to Weyl reflections which generate the reflective subgroup of the T-duality group of the theory, which in turn is the automorphism group of the lattice of electric charges.\footnote{Non-reflective operations, i.e. outer automorphisms, correspond to symmetries of the diagram itself.} This allows a clean determination of the Coxeter diagram in each case \cite{DeFreitas:2024yzr}. As we will now show, the rank reduced theories also have tree-level moduli spaces with such a description, but their determination is not as straightforward. 

With these diagrams at hand we can determine with complete control every non-Abelian symmetry enhancement in the 9D theory. Moreover, these diagrams also encode in a clean manner the different decompactification limits in the form of affine Dynkin subdiagrams, providing a neat visualization of the T-duality relations among the different freely acting constructions.

\subsection{\BIII{}}
Let us start with the $S^1$ compactification of the $E_8$ string, i.e. the \BIII{} string. The simplest way to obtain the Coxeter diagram is by folding the one in Figure \ref{fig:CoxeterKnown} (a), resulting in the diagram for the  $CE_{10}$ Coxeter group:
\begin{eqnarray}\label{diag:e8s1}
	\begin{tikzpicture}[scale = 1.5]
		\usetikzlibrary{arrows.meta}
		\draw(0.5,0)--(0.5,1);
		\draw(0,0)--(3,0);
		\draw[double](3,0)--(3.5,0);
		\draw[double,Straight Barb-](3.2,0)--(3.5,0);
		
		\draw[fill=white](0,0)circle(0.07)node[below=0.05in]{\scriptsize{\textcolor{Red}{\texttt{1}}}};
		\draw[fill=white](0.5,0)circle(0.07)node[below=0.05in]{\scriptsize{\textcolor{Red}{\texttt{2}}}};
		\draw[fill=white](1,0)circle(0.07)node[below=0.05in]{\scriptsize{\textcolor{Red}{\texttt{3}}}};
		\draw[fill=white](1.5,0)circle(0.07)node[below=0.05in]{\scriptsize{\textcolor{Red}{\texttt{4}}}};
		\draw[fill=white](2,0)circle(0.07)node[below=0.05in]{\scriptsize{\textcolor{Red}{\texttt{5}}}};
		\draw[fill=white](2.5,0)circle(0.07)node[below=0.05in]{\scriptsize{\textcolor{Red}{\texttt{6}}}};
		\draw[fill=white](3,0)circle(0.07)node[below=0.05in]{\scriptsize{\textcolor{Red}{\texttt{0}}}};
		\draw[fill=white](3.5,0)circle(0.07)node[below=0.05in]{\scriptsize{\textcolor{Red}{\texttt{c}}}};
		
		\draw[fill=white](0.5,0.5)circle(0.07)node[left=0.05in]{\scriptsize{\textcolor{Red}{\texttt{7}}}};
		\draw[fill=white](0.5,1)circle(0.07)node[left=0.05in]{\scriptsize{\textcolor{Red}{\texttt{8}}}};
	\end{tikzpicture}
\end{eqnarray}
The node \texttt{c} corresponds to a long root, invariant under the folding. We see that the rank 9 Dynkin subdiagrams correspond exactly with the results obtained with the exploration algorithm in Section \ref{ss:exploration}, see Table \ref{tab:BIII9D}. 

From the Coxeter diagram we can infer that the T-duality group of the theory is exactly the automorphism group of the charge lattice $\Gamma_{1,1}\oplus E_8(\tfrac12)$. First we scale the lattice by 2 and use the isomorphism
\begin{equation}
	\Gamma_{1,1}(2)\oplus E_8 \simeq \Gamma_{1,1}\oplus D_8\,.
\end{equation} 
The group of automorphisms of $D_8$ acts on the weight lattice $D_8^*$ mapping the vector class to itself and possibly trading the spinor and cospinor classes. It follows that extending $D_8$ to $\mathbb{Z}^8$ by adding sites in the vector class preserves the automorphism group. The full charge lattice is correspondingly extended to the odd self-dual lattice $\text{I}_{1,9}$, and it turns out that its automorphism group is the Coxeter group encoded in the diagram above \cite{Vinberg:1972}. The T-duality group is thus
\begin{equation}
	\Theta(1,9;\mathbb{Z}) = \text{Aut}(\Gamma^v_{1,9})\,,
\end{equation}
where we emphasize the use of the vector class lattice, as it encodes the symmetry enhancements appearing at points fixed under T-duality reflections. This form of the T-duality group is kept in compactifications to lower dimensions just as for the CHL string \cite{Mikhailov:1998si}.

The Coxeter diagram \eqref{diag:e8s1} has two affine Dynkin subdiagrams associated to decompactification limits. Deleting note \texttt{c} we obtain the diagram $\widehat E_8$, corresponding to decompactification to the 10D $E_8$ string. Deleting node \texttt{8} instead we get the diagram $\widehat{B}_8^\vee$, where $\vee$ denotes an exchange of long roots with short roots (Langlands dual). It corresponds to the decompactification to the $U(16)$ heterotic string, with twisted affine algebra $A_{15}^{(2)}$ as can be seen by a folding procedure \cite{kac}. This gives a clean example of a twisted affine Lie algebra corresponding to decompactification with rank enhancement \cite{Collazuol:2024kzl}, with the difference that the twist is visible at the level of the Dynkin diagram.

Deleting both nodes \texttt{c} and \texttt{8} we fix a 1-parameter moduli space interpolating between the two decompactification limits. As we approach the $U(16)$ limit, the circle compactification of the $E_8$ string is T-dualized to the $U(16)$ string on $S^1$ with a twist $U(16)\to Sp(8)$ (we omit gauge group topology).

It is also instructive to see how the diagram \eqref{diag:e8s1} arises through the following naive procedure. Start with the Dynkin diagram of the 10D gauge group $E_8$ and make it affine, with the lowest root having a negative unit KK momentum charge. Since $E_8$ is at level 2, all of these roots are short. Then add a long root corresponding to the level 1 $SU(2)$ enhancement at self-dual radius. In other words, the diagram encodes the combined effect of symmetry breaking by Wilson lines and stringy symmetry enhancement.

Can we transpose this procedure starting from the $U(16)$ string? Yes --- if we account for an important caveat. In the $U(16)$ orbifold frame, one can think of the canonical gauge symmetry group as $Sp(8)/\mathbb{Z}_2$. However, the \textit{structure} group of the gauge bundle over $S^1$, which governs the symmetry enhancement patterns, is the Langlands dual $Spin(17)$ \cite{Lerche:1997rr}. Hence we affinize the $B_8$ diagram and add a long root, and then take the Langlands dual, and get the correct diagram.   

\subsection{\BIIb{}}
There are various ways of obtaining the Coxeter diagram for the \BIIb{} theory. The easiest is by starting in the $SO(16)\times SO(16)$ orbifold frame where the canonical gauge group is $Spin(16)$ at level 2. Since this gauge group is simply-laced, we take the affine $\widehat{D}_8$ with long roots, add an extra long root to the lowest root and then make all the roots short. This produces the root system of the $DE_{10}$ Coxeter group, with diagram
\begin{eqnarray}\label{diag:BIIbs1}
	\begin{tikzpicture}[scale = 1.5]
		\usetikzlibrary{arrows.meta}
		\draw(0.5,0)--(0.5,1);
		\draw(0,0)--(3,0);
		\draw(2.5,0)--(2.5,0.5);
		\draw[double](3,0)--(3,0);
		
		\draw[fill=white](0,0)circle(0.07)node[below=0.05in]{\scriptsize{\textcolor{Red}{\texttt{1}}}};
		\draw[fill=white](0.5,0)circle(0.07)node[below=0.05in]{\scriptsize{\textcolor{Red}{\texttt{2}}}};
		\draw[fill=white](1,0)circle(0.07)node[below=0.05in]{\scriptsize{\textcolor{Red}{\texttt{3}}}};
		\draw[fill=white](1.5,0)circle(0.07)node[below=0.05in]{\scriptsize{\textcolor{Red}{\texttt{4}}}};
		\draw[fill=white](2,0)circle(0.07)node[below=0.05in]{\scriptsize{\textcolor{Red}{\texttt{5}}}};
		\draw[fill=white](2.5,0)circle(0.07)node[below=0.05in]{\scriptsize{\textcolor{Red}{\texttt{6}}}};
		\draw[fill=white](3,0)circle(0.07)node[below=0.05in]{\scriptsize{\textcolor{Red}{\texttt{0}}}};
		\draw[fill=white](2.5,0.5)circle(0.07)node[right=0.05in]{\scriptsize{\textcolor{Red}{\texttt{c}}}};
		
		\draw[fill=white](0.5,0.5)circle(0.07)node[left=0.05in]{\scriptsize{\textcolor{Red}{\texttt{7}}}};
		\draw[fill=white](0.5,1)circle(0.07)node[left=0.05in]{\scriptsize{\textcolor{Red}{\texttt{8}}}};
	\end{tikzpicture}
\end{eqnarray}
Again, we check that the rank 9 Dynkin diagrams match the results of the exploration algorithm, cf. Table \ref{tab:BIIb9D}. 

The T-duality group of the theory is not, in this case, the automorphism group of the vector class lattice $\Gamma_{1,1}\oplus E_8(\tfrac12)$. This is because the norm 2 vector in $\Gamma_{1,1}$ generates a Weyl reflection which, unlike in the \BIII{} theory above, does not leave the spectrum invariant. Even though this reflection is an automorphism of the vector class lattice, it is not an automorphism of the full charge lattice $\Gamma_{1,1}(\tfrac12)\oplus E_8(\tfrac12)$. We write it as
\begin{equation}\label{tdualityBIIb}
	\Theta(1,9;\mathbb{Z}) = \text{Aut}^+(\Gamma_{1,9}^v) \equiv \text{Aut}(\Gamma_{1,1}\oplus E_8) \cap \text{Aut}(\Gamma_{1,1}(\tfrac12)\oplus E_8)\,.
\end{equation} 
This particular property of the theory is traced back to the fact that there are no reflective vectors in the twisted sector of the theory when constructed as an orbifold of either the $E_8\times E_8$ or the $E_8$ string on $S^1$, while the $SU(2)$ enhancement at self-dual radius (and its associated T-duality symmetry) is projected out in both circle compactifications.

The group in \eqref{tdualityBIIb} is exactly the Coxeter group encoded in the diagram \eqref{diag:BIIbs1}, with a caveat. This diagram has an outer automorphism exchanging nodes \texttt{c} and \texttt{0}, which is in fact the same transformation generated by the Weyl reflection that the orbifold projected out. This is precisely as it should be, since this operation trades the two $\widehat{E}_8$ affine subdiagrams corresponding to decompactification to the $E_8\times E_8$ string and the $E_8$ string in 10D. We must then be careful when using the Coxeter diagram; one should mark it to break the diagram symmetry explicitly. The $\widehat{D}_8$ subdiagram of course corresponds to decompactification to the $SO(16)\times SO(16)$ string. 

There are in this theory three 1-parameter moduli spaces interpolating between distinct 10D theories at infinite distance. These are obtained by deleting two out of the three nodes \texttt{0}, \texttt{t} and \texttt{8}. 

\subsection{\BIIa{}}
Finally, the diagram for the \BIIa{} theory can be constructed in the $(E_7\times SU(2))^2$ string orbifold frame, joining the diagrams $\widehat{E}_7$ and $\widehat{A}_1$ by an extra node associated to a short root:
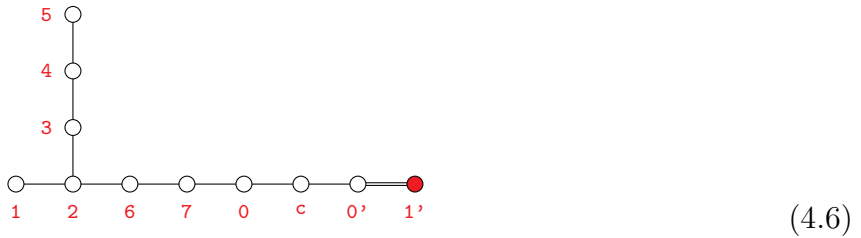
\begin{eqnarray}\label{diag:BIIas1}
	\begin{tikzpicture}[scale = 1.5]
		\usetikzlibrary{arrows.meta}
		\draw(0.5,0)--(0.5,1.5);
		\draw(0,0)--(3,0);
		\draw[double](3,0)--(3.5,0);
	
		\draw[fill=white](0,0)circle(0.07)node[below=0.05in]{\scriptsize{\textcolor{Red}{\texttt{1}}}};
		\draw[fill=white](0.5,0)circle(0.07)node[below=0.05in]{\scriptsize{\textcolor{Red}{\texttt{2}}}};
		\draw[fill=white](1,0)circle(0.07)node[below=0.05in]{\scriptsize{\textcolor{Red}{\texttt{6}}}};
		\draw[fill=white](1.5,0)circle(0.07)node[below=0.05in]{\scriptsize{\textcolor{Red}{\texttt{7}}}};
		\draw[fill=white](2,0)circle(0.07)node[below=0.05in]{\scriptsize{\textcolor{Red}{\texttt{0}}}};
		\draw[fill=white](2.5,0)circle(0.07)node[below=0.05in]{\scriptsize{\textcolor{Red}{\texttt{c}}}};
		\draw[fill=white](3,0)circle(0.07)node[below=0.05in]{\scriptsize{\textcolor{Red}{\texttt{0'}}}};
		\draw[fill=Red](3.5,0)circle(0.07)node[below=0.05in]{\scriptsize{\textcolor{Red}{\texttt{1'}}}};
		
		\draw[fill=white](0.5,0.5)circle(0.07)node[left=0.05in]{\scriptsize{\textcolor{Red}{\texttt{3}}}};
		\draw[fill=white](0.5,1)circle(0.07)node[left=0.05in]{\scriptsize{\textcolor{Red}{\texttt{4}}}};
		\draw[fill=white](0.5,1.5)circle(0.07)node[left=0.05in]{\scriptsize{\textcolor{Red}{\texttt{5}}}};
	\end{tikzpicture}
\end{eqnarray}
The node representing the $SU(2)$ in the canonical gauge group $E_7\times SU(2)$ is colored in red to signify that this enhancement comes with two extra tachyonic states (cf. Section \ref{sss:BIIa}). Note that to get a valid symmetry enhancement we are forced to delete either the node \texttt{0'} or the node \texttt{1'}. We also see an affine subdiagram $\widehat{E}_8$, corresponding to decompactification to the $E_8$ string. Again, from the $E_8$ string orbifold frame the naive diagram construction is not sufficient. This Coxeter diagram is of pyramid type with $n+2 = 11$ facets in $H_n = H_9$ hyperbolic space, corresponding to the last entry in Table 8 of \cite{Tumarkin} with $k = 2$. 

The T-duality group of this theory is again a congruence subgroup of that of the CHL string, since there are no twisted states associated to new $\mathbb{Z}_2$ symmetry enhancements. The enhancements match those obtained with the exploration algorithm, see Table \ref{tab:BIIa9D}.

\subsection{T-dual string frames}
\label{ss:frames}
From the above Coxeter diagrams we have seen what are the decompactification limits from $D = 9$ to $D = 10$ in three non-supersymmetric theories. In each asymptotic regime, the theory is described as a freely acting orbifold of the limit theory, and for each moduli space the set of these orbifolds are T-dual. We confirm in particular the T-duality between the $\delta \cdot \theta_L$ orbifold of the $SO(16)\times SO(16)$ string and the $\delta\cdot \theta_L (-1)^F$ orbifold of the $E_8\times E_8$ string, as well as the T-duality between the $\delta\cdot \theta_L$ orbifold of the $E_7\times SU(2)\times E_7 \times SU(2)$ string and the Scherk-Schwarz reduction of the $E_8$ string, which were argued for by comparing partition functions in \cite{DeFreitas:2024ztt}. We learn also that the $\delta \cdot \theta_L$ orbifold of the $U(16)$ string is T-dual to the $S^1$ compactification of the $E_8$ string. 

Strikingly, the three $D = 9$ are all realized as freely acting $\delta \cdot \theta_L$ orbifolds of non-supersymmetric strings! In fact, we can also define the \BI{} theory in $D = 8$ as the $\delta \cdot \theta_L$ orbifold of the Scherk-Schwarz reduction of the $E_8\times E_8$ string. The four theories thus descend from the $\mathcal{A}_I$ by a construction completely analogous to that of the CHL string, giving yet another argument for treating them as cousins. More concretely, there are four inequivalent outer automorphisms $\theta_L$ of the charge lattice of the $\mathcal{A}_I$ string, each of which defines one of the four rank reduced theories.

One can of course ask what are the different T-duality frames in lower dimensions. In the supersymmetric case, we find that the CHL string in $D = 8$ is T-dual to the $Spin(32)/\mathbb{Z}_2$ string on a $T^2$ without vector structure \cite{Witten:1997bs}, i.e. with a pair of flat holonomies commuting to $\pi_1(G) = \mathbb{Z}_2$. We can also play this game with the 10D non-supersymmetric theories with $\pi_1(G) = \mathbb{Z}_2$, namely the $SO(16)\times SO(16)$ string, the $E_7\times SU(2)\times E_7 \times SU(2)$ string, the $U(16)$ string and the $SO(8)\times SO(24)$ string \cite{Fraiman:2023cpa}. The action of the holonomies breaks the gauge groups to $Sp(4)\times Sp(4)$, $F_4\times F_4$, $U(8)$ and $Sp(2)\times Sp(6)$. Comparing with our results, these theories must be respectively \BIIb{} (long roots, no generic tachyon), \BIII{} ($F_4$ enhancement), \BI{} (no generic tachyon, two tachyons charged under $U(1) \subset U(8)$) and \BIIa{} ($Sp(2) \simeq SO(5)$ with tachyons in the \textbf{5}). 

We see again that the four theories can be all constructed in a democratic way, this time using holonomy doubles. Curiously, we can interpret these results as ``explaining" why there are four non-supersymmetric theories in 10D with non-trivial $\pi_1(G)$, but only three with non-trivial $\pi_0(G)$, since these homotopy groups are correlated respectively with $D = 8$ and $D = 9$ orbifolds. There are potentially many more constructions allowed, even more so as we go to lower dimensions, and we expect all of them to fall again into one of the four theories as above.

\section{S-duality}
\label{s:duality}
In Section \ref{s:intro} we have anticipated the existence of an orientifold dual for the \BIIb{} theory. Here we show how this duality follows from the adiabatic argument of \cite{Vafa:1995gm}, and verify explicitly that the perturbative spectrum in the orientifold matches exactly with that of the heterotic string. We then use this duality to explain why certain states in the \BIIb{} theory are seemingly arranged into $\mathcal{N} = 1$ vectormultiplets.

\subsection{Review of supersymmetric case in 8D}
We start with the supersymmetric duality of HO and type I. Compactify both sides on $T^2$ and turn on a flat connection characterized by a non-trivial Stiefel-Whitey class valued on the $-1$ element in the fundamental group of $Spin(32)/\mathbb{Z}_2$. On the heterotic side this connection is realized by a pair of anticommuting holonomies, while on the type I side it is realized by turning on the 2-torsional NSNS B-field resulting from the orientifold operation \cite{Bianchi:1991eu}. In other words, compactify both HO and type I on a $T^2$ without vector structure \cite{Witten:1997bs}. 

The heterotic string on $T^2$ is described at large 8D coupling by type IIB on a $T^2/\mathbb{Z}_2$ orientifold, obtained by T-dualizing both directions in type I on $T^2$. Turning on the torsional 2-form in the type I torus, T-duality yields type IIB on $T^2/\mathbb{Z}_2$ with three $O7^-$-planes and one $O7^+$-plane as well as eight $D7$-branes plus their reflections. 

The heterotic Wilson lines are mapped exactly to the type I Wilson lines, which are in turn mapped under T-duality to the positions of the $D7$-branes. For generic values of the $T^2$ moduli and with zero Wilson lines, the gauge algebra is $\mathfrak{sp}(8)$. Turning on a Wilson line
\begin{equation}
	A_1 = (\tfrac12^n,0^{8-n})
\end{equation} 
projects out the massless short roots in the $Spin(17)$ structure group, breaking it to $Spin(2n)\times Spin(2(8-n)+1)$. The gauge algebra is hence $\mathfrak{so}(2n)_2\oplus \mathfrak{sp}(8-n)_1$ where the subscript denotes the current algebra level. The effect of the second Wilson line $A_2$ is analogous, and we see that the components of both Wilson lines correspond to the positions of the $D7$-branes with fixed points at $(0,0), ~(\tfrac12,0),~(0,\tfrac12),~(\tfrac12,\tfrac12)$; the $O7^+$ sits of course at $(0,0)$. 

The full pattern of heterotic gauge symmetry enhancements, which involves winding states, is reproduced in the orientifold dual by lifting it to F-theory on an elliptically fibered K3 surface with a frozen singularity, or equivalently in terms of type IIB string junctions \cite{Cvetic:2022uuu}.

\subsection{Supersymmetry breaking}
Let us now come back to the duality between HO and type I, and now take the $T^2$ without vector structure to have antiperiodic Spin structure along one of the 1-cycles. This flip can be realized as a $(-1)^F$ holonomy, and so by the adiabatic argument both theories are still dual to each other although supersymmetry is broken. 

On the heterotic side we have a torus supporting two holonomies, $g_1 = g$ and $g_2 = g' (-1)^F$ such that $gg' = -g'g$ (note that also $g_1g_2 = -g_2g_1$). We realize the holonomies $g$ and $g'$ as usual, with $g$ breaking $\mathfrak{so}(32)$ to $\mathfrak{u}(16)$ and $g'$ twisting $\mathfrak{u}(16)$ into $\mathfrak{sp}(8)$. The first holonomy corresponds to a discrete jump in moduli space towards a locus with worldsheet gobal symmetry $\theta_L = g'$, and so $g_2$ is realized by orbifolding the theory by $\theta_L (-1)^F$ together with a half-period shift along the second torus direction. It follows that turning the $(-1)^F$ holonomy yields the \BIIb{} theory described as an orbifold of the HO theory. 

We focus on the untwisted sector, since stringy symmetry enhancements are not visible in the orientifold. Bosonic states behave exactly as in the supersymmetric case since $(-1)^F = 1$ on them. For fermionic states, consider first the case with $\mathfrak{sp}(8)$ gauge algebra.  The projection $(1+g_2)/2$ preserves the antisymmetric combinations of $\mathfrak{u}(16)$ states and furnishes the antisymmetric rep of $\mathfrak{sp}(8)$, splitting into its irreducible traceless part and its trace. Turning on a Wilson line in the direction carrying $(-1)^F$, of the form
\begin{equation}
	A_2 = (\tfrac12^n,0^{8-n})\,,
\end{equation}
flips the sign of the projection on the long roots, so that the gauge algebra is now $\mathfrak{so}(2n)_2\oplus \mathfrak{sp}(8-n)_1$ with the orthogonal group carrying a fermion in the symmetric representation.  Turning on 
\begin{equation}
	A_1 = (\tfrac12^n,0^{8-n})\,,
\end{equation}
on the other hand, gives  $\mathfrak{so}(2n)_2\oplus \mathfrak{sp}(8-n)_1$ with the orthogonal part supporting a spinor in the adjoint. 

The combined effect of the holonomies is reproduced in the open string spectrum of the orientifold if one flips the sign of the projection on fermions for the $O7^+$ and one $O7^-$. This corresponds to conjugating them to anti-O-planes $\overline{O7}^+$ and $\overline{O7}^-$. The resulting model was studied rather recently in \cite{Coudarchet:2021qwc}, motivated by the fact that the conjugation of a pair $Op^+$-$Op^-$ preserves the overall NSNS and RR tadpole, thus unlike the standard models with brane-supersymmetry-breaking, both of these tadpoles vanish.  

In \cite{Coudarchet:2021qwc} the above orientifold was constructed by observing that conjugating the charges of O-planes in that way one preserved the cancellation of local tadpoles. One can do the same without an $O7^+$-plane if one conjugates eight $D7$-branes.\footnote{The resulting model is dual to those of \cite{Horava:2007hg} and \cite{Antoniadis:1998ki}.} Although we expect the branes and antibranes to come together and decay, we can still ask if there exists is a suitable dual heterotic string to this configuration. 

In a ``well-behaved" dual pair we would expect the D3-brane wrapping the $T^2/\mathbb{Z}_2$ to reduce to a heterotic string soliton.\footnote{In the non-supersymmetric setup of the Scherk-Schwarz reduction of the type I string this was worked out in detail in \cite{Blum:1997cs,Blum:1997gw}.} Now recall that open strings going from $D7$-branes to D3-branes furnish holomorphic fermions on the soliton's worldsheet, by T-dualizing the D1-D9 brane analysis of \cite{Polchinski:1995df}. Anti-$D7$-branes however lead to anti-holomorphic fermions, and so the soliton does not correspond to a critical heterotic string. However if we bring the anti-$D7$-branes to an anti-$O7^-$-plane and trade the stack for an anti-$O7^+$-plane, this problem disappears. Moreover we freeze the degrees of freedom associated to the instability from having both branes and anti-branes. 

It is also instructive to compare this setup with the Sugimoto string \cite{Sugimoto:1999tx}, realized as an orientifold of the type IIB string in 10D with an $O9^+$-plane and 32 anti-D9-branes. The gauge symmetry algebra is $\mathfrak{sp}(16)$ (the full gauge group is likely $Sp(16)/\mathbb{Z}_2$ \cite{Larotonda:2024thv}) and one can ask if, as for the usual type I string, there is a heterotic dual. It is straightforward to see that this cannot work, because the central charge of an $\mathfrak{sp}(16)$ worldsheet current is too large, $c_L = 29+\tfrac13 > 16$. By having a reduced gauge symmetry rank, the above orientifold also circumvents this problem, since for $\mathfrak{sp}(8)$ we have $c_L = 13.6 < 16$. 

As we have explained the adiabatic argument automatically matches the orientifold perturbative states with the $w = 0$ sector of the heterotic string. It is natural to ask if this matching extends to the winding - non-perturbative sector, as happens in the supersymmetric case. We leave this problem for future work. It is very interesting to note however that this duality gives a nice explanation for the presence of a Bose-Fermi degenerate subsector in the heterotic string thanks to the mutually BPS branes in the orientifold. We should emphasize that in this sector the degeneracy carries to winding states which are not visible perturbatively in the orientifold. It would be quite interesting to explore the consequences of having this sector for the overall theory and if it is related to its stability properties.

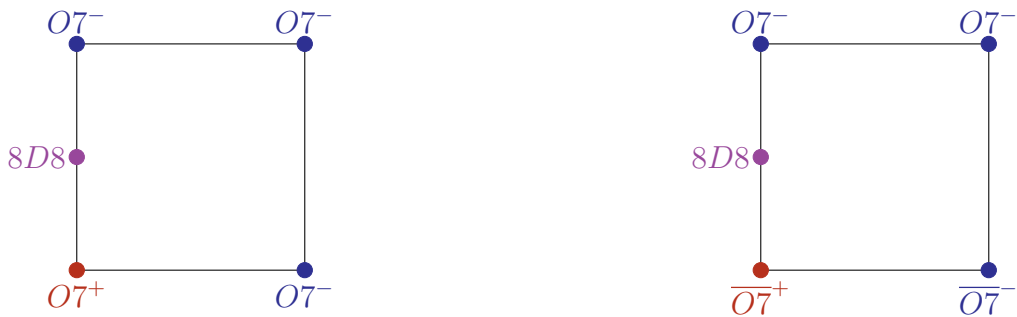
\begin{figure}
	\centering
	\begin{tikzpicture}
		\draw(0,0)--(0,3)--(3,3)--(3,0)--(0,0);
		\draw(9,0)--(9,3)--(12,3)--(12,0)--(9,0);

		\draw[fill=BrickRed,BrickRed](9,0)circle(0.1)node[below]{$\overline{O7}^+$};
		\draw[fill=Blue,Blue](12,0)circle(0.1)node[below]{$\overline{O7}^-$};
		\draw[fill=Blue,Blue](12,3)circle(0.1)node[above]{$O7^-$};
		\draw[fill=Blue,Blue](9,3)circle(0.1)node[above]{$O7^-$};

		\draw[fill=BrickRed,BrickRed](0,0)circle(0.1)node[below]{${O7}^+$};
		\draw[fill=Blue,Blue](3,0)circle(0.1)node[below]{${O7}^-$};
		\draw[fill=Blue,Blue](3,3)circle(0.1)node[above]{$O7^-$};
		\draw[fill=Blue,Blue](0,3)circle(0.1)node[above]{$O7^-$};

		\draw[Purple, fill = Purple](0,1.5)circle(0.1)node[left]{$8D8$};
		\draw[Purple, fill = Purple](9,1.5)circle(0.1)node[left]{$8D8$};

	\end{tikzpicture}
	\caption{Orientifold S-duals of the supersymmetric CHL string and the non-supersymmetric \BIIb{} string in $D = 8$. Supersymmetry breaking in this frame corresponds to conjugation of the bottom $Op$-plane RR charges.}
	\label{fig:Sdual}
\end{figure}

\section{Discussion}
\label{s:disc}
In this paper we have analysed four different tree-level moduli spaces associated to non-supersymmetric heterotic strings with rank reduced by 8, which can be thought of as the non-supersymmetric cousins of the CHL string. We have determined their 1-loop partition functions in a canonical form which facilitates studying their spectra, lattice structures and T-duality groups. We have then used an exploration algorithm to determine their maximal symmetry enhancements in $D = 9,8$, computing as well the fundamental group $\pi_1(G)$ for each enhanced gauge group $G$ and the rest of the massless spectrum. We have checked that the $\pi_1(G)$'s in every theory satisfy the anomaly cancellation constraint of \cite{Cvetic:2020kuw}. 

Specializing to the $D = 9$ case we have also determined the Coxeter diagrams that encode the global structure of the three corresponding moduli spaces, making transparent the allowed symmetry enhancements as well as decompactification limits, veryfing various T-duality pairs proposed in \cite{DeFreitas:2024ztt} and finding others.

Finally we have used the adiabatic argument \cite{Vafa:1995gm} together with T-duality to construct an orientifold dual to one of the four theories in $D = 8$, and shown that this theory enjoys many properties that single it out as particularly well behaved in terms of duality and stability. We have also used this S-duality to interpret a Bose-Fermi degenerate subsector in the heterotic string as corresponding to open strings ending on mutually BPS $D7$-branes.

The tools we have developed in this paper may be adapted to other types of heterotic orbifolds, specially those obtained by gauging order 2 symmetries. For example there is a $\mathbb{Z}_2$ orbifold in $D = 6$ where the lattice automorphism is anomalous \cite{Aldazabal:2025zht}, for which there are a few non-supersymmetric cousins in the same dimension predicted in \cite{DeFreitas:2024ztt}. One may also consider a $\mathbb{Z}_2$ right-moving operation in $D = 6$ and combine it with the usual CHL operation used in this paper, obtaining models with eight \cite{Baykara:2023plc} or zero \cite{Angelantonj:2024jtu} supercharges in $D \leq 6$. In both cases the spectrum splits into different classes which can be treated as we have done here. $\mathbb{Z}_2$ orbifolds are singled out in that they are compatible with fermionic formulations such as in the original work of CHL \cite{Chaudhuri:1995bf}, as well as orientifold descriptions such as in \cite{Witten:1997bs}. 

Another avenue for research is in understanding the role of RR charge conjugation in supersymmetry breaking in the proposed orientifold dual. In a more general setup given by type IIB with $(p,q)$-7-branes this procedure might correspond to conjugating the uplifts of the two $O_p$-planes, which suggests that reflection 7-branes \cite{Dierigl:2022reg,Heckman:2025wqd} could play a role in understanding this particular background.  

Finally, it would certainly be interesting to see how the results and techniques of \cite{Basile:2023knk,Tachikawa:2024ucm,Basile:2025mnj}, concerning topological aspects of the $SO(16)\times SO(16)$, could be applied in the setups we have studied, specially in the \BIIb{} theory.

\section*{Acknowledgements}
We thank Anamaria Font, Miguel Montero, Salvatore Raucci, Cumrun Vafa and Fengjun Xu for useful discussions. BF is supported by a Juan de la Cierva contract (JDC2023-050850-I) from Spain’s Ministry of Science, Innovation and Universities. HPF is supported by a grant from the Simons Foundation (602883,CV), the DellaPietra Foundation and by the NSF grant PHY2013858.

\appendix

\section{Details on 1-loop partition functions}\label{app:Z}

In this appendix we derive the master formula  \eqref{master} for the 1-loop partition functions of the heterotic theories with rank reduced by 8. 

\subsection{Supersymmetric CHL string}
The supersymmetric CHL string is constructed by orbifolding the $E_8\times E_8$ heterotic string on $S^1$ by $g = \theta_L \delta$, and its 1-loop partition function takes the standard form
\begin{equation}\label{std}
	Z_\text{unt}(\tau, \bar \tau) \equiv \frac{1}{2}(Z_{1,1}(\tau, \bar \tau)+Z_{1,g}(\tau, \bar \tau))\,, ~~~~~ Z_\text{twi} \equiv \frac{1}{2}(Z_{g,1}(\tau, \bar \tau)+Z_{g,g}(\tau, \bar \tau))\,,
\end{equation}
where $Z_{g^i,g^j}$ is the trace over $g^{i}$-twisted states with $g^j$ insertion, $i,j = 0,1$. For vanishing wilson line $A = 0$, the different blocks read
\begin{align}
	Z_{1,1} &= \frac{1}{\tau_2^{7/2}\eta^{17}\bar \eta}\sum_{\begin{smallmatrix}
			w\in\mathbb{Z}\\
			n \in \mathbb{Z}\end{smallmatrix}}q^{\tfrac12 p_L^2}\bar q^{\tfrac12 p_R^2}\sum_{\pi \in E_8\oplus E_8}q^{\tfrac12 \pi^2}(\bar V_8 - \bar S_8)\,,\label{Z11CHL}\\
	Z_{1,g} &= \frac{f_{01}}{\tau_2^{7/2}\eta^{17}\bar \eta}\sum_{\begin{smallmatrix}
			w\in\mathbb{Z}\\
			n \in \mathbb{Z}\end{smallmatrix}}(-1)^nq^{\tfrac12 p_L^2}\bar q^{\tfrac12 p_R^2}\sum_{\pi \in E_8(2)}q^{\tfrac12 \pi^2}(\bar V_8 - \bar S_8)\,,\label{Z1gCHL}\\
	Z_{g,1} &= \frac{f_{10}}{\tau_2^{7/2}\eta^{17}\bar \eta}\sum_{\begin{smallmatrix}
			w\in\mathbb{Z}+\tfrac12\\
			n \in \mathbb{Z}\end{smallmatrix}}q^{\tfrac12 p_L^2}\bar q^{\tfrac12 p_R^2}\sum_{\pi \in E_8(\tfrac12)}q^{\tfrac12 \pi^2}(\bar V_8 - \bar S_8)\,,\label{Zg1CHL}\\
	Z_{g,g} &= \frac{-f_{11}}{\tau_2^{7/2}\eta^{17}\bar \eta}\sum_{\begin{smallmatrix}
			w\in\mathbb{Z}+\tfrac12\\
			n \in \mathbb{Z}\end{smallmatrix}}(-1)^nq^{\tfrac12 p_L^2}\bar q^{\tfrac12 p_R^2}\sum_{\pi \in E_8(\tfrac12)}(-1)^{P^2}q^{\tfrac12 \pi^2}(\bar V_8 - \bar S_8)\label{ZggCHL}\,,
\end{align}
cf. Section \ref{ss:partition} for notation. See Appendix A of \cite{Font:2021uyw} for detailed explanations on computation of these blocks.
 
It is well known that orbifolding by $\theta_L$ alone gives produces an orbifold CFT equivalent to the parent theory.
As explained in Appendix A of \cite{Bossard:2017wum},\footnote{We thank A. Font for bringing this paper to our attention.} one can use this fact to derive the identity
\begin{equation}\label{id}
	\sum_{\pi \in E_8\oplus E_8}q^{\tfrac12 \pi^2} =  f_{01}\sum_{\pi \in E_8(2)}q^{\tfrac12 \pi^2} +  f_{10}\sum_{\pi \in E_8(\tfrac12)}q^{\tfrac12 \pi^2} - f_{11}\sum_{\pi\in E_8(\tfrac12)}(-1)^{\pi^2}q^{\tfrac12\pi^2}\,,
\end{equation}
which can be substituted back in $Z_{1,1}$ in \eqref{Z11CHL}. The first term in this substitution combines with $Z_{1,g}$ in \eqref{Z1gCHL} into
\begin{equation}\label{comb1}
	\frac{f_{01}}{\tau_2^{7/2}\eta^{17}\bar \eta}\sum_{\begin{smallmatrix}
			w\in\mathbb{Z}\\
			n \in 2\mathbb{Z}\end{smallmatrix}}q^{\tfrac12 p_L^2}\bar q^{\tfrac12 p_R^2}\sum_{\pi \in E_8(2)}q^{\tfrac12 \pi^2}(\bar V_8 - \bar S_8)\,,
\end{equation}
where $n \in 2\mathbb{Z}$ is due to the projector $(1+(-1)^n)/2$, using the prefactor $1/2$ in \eqref{std}. For arbitrary values of the Wilson line $A$ the lattice $\Gamma_{1,1}(2)\oplus E_8(2)$ is not orthogonally split; $p_{L,R}$ depend on $\pi$ and $P$ depends on $w$, cf. eq. \eqref{momenta}. Pulling together the summands, we obtain the second line of \eqref{master} with $d = 1$ and $(J, K, M)=(1,0,0)$. 

Combining the insertion of the second term in \eqref{id} into \eqref{Z11CHL} with $Z_{1,g}$ in \eqref{Zg1CHL} simply extends $w \in \mathbb{Z} + \tfrac12$ to $2w \in \mathbb{Z}$ in the latter, making it into an unshifted lattice sum. The analogous result for $Z_{g,g}$ in \eqref{ZggCHL} is obtained by using $(-1)^{p_L^2 - p_R^2} = (-1)^{2nw} = (-1)^n$ for $w\in \mathbb{Z}+1/2$ and $(-1)^{p_L^2 - p_R^2} = 1$ for $w \in \mathbb{Z}$. $Z_{g,g}$ is then modified by the same extension to $2w \in \mathbb{Z}$ and replacing $(-1)^n \to (-1)^{p_L^2-p_R^2}$. These two expressions can be alternatively obtained by applying the $S$ and $TS$-modular transformations to \eqref{comb1}. Putting them together and allowing $A \neq 0$ we get
\begin{equation}
	\frac{1}{\tau_2^{7/2}\eta^{17}\bar \eta}\sum_{\begin{smallmatrix}
			2w\in\mathbb{Z}\\
			n \in \mathbb{Z}\end{smallmatrix}}\sum_{\pi \in E_8(\tfrac12)}\frac{1}{2}\left[f_{10}-(-1)^{p_L^2-p_R^2}f_{11} \right]q^{\tfrac12 p_L^2 + \tfrac12}\bar q^{\tfrac12 p_R^2}(\bar V_8 - \bar S_8)\,,
\end{equation}
matching the first line of \eqref{master}.

There are three important properties of this form of the partition function. (1) It is written manifestly in terms of the charge lattice of the theory, (2) it does not involve shift-phases and (3) it consists of one modular orbit rather than two. In particular, property (1) makes it clear that the automorphisms of the charge lattice are symmetries of the partition function.   

\subsection{Non-supersymmetric strings}
For the four non-supersymmetric strings, the strategy is the same: compute the standard 1-loop partition function and use identity \eqref{id} to rewrite it.

\subsubsection{\BIII{}}
The partition function of the $E_8$ string has standard blocks
\begin{align}
	Z_{1,1} &= \frac{1}{\tau_2^{4}\eta^{16}}\sum_{\pi \in E_8\oplus E_8}q^{\tfrac12 \pi^2}(\bar V_8 - \bar S_8)\,,\label{Z11BIII}\\
	Z_{1,g} &= \frac{f_{01}}{\tau_2^{4}\eta^{16}}\sum_{\pi \in E_8(2)}q^{\tfrac12 \pi^2}(\bar V_8 + \bar S_8)\,,\label{Z1gBIII}\\
	Z_{g,1} &= \frac{f_{10}}{\tau_2^{4}\eta^{16}}\sum_{\pi \in E_8(\tfrac12)}q^{\tfrac12 \pi^2}(\bar O_8 - \bar C_8)\,,\label{Zg1BIII}\\
	Z_{g,g} &= \frac{f_{11}}{\tau_2^{4}\eta^{16}}\sum_{\pi \in E_8(\tfrac12)}(-1)^{\pi^2}q^{\tfrac12 \pi^2}(\bar O_8 + \bar C_8)\,.\label{ZggBIII}
\end{align}
Substituting \eqref{id} into \eqref{Z11BIII} and combining with \eqref{Z1gBIII} we get
\begin{align}
	Z_v = f_{01}\sum_{\pi \in E_8(2)}q^{\tfrac12 \pi^2} +  \sum_{\pi \in E_8(\tfrac12)}\left[f_{10}-(-1)^{\pi^2}f_{11} \right]q^{\tfrac12 \pi^2}\,,
\end{align}
matching \eqref{master} with $d = 0$ and $F = T = 0$. We also obtain $Z_s$ with a similar form, but we use instead the identity $Z_s = Z_c$ and compute $Z_c$ from $Z_{g,1}$ and $Z_{g,g}$. We obtain
\begin{equation}
	Z_s = Z_c = \sum_{\pi \in E_8(\tfrac12)}\frac12\left[f_{10}-(-1)^{\pi^2}f_{11} \right]q^{\tfrac12 \pi^2}\,,
\end{equation}
giving \eqref{master} with $F = 1$ and $T = 0,1$. Finally, the case $F = 0$ and $T = 1$ is matched with
\begin{equation}
	Z_o =  \sum_{\pi \in E_8(\tfrac12)}\frac12\left[f_{10}+(-1)^{\pi^2}f_{11} \right]q^{\tfrac12 \pi^2}\,.
\end{equation}

\subsubsection{\BIIb{}}
The partition function of the \BIIb{} theory in $D = 9$ is given by 
\begin{align}
	Z_{1,1} &= \frac{1}{\tau_2^{7/2}\eta^{17}\bar \eta}\sum_{\begin{smallmatrix}
			w\in\mathbb{Z}\\
			n \in \mathbb{Z}\end{smallmatrix}}q^{\tfrac12 p_L^2}\bar q^{\tfrac12 p_R^2}\sum_{\pi \in E_8\oplus E_8}q^{\tfrac12 \pi^2}(\bar V_8 - \bar S_8)\,,\label{Z11BIIb}\\
	Z_{1,g} &= \frac{f_{01}}{\tau_2^{7/2}\eta^{17}\bar \eta}\sum_{\begin{smallmatrix}
			w\in\mathbb{Z}\\
			n \in \mathbb{Z}\end{smallmatrix}}(-1)^nq^{\tfrac12 p_L^2}\bar q^{\tfrac12 p_R^2}\sum_{\pi \in E_8(2)}q^{\tfrac12 \pi^2}(\bar V_8 + \bar S_8)\,,\label{Z1gBIIb}\\
	Z_{g,1} &= \frac{f_{10}}{\tau_2^{7/2}\eta^{17}\bar \eta}\sum_{\begin{smallmatrix}
			w\in\mathbb{Z}+\tfrac12\\
			n \in \mathbb{Z}\end{smallmatrix}}q^{\tfrac12 p_L^2}\bar q^{\tfrac12 p_R^2}\sum_{\pi \in E_8(\tfrac12)}q^{\tfrac12 \pi^2}(\bar O_8 - \bar C_8)\,,\label{Zg1BIIb}\\
	Z_{g,g} &= \frac{f_{11}}{\tau_2^{7/2}\eta^{17}\bar \eta}\sum_{\begin{smallmatrix}
			w\in\mathbb{Z}+\tfrac12\\
			n \in \mathbb{Z}\end{smallmatrix}}(-1)^nq^{\tfrac12 p_L^2}\bar q^{\tfrac12 p_R^2}\sum_{\pi \in E_8(\tfrac12)}(-1)^{\pi^2}q^{\tfrac12 P^2}(\bar O_8 + \bar C_8)\,.\label{ZggBIIb}
\end{align}
Here and in what follows it should be clear from the context whether $p_L$ includes the gauge contribution $\pi + A_i w^i$ (in this case it does not). Substituting \eqref{id} into $Z_{1,1}$ in \eqref{Z11BIIb} and combining with $Z_{1,g}$ in \eqref{Z1gBIIb}, we obtain exactly \eqref{master} with $F = 0,1$ and $T = 0$ for $(J,K,M) = (1,0,1)$. Combining $Z_{g,1}$ with $Z_{g,g}$ we similarly get the cases with $T = 1$. 

\subsubsection{\BIIa{}}
The partition function of the Scherk-Schwarz reduction of the $E_8$ string (\BIIa{} theory) is derived in \cite{DeFreitas:2024ztt}. Its vector class reads
\begin{equation}
	\begin{split}
	Z_v &= \sum_{\begin{smallmatrix}
		w\in\mathbb{Z}\\
		n \in 2\mathbb{Z}\end{smallmatrix}} q^{\tfrac12 p_L^2}\bar q^{\tfrac12 p_R^2}\frac12\left\{\sum_{\pi \in E_8\oplus E_8}q^{\tfrac12 P^2} + f_{01}\sum_{\pi \in E_8(2)}q^{\tfrac12 P^2} \right\}\\
		&+ \sum_{\begin{smallmatrix}
				w\in\mathbb{Z}+\tfrac12\\
				n \in 2\mathbb{Z}+1\end{smallmatrix}} q^{\tfrac12 p_L^2}\bar q^{\tfrac12 p_R^2}\sum_{\pi \in E_8(\tfrac12)}\frac12\left[f_{10}-(-1)^{\pi^2}f_{11} \right]q^{\tfrac12 \pi^2}\,.			
	\end{split}
\end{equation}
Substituting \eqref{id} we obtain
\begin{equation}
	\begin{split}
		Z_v &= f_{01}\sum_{\begin{smallmatrix}
				w\in\mathbb{Z}\\
				n \in 2\mathbb{Z}\end{smallmatrix}} q^{\tfrac12 p_L^2}\bar q^{\tfrac12 p_R^2}\sum_{P \in E_8(2)}q^{\tfrac12 \pi^2}\\
		&+ \sum_{\begin{smallmatrix}
				w\in\mathbb{Z}+\tfrac12\\
				n \in 2\mathbb{Z}+1\end{smallmatrix}} q^{\tfrac12 p_L^2}\bar q^{\tfrac12 p_R^2}\sum_{\pi \in E_8(\tfrac12)}\frac12\left[f_{10}+(-1)^{\pi^2}f_{11} \right]q^{\tfrac12 \pi^2}\\		
		&+ \sum_{\begin{smallmatrix}
				w\in\mathbb{Z}\\
				n \in 2\mathbb{Z}\end{smallmatrix}} q^{\tfrac12 p_L^2}\bar q^{\tfrac12 p_R^2}\sum_{\pi \in E_8(\tfrac12)}\frac12\left[f_{10}-(-1)^{\pi^2}f_{11} \right]q^{\tfrac12 \pi^2}\,.				
	\end{split}
\end{equation}
We use the fact that $p_L^2-p_R^2$ is respectively odd and even in the second and third lines to write both $\pm (-1)^{P^2}$ as $-(-1)^{P^2+p_L^2-p_R^2}$ and combine both lines into one sum, obtaining
\begin{equation}
	\begin{split}
		Z_v &= f_{01}\sum_{\begin{smallmatrix}
				w\in\mathbb{Z}\\
				n \in 2\mathbb{Z}\end{smallmatrix}} q^{\tfrac12 p_L^2}\bar q^{\tfrac12 p_R^2}\sum_{\pi \in E_8(2)}q^{\tfrac12 \pi^2}\\
		&+ \sum_{\begin{smallmatrix}
				2w\in\mathbb{Z}\\
				n \in 2\mathbb{Z}+2w\end{smallmatrix}} q^{\tfrac12 p_L^2}\bar q^{\tfrac12 p_R^2}\sum_{\pi \in E_8(\tfrac12)}\frac12\left[f_{10}-(-1)^{\pi^2}f_{11} \right]q^{\tfrac12 \pi^2}\,,			
	\end{split}
\end{equation}
where in the second line we have both states with $(w,n)\in (\mathbb{Z}+\tfrac12)\times (2\mathbb{Z}+1)$ and $(w,n)\in \mathbb{Z}\times 2\mathbb{Z}$, matching \eqref{master} with $F  = T = 0$.

The spinor class reads
\begin{equation}
	\begin{split}
		Z_s &= \sum_{\begin{smallmatrix}
				w\in\mathbb{Z}\\
				n \in 2\mathbb{Z}+1\end{smallmatrix}} q^{\tfrac12 p_L^2}\bar q^{\tfrac12 p_R^2}\frac12\left\{\sum_{\pi \in E_8\oplus E_8}q^{\tfrac12 \pi^2} - f_{01}\sum_{\pi \in E_8(2)}q^{\tfrac12 \pi^2} \right\}\\
		&+ \sum_{\begin{smallmatrix}
				w\in\mathbb{Z}+\tfrac12\\
				n \in 2\mathbb{Z}\end{smallmatrix}} q^{\tfrac12 p_L^2}\bar q^{\tfrac12 p_R^2}\sum_{\pi \in E_8(\tfrac12)}\frac12\left[f_{10}-(-1)^{\pi^2}f_{11} \right]q^{\tfrac12 \pi^2}\,.			
	\end{split}
\end{equation}
Substituting \eqref{id} into the first line transforms the terms in curly brackets into the second sum in the second line, and combining both expressions we find
\begin{equation}
	Z_s = \sum_{\begin{smallmatrix}
			2w\in\mathbb{Z}\\
			n \in 2\mathbb{Z}+2w+1\end{smallmatrix}} q^{\tfrac12 p_L^2}\bar q^{\tfrac12 p_R^2}\sum_{\pi \in E_8(\tfrac12)}\frac12\left[f_{10}-(-1)^{\pi^2}f_{11} \right]q^{\tfrac12 \pi^2}\,,
\end{equation}
matching \eqref{master} with $F = 1$ and $T = 0$. On the other hand, the cospinor class reads
\begin{equation}
	\begin{split}
		Z_c &= \sum_{\begin{smallmatrix}
				w\in\mathbb{Z}+\tfrac12\\
				n \in 2\mathbb{Z}\end{smallmatrix}} q^{\tfrac12 p_L^2}\bar q^{\tfrac12 p_R^2}\frac12\left\{\sum_{\pi \in E_8\oplus E_8}q^{\tfrac12 \pi^2} - f_{01}\sum_{\pi \in E_8(2)}q^{\tfrac12 \pi^2} \right\}\\
		&+ \sum_{\begin{smallmatrix}
				w\in\mathbb{Z}\\
				n \in 2\mathbb{Z}+1\end{smallmatrix}} q^{\tfrac12 p_L^2}\bar q^{\tfrac12 p_R^2}\sum_{\pi \in E_8(\tfrac12)}\frac12\left[f_{10}-(-1)^{\pi^2}f_{11} \right]q^{\tfrac12 \pi^2}\,,			
	\end{split}
\end{equation}
but using \eqref{id} we obtain the same expression as before, in accordance with the condition $Z_s = Z_c$ inherited from the parent $E_8$ string. 

Finally, the scalar class reads
\begin{equation}
	\begin{split}
		Z_o &= \sum_{\begin{smallmatrix}
				w\in\mathbb{Z}+\tfrac12\\
				n \in 2\mathbb{Z}+1\end{smallmatrix}} q^{\tfrac12 p_L^2}\bar q^{\tfrac12 p_R^2}\frac12\left\{\sum_{\pi \in E_8\oplus E_8}q^{\tfrac12 \pi^2} + f_{01}\sum_{\pi \in E_8(2)}q^{\tfrac12 \pi^2} \right\}\\
		&+ \sum_{\begin{smallmatrix}
				w\in\mathbb{Z}\\
				n \in 2\mathbb{Z}\end{smallmatrix}} q^{\tfrac12 p_L^2}\bar q^{\tfrac12 p_R^2}\sum_{\pi \in E_8(\tfrac12)}\frac12\left[f_{10}+(-1)^{P^2}f_{11} \right]q^{\tfrac12 \pi^2}\,.			
	\end{split}
\end{equation}
Substituting \eqref{id} into the first line and proceding as with the vector class we obtain 
\begin{equation}
	\begin{split}
		Z_o &= f_{01}\sum_{\begin{smallmatrix}
				w\in\mathbb{Z}+\tfrac12\\
				n \in 2\mathbb{Z}+1\end{smallmatrix}} q^{\tfrac12 p_L^2}\bar q^{\tfrac12 p_R^2}\sum_{\pi \in E_8(2)}q^{\tfrac12 \pi^2}\\
		&+ \sum_{\begin{smallmatrix}
				2w\in\mathbb{Z}\\
				n \in 2\mathbb{Z}+2w\end{smallmatrix}} q^{\tfrac12 p_L^2}\bar q^{\tfrac12 p_R^2}\sum_{\pi \in E_8(\tfrac12)}\frac12\left[f_{10}-(-1)^{\pi^2}f_{11} \right]q^{\tfrac12 \pi^2}\,,			
	\end{split}
\end{equation}
matching \eqref{master} with $F = 0$ and $T = 1$. 

\subsubsection{\BI{}}
The partition function of the Scherk-Schwarz reduction of the CHL string (\BI{} theory) is derived in \cite{DeFreitas:2024ztt}. Each class is schematically given by the product of the CHL string partition function and the ordinary SS reduction blocks
\begin{equation}
	v\sim \sum_{\begin{smallmatrix}
			w\in\mathbb{Z}\\
			n \in 2\mathbb{Z}\end{smallmatrix}} q^{\tfrac12 p_L^2}\bar q^{\tfrac12 p_R^2}\,, ~~~ s\sim \sum_{\begin{smallmatrix}
			w\in\mathbb{Z}\\
			n \in 2\mathbb{Z}+1\end{smallmatrix}} q^{\tfrac12 p_L^2}\bar q^{\tfrac12 p_R^2}\,, ~~~
			c\sim \sum_{\begin{smallmatrix}
			w\in\mathbb{Z}+\tfrac12\\
			n \in 2\mathbb{Z}\end{smallmatrix}} q^{\tfrac12 p_L^2}\bar q^{\tfrac12 p_R^2}\,, ~~~
			o\sim \sum_{\begin{smallmatrix}
			w\in\mathbb{Z}+\tfrac12\\
			n \in 2\mathbb{Z}+1\end{smallmatrix}} q^{\tfrac12 p_L^2}\bar q^{\tfrac12 p_R^2}\,,	
\end{equation} 
giving the case $(J,K,M) = (1,1,1)$ of \eqref{master} as required. From the factorization of the blocks, the \BI{} theory is then obtained using the two parameter combinations $(J,K,M) = (1,1,1), (1,0,0)$ as in Table \ref{tab:quant}.

\newpage

\section{Maximal enhancements}
\label{app:max}
Here we record the maximal enhancements obtained with the exploration algorithm as explained in Section \ref{ss:exploration}. We specify the fundamental groups $\pi_1(G)$ by giving a set of generators $\{k\}$ where the $k$'s are elements of the center $Z(\tilde G)$ of the universal cover $\tilde G$ of $G$. We use this same notation to write down the representations of $G$ in which massless states transform in the case that they are minuscule. TF means tachyon-free. The rest of the conventions are explained in the main text in Section \ref{ss:analysis}. The accidental representations $a_i$ are recorded in Table \ref{tab:acc}, and the exceptional representations $e_i$ in Table \ref{tab:exc}. For the \BIIa{} and \BI{} strings, in the special cases where there are two different representations of the type of $\tilde v$, we label them as $v'_i$ and $v''_i$ and record them explicitly in Table \ref{tab:altv}. For these theories we specify the right-moving symmetry enhancements in the column $L'$. In Tables \ref{tab:BIIa9D} and \ref{tab:BI8D} we have written the accidental representations directly for simplicity  . 

Non-minuscule representations are always left implicit in the notation. In the \BIII{} and \BIIb{} strings we find massless spinors in such representations and they are read off from the entries in the $L$ column as explained in Section \ref{ss:analysis}. The same notation is used for the $\tilde v$ representations for massless scalars in the \BIIa{} and \BI{} theories. In the \BIIa{} theory we furthermore use $A_1^t$ and $C_2^t$ instead of $A_1$ and $C_2$ in the cases where there are tachyons charged in the vector representation of $SO(3)$, $SO(5)$. Underlining in Tables \ref{tab:exc} and \ref{tab:altv} means the sum of permutations, e.g. $(a,\underline{b,c}) = (a,b,c) + (a,c,b)$.

\begin{table}[H]\renewcommand{\arraystretch}{0.85}
	\centering 
	\begin{tabular}{|c|c|c|c|c|c|}
		\hline
		\# &$L$ & $H$& $\{k\}$ & $o$ &Node\\ \hline\hline
		1 &$C_1E_8$ & 1 & -  &- &0 \\ 
		2 &$C_3E_6$ & 1 & - & - &5 \\ 
		3 &$C_2E_7$ & $\Z_2$ & $(11)$ & $[s]$&6  \\ 
		4 &$C_4D_5$ & $\Z_2$ & $(21)$ & - &4 \\ 
		5 &$C_9$ & 1 & - & - &1 \\ 
		6 &$A_1C_8$ & $\Z_2$ & $(01)$ & $[s]$ &7 \\ 
		7 &$A_1A_2C_6$ & $\Z_2$ & $(101)$ & -&2  \\ 
		8 &$A_4C_5$ & 1 & - & -&3  \\ 
		\hline
	\end{tabular}
	\caption{Maximal enhancements in the \BIII{} theory in $D = 9$.}
	\label{tab:BIII9D}
\end{table}

\begin{table}[H]\renewcommand{\arraystretch}{0.85}
	\centering
	\begin{tabular}{|c|c|c|c|c|c|c|}
		\hline
		\# &$L$ & $H$& $\{k\} \simeq c$ & $o$ &Node & TF\\ \hline\hline
		1 &$D_2'E_7$ & $\mathbb{Z}_2$ &  $(s1)$&$[c]$& 6& \xmark\\ 
		2 &$D_3'E_6$ & 1 & -  &  -&5 &\xmark \\
		3 &$D_9'$ & 1 & -   &-&1 &  \cmark\\ 
		4 &$A_1D_8'$ & $\Z_2$ & $(0s)$  &$[c]$&7 & \cmark \\
		5 &$A_1A_2D_6'$ & $\mathbb{Z}_2$  & $(10s)$ &-& 2& \xmark\\
		6 &$D_4'D_5$ & $\mathbb{Z}_2$& $(v2)$ & -&4&\xmark\\
		7 &$A_4D_5'$ & 1 & -  & - &3 &\xmark\\
		\hline
	\end{tabular}
	\caption{Maximal enhancements in the \BIIb{} theory in $D = 9$.}
	\label{tab:BIIb9D}
\end{table}

\begin{table}[H]\renewcommand{\arraystretch}{0.85}
	\centering
	\begin{tabular}{|c|c|c|c|c|c|}
		\hline
		\# &$L$ & $H$& $\{k\} \simeq s,c$&$o$&Node \\ \hline\hline
		1 &$A_1^t E_8$ & 1 & -& 	$[v]$ & (5,0')\\ 
		2 &$A_2 E_7$ & 1 & - & - &(0,1')\\ 
		3 &$(A_1A_1^t)' E_7$ & $\Z_2$ & $(011)$&$[\tilde v]$ & (0,0')\\
		4 &$A_1 D_8'$ & $\Z_2$ & $(0s)$ &$[\tilde v]$& (4,1')\\
		5 &$A_1A_1^tD_7$ & $\Z_2$ & $(112)$ &- & (4,0')\\
		6 &$A_3D_6$ & $\Z_2$ & $(2v)$ &- & (7,1')\\
		7 &$A_1A_2D_6$ & $\Z_2$ & $(10c)$ &- & (7,0')\\
		8 &$A_9$ & 1 & -  & -&(1,1')\\		 
		9 &$A_1^tA_8$ & 1 & - & $[(03)]$&(1,0') \\
		10 &$A_2A_7$ & $\Z_2$ & $(04)$ & - & (3,1')\\
		11 &$A_1^tA_2A_6$ & 1 & -  & -&(3,0')\\
		12 &$A_4A_5$ & 1 & - &-  & (6,1')\\
		13 &$A_1^tA_3A_5$ & $\Z_2$ & $(103)$ &- &(6,0') \\
		14 &$A_1A_3A_5$ & $\Z_2$ & $(103)$ &- & (2,1')\\
		15 &$A_1^tA_1A_3A_4$ & $\Z_2$ & $(1120)$ &- & (2,0')\\
		\hline
	\end{tabular}
	\caption{Maximal enhancements in the \BIIa{} theory in $D = 9$.}
	\label{tab:BIIa9D}
\end{table}
\newpage

\def\tabscale{0.65}
\def\tabscaleirrep{0.75}
\def\spinorsSC{s \simeq c}
\def\spinorsS{s}
\def\spinorsC{c}
\def\scalars{o}

\clearpage

\begin{table}[H]\renewcommand{\arraystretch}{\tabscale}\centering\begin{tabular}{|c|>{$}c<{$}|>{$}c<{$}|>{$}c<{$}|>{$}c<{$}|}\hline \# & L & H & \{k\} & \scalars\\ \hline\hline 
		1&A_2F_4F_4 & 1 & & \left[s,s,s\right] \\\hline 
		2&A_1A_1F_4F_4 & 1 & & \left[s,s\right] \\\hline 
		3& E_6F_4 & 1 & & - \\\hline 
		4&C_1D_5F_4 & 1 & & \left[a_1,a_1\right] \\\hline
		5&A_2^2D_4F_4 & 1 & & - \\\hline 
		6&C_6F_4 & 1 & & \left[s,s\right] \\\hline 
		7&A_1C_5F_4 & 1 & & \left[s\right] \\\hline 
		8&A_1A_2C_3F_4 & 1 & & - \\\hline 
		9&A_4C_2F_4 & 1 & & - \\\hline 
		10&A_2A_2C_2F_4 & 1 & & - \\\hline		
		11&A_5C_1F_4 & 1 & & \left[s\right] \\\hline 
		12&A_1A_4C_1F_4 & 1 & & \left[s\right] \\\hline 
		13&A_3B_3F_4 & 1 & & \left[s\right] \\\hline 
		14&A_1A_2B_3F_4 & 1 & & \left[a_2,s,a_2\right] \\\hline 
		15&A_2^2A_4F_4 & 1 & & \left[s\right] \\\hline 
		16&A_1A_2^2A_3F_4 & 1 & & \left[s\right] \\\hline
		17&C_1C_1E_8 & 1 & & \left[a_3,a_3\right] \\\hline 
		18&A_2^2E_8 & 1 & & - \\\hline 
		19&C_1C_2E_7 & \mathbb{Z}_2 & \begin{smallmatrix} 011 \\\end{smallmatrix} & \left[s\right] \\\hline 
		20&A_2^2C_1E_7 & 1 & & \left[s\right] \\\hline 
		21&B_3E_7 & \mathbb{Z}_2 & \begin{smallmatrix} 11 \\\end{smallmatrix} & \left[s,s\right] \\\hline 
		22&C_1C_3E_6 & 1 & & - \\\hline
		23&A_2^2C_2E_6 & 1 & & - \\\hline 
		24&C_1B_3E_6 & 1 & & \left[s\right] \\\hline 
		25&A_2^2A_2^2E_6 & \mathbb{Z}_3 & \begin{smallmatrix} 121 \\\end{smallmatrix} & \left[s,s,s\right] \\\hline 
		26&C_1C_1D_8 & \mathbb{Z}_2 & \begin{smallmatrix} 00c \\\end{smallmatrix} & \left[a_4,a_4\right] \\\hline 
		27&A_2^2C_1D_7 & 1 & & - \\\hline 
		28&C_1C_4D_5 & \mathbb{Z}_2 & \begin{smallmatrix} 012 \\\end{smallmatrix} & \left[a_1,a_1\right] \\\hline 
		29&C_2C_2D_6 & \mathbb{Z}_2^2 & \begin{smallmatrix} 01c \\ 10s \\\end{smallmatrix} & \left[s,s\right] \\\hline 
		30&B_3C_1D_6 & \mathbb{Z}_2 & \begin{smallmatrix} 10s \\\end{smallmatrix} & \left[s\right] \\\hline 
		31&A_2^2A_2^2D_6 & 1 & & - \\\hline  
		32&A_2^2C_3D_5 & 1 & & - \\\hline 
		33&B_3C_2D_5 & \mathbb{Z}_2 & \begin{smallmatrix} 020 \\\end{smallmatrix} & \left[s\right] \\\hline 
		34&A_2^2B_3D_5 & 1 & & \left[s\right] \\\hline 
		35&B_3B_3D_4 & \mathbb{Z}_2 & \begin{smallmatrix} 11s \\\end{smallmatrix} & \left[s,s\right] \\\hline 
		36&C_1C_9 & 1 & & \left[a_5,a_5\right] \\\hline 
		37&C_2C_8 & \mathbb{Z}_2 & \begin{smallmatrix} 10 \\\end{smallmatrix} & \left[s,s\right] \\\hline 
		38&A_1C_1C_8 & \mathbb{Z}_2 & \begin{smallmatrix} 001 \\\end{smallmatrix} & \left[s\right] \\\hline 
		39&A_2^2C_8 & 1 & & - \\\hline 
		40&B_3C_7 & 1 & & \left[s\right] \\\hline 
		41&A_1A_2^2C_7 & 1 & & \left[s\right] \\\hline
		42&A_1C_3C_6 & \mathbb{Z}_2 & \begin{smallmatrix} 101 \\\end{smallmatrix} & \left[s\right] \\\hline 
		43&A_2C_2C_6 & \mathbb{Z}_2 & \begin{smallmatrix} 010 \\\end{smallmatrix} & \left[a_6,s,a_6\right] \\\hline 
	\end{tabular}\hspace{1em}\begin{tabular}{|c|>{$}c<{$}|>{$}c<{$}|>{$}c<{$}|>{$}c<{$}|}\hline  \# & L & H & \{k\} & \scalars\\ \hline\hline
		44&A_1A_2C_1C_6 & \mathbb{Z}_2 & \begin{smallmatrix} 0011 \\\end{smallmatrix} & - \\\hline 
		45&A_1B_3C_6 & \mathbb{Z}_2 & \begin{smallmatrix} 010 \\\end{smallmatrix} & \left[s,s\right] \\\hline 
		46&C_5C_5 & 1 & & \left[a_7,a_7\right] \\\hline 
		47&A_4C_1C_5 & 1 & & - \\\hline 
		48&A_2B_3C_5 & 1 & & - \\\hline 
		59&A_2^2A_3C_5 & 1 & & - \\\hline 
		50&A_1A_2A_2^2C_5 & 1 & & e_1 \\\hline 
		51&A_1A_1C_4C_4 & \mathbb{Z}_2^2 & \begin{smallmatrix} 0011 \\ 1101 \\\end{smallmatrix} & \left[s,s\right] \\\hline 
		52&A_1A_3C_2C_4 & \mathbb{Z}_2^2 & \begin{smallmatrix} 0210 \\ 1010 \\\end{smallmatrix} & \left[s\right] \\\hline 
		53&A_1A_2B_3C_4 & \mathbb{Z}_2 & \begin{smallmatrix} 1010 \\\end{smallmatrix} & \left[a_2,s,a_2\right] \\\hline 
		54&A_2^2A_4C_4 & 1 & & - \\\hline 
		55&A_2A_2C_3C_3 & 1 & & - \\\hline
		56&A_5C_2C_3 & \mathbb{Z}_2 & \begin{smallmatrix} 300 \\\end{smallmatrix} & \left[s\right] \\\hline 
		57&A_1A_5C_1C_3 & \mathbb{Z}_2 & \begin{smallmatrix} 0310 \\\end{smallmatrix} & - \\\hline
		58&A_4B_3C_3 & 1 & & \left[s\right] \\\hline 
		59&A_1A_3B_3C_3 & \mathbb{Z}_2 & \begin{smallmatrix} 1201 \\\end{smallmatrix} & \left[a_8,a_8\right] \\\hline 
		60&A_1A_2^2A_4C_3 & 1 & & - \\\hline  
		61&A_3A_3C_2C_2 & \mathbb{Z}_2^2 & \begin{smallmatrix} 0211 \\ 2011 \\\end{smallmatrix} & \left[a_{9},a_{9}\right] \\\hline 
		62&A_7C_1C_2 & \mathbb{Z}_2 & \begin{smallmatrix} 400 \\\end{smallmatrix} & \left[a_{10}\right] \\\hline
		63&A_2A_5C_1C_2 & \mathbb{Z}_2 & \begin{smallmatrix} 0301 \\\end{smallmatrix} & e_2 \\\hline 
		64&A_5B_3C_2 & \mathbb{Z}_2 & \begin{smallmatrix} 310 \\\end{smallmatrix} & - \\\hline 
		65&A_2A_3B_3C_2 & \mathbb{Z}_2 & \begin{smallmatrix} 0200 \\\end{smallmatrix} & - \\\hline 
		66&A_2^2A_6C_2 & 1 & & - \\\hline 
		67&A_2A_2^2A_4C_2 & 1 & & - \\\hline 
		68&A_8C_1C_1 & 1 & & \left[ a_3\right] \\\hline 
		69&A_1A_7C_1C_1 & \mathbb{Z}_4 & \begin{smallmatrix} 1201 \\\end{smallmatrix} & \left[a_{11},s,a_{11}\right] \\\hline
		70&A_4A_4C_1C_1 & 1 & & - \\\hline 
		71& A_6B_3C_1 & 1 & & - \\\hline 
		72&A_1A_5B_3C_1 & \mathbb{Z}_2 & \begin{smallmatrix} 0310 \\\end{smallmatrix} & \left[s\right] \\\hline 
		73&A_2A_4B_3C_1 & 1 & & - \\\hline
		74&A_2^2A_7C_1 & 1 & & \left[a_{12}\right] \\\hline 
		75&A_1A_2^2A_6C_1 & 1 & & \left[s\right] \\\hline
		76&A_2^2A_3A_4C_1 & 1 & & - \\\hline 
		77&A_4B_3B_3 & 1 & & - \\\hline 
		78&A_1A_3B_3B_3 & \mathbb{Z}_2 & \begin{smallmatrix} 0211 \\\end{smallmatrix} & \left[s\right] \\\hline 
		79&A_2A_2B_3B_3 & 1 & & - \\\hline 
		80&A_2^2A_5B_3 & 1 & & - \\\hline 
		81&A_1A_2^2A_4B_3 & 1 & & \left[s\right] \\\hline 
		82&A_2A_2^2A_3B_3 & 1 & & e_3 \\\hline 
		83&A_2^2A_2^2A_6 & 1 & & - \\\hline 
		84&A_1A_2^2A_2^2A_5 & \mathbb{Z}_3 & \begin{smallmatrix} 0122 \\\end{smallmatrix} & \left[a_{13},s,a_{13}\right] \\\hline 
		85&A_2^2A_2^2A_3A_3 & 1 & & - \\\hline
		\multicolumn{5}{c}{}\end{tabular}\caption{Maximal symmetry enhancements in the \BIII{} theory in $D = 8$.}\label{tab:BIII8D}\end{table}

\begin{table}[H]\renewcommand{\arraystretch}{\tabscale}\centering\begin{tabular}{|c|@{}>{$}c<{$}@{}|>{$}c<{$}|@{}>{$}c<{$}@{}|@{}>{$}c<{$}@{}|@{}>{$}c<{$}@{}|@{}>{}c<{}@{}|}\hline \# & L & H & \{k\} & \spinorsC  & \scalars & \,TF\, \\ \hline\hline 
		1& C_1C_1E_8 & 1 & & \begin{smallmatrix} 110 \\\end{smallmatrix} & \left[s,s\right] & \xmark \\\hline
		2& C_1C_2E_7 & 1 & & \begin{smallmatrix} 110 \\\end{smallmatrix} & \left[c\right] & \xmark \\\hline
		3& (A_1C_2)'E_7 & \mathbb{Z}_2 & \begin{smallmatrix} 011 \\\end{smallmatrix} & & - & \xmark \\\hline
		4& (A_1C_2)'E_7 & \mathbb{Z}_2 & \begin{smallmatrix} 101 \\\end{smallmatrix} & \begin{smallmatrix} 101 \\\end{smallmatrix} & \left[c,c\right] & \xmark \\\hline
		5& (A_1A_1)'C_1E_7 & \mathbb{Z}_2 & \begin{smallmatrix} 0011 \\\end{smallmatrix} & \begin{smallmatrix} 0101 \\\end{smallmatrix} & \left[c\right] & \xmark \\\hline
		6& C_4'E_6 & 1 & & & - & \xmark \\\hline
		7& C_1C_3E_6 & 1 & & \begin{smallmatrix} 110 \\\end{smallmatrix} & - & \xmark \\\hline
		8& C_1(A_1C_2)'E_6 & 1 & & \begin{smallmatrix} 1010 \\\end{smallmatrix} & \left[c\right] & \xmark \\\hline
		9& A_3'C_1E_6 & 1 & & & - & \xmark \\\hline
		10& C_1D_9' & 1 & & & - & \cmark \\\hline
		11& C_2D_8' & \mathbb{Z}_2 & \begin{smallmatrix} 0c \\\end{smallmatrix} & \begin{smallmatrix} 0c \\\end{smallmatrix} & \left[c,c\right] & \cmark \\\hline
		12& C_1C_1D_8 & \mathbb{Z}_2 & \begin{smallmatrix} 00c \\\end{smallmatrix} & \begin{smallmatrix} 00c \\ 110 \\\end{smallmatrix} & \left[s,s\right] & \xmark \\\hline
		13& A_1C_1D_8' & \mathbb{Z}_2 & \begin{smallmatrix} 00c \\\end{smallmatrix} & \begin{smallmatrix} 00c \\\end{smallmatrix} & \left[c\right] & \cmark \\\hline
		14& (A_1A_1)'D_8' & \mathbb{Z}_2^2 & \begin{smallmatrix} 00s \\ 11c \\\end{smallmatrix} & & - & \cmark \\\hline
		15& (A_1C_2)'D_7' & \mathbb{Z}_2 & \begin{smallmatrix} 112 \\\end{smallmatrix} & & - & \cmark \\\hline
		16& C_4'D_6' & \mathbb{Z}_2 & \begin{smallmatrix} 1v \\\end{smallmatrix} & & - & \cmark \\\hline
		17& A_1C_3D_6' & \mathbb{Z}_2 & \begin{smallmatrix} 10s \\\end{smallmatrix} & \begin{smallmatrix} 10s \\\end{smallmatrix} & \left[c\right] & \xmark \\\hline
		18& C_2C_2D_6 & \mathbb{Z}_2 & \begin{smallmatrix} 11v \\\end{smallmatrix} & \begin{smallmatrix} 110 \\\end{smallmatrix} & \left[c,c\right] & \xmark \\\hline
		19& A_2C_2D_6' & \mathbb{Z}_2 & \begin{smallmatrix} 01s \\\end{smallmatrix} & & - & \xmark \\\hline
		20& C_1(A_1C_2)'D_6 & \mathbb{Z}_2 & \begin{smallmatrix} 100s \\\end{smallmatrix} & \begin{smallmatrix} 1010 \\ 010s \\\end{smallmatrix} & \left[c\right] & \xmark \\\hline
		21& A_1(A_1C_2)'D_6' & \mathbb{Z}_2^2 & \begin{smallmatrix} 001c \\ 010s \\\end{smallmatrix} & \begin{smallmatrix} 010s \\\end{smallmatrix} & \left[c,c\right] & \cmark \\\hline
		22& (A_1A_1)'C_2D_6 & \mathbb{Z}_2^2 & \begin{smallmatrix} 001s \\ 010c \\\end{smallmatrix} & \begin{smallmatrix} 010c \\\end{smallmatrix} & \left[c,c\right] & \xmark \\\hline
		23& A_1A_2C_1D_6' & \mathbb{Z}_2 & \begin{smallmatrix} 001s \\\end{smallmatrix} & \begin{smallmatrix} 100s \\\end{smallmatrix} & - & \xmark \\\hline
		24& A_1A_3'D_6' & \mathbb{Z}_2^2 & \begin{smallmatrix} 02v \\ 10c \\\end{smallmatrix} & & - & \cmark \\\hline
		25& (A_1A_1)'A_2D_6' & \mathbb{Z}_2^2 & \begin{smallmatrix} 010s \\ 100c \\\end{smallmatrix} & \begin{smallmatrix} 100c \\\end{smallmatrix} & \left[a_{14},c,a_{14}\right] & \cmark \\\hline
		26& (A_1A_1)'(A_1A_1)'D_6 & \mathbb{Z}_2^3 & \begin{smallmatrix} 0100s \\ 0001c \\ 1010v \\\end{smallmatrix} & \begin{smallmatrix} 0101v \\\end{smallmatrix} & \left[c,c\right] & \xmark \\\hline
		27& D_5'D_5' & \mathbb{Z}_2 & \begin{smallmatrix} 22 \\\end{smallmatrix} & & - & \cmark \\\hline
		28& C_1D_4^s D_5 & \mathbb{Z}_2 & \begin{smallmatrix} 0c2 \\\end{smallmatrix} & \begin{smallmatrix} 0c2 \\\end{smallmatrix} & \left[a_{15},a_{15}\right] & \xmark \\\hline
		29& C_5D_5' & 1 & & & - & \xmark \\\hline
		30& C_1C_4D_5 & 1 & & \begin{smallmatrix} 110 \\\end{smallmatrix} & \left[a_{16},a_{16}\right] & \xmark \\\hline
		31& A_1C_4'D_5' & \mathbb{Z}_2 & \begin{smallmatrix} 012 \\\end{smallmatrix} & & - & \cmark \\\hline
		32& (A_1C_2)'C_2D_5 & \mathbb{Z}_2 & \begin{smallmatrix} 0112 \\\end{smallmatrix} & \begin{smallmatrix} 0110 \\\end{smallmatrix} & \left[c\right] & \xmark \\\hline
		33& (A_1A_1)'(A_1C_2)'D_5 & \mathbb{Z}_2^2 & \begin{smallmatrix} 01102 \\ 10012 \\\end{smallmatrix} & \begin{smallmatrix} 01102 \\\end{smallmatrix} & \left[c\right] & \xmark \\\hline
		34& A_2(A_1C_2)'D_5' & \mathbb{Z}_2 & \begin{smallmatrix} 0112 \\\end{smallmatrix} & & - & \xmark \\\hline
		35& A_4C_1D_5' & 1 & & & - & \cmark \\\hline
		36& A_4C_1D_5' & 1 & & & - & \xmark \\\hline
		37& A_1A_1D_4^s D_4^s & \mathbb{Z}_2^3 & \begin{smallmatrix} 00cv \\ 00ss \\ 110v \\\end{smallmatrix} & \begin{smallmatrix} 00cv \\\end{smallmatrix} & \left[c,c\right] & \cmark \\\hline
		38& A_1A_1C_4D_4^s & \mathbb{Z}_2^2 & \begin{smallmatrix} 001c \\ 110c \\\end{smallmatrix} & \begin{smallmatrix} 110c \\\end{smallmatrix} & \left[c,c\right] & \xmark \\\hline
		39& A_1A_3C_2D_4^s & \mathbb{Z}_2^2 & \begin{smallmatrix} 020c \\ 101c \\\end{smallmatrix} & \begin{smallmatrix} 020c \\\end{smallmatrix} & \left[c\right] & \xmark \\\hline
		40& (A_1C_2)'(A_1C_2)'D_4 & \mathbb{Z}_2^2 & \begin{smallmatrix} 0101s \\ 1010s \\\end{smallmatrix} & \begin{smallmatrix} 01010 \\ 1010s \\\end{smallmatrix} & \left[c,c\right] & \xmark \\\hline
		41& A_1A_2(A_1C_2)'D_4^c & \mathbb{Z}_2^2 & \begin{smallmatrix} 1001v \\ 0011c \\\end{smallmatrix} & \begin{smallmatrix} 1010s \\\end{smallmatrix} & \left[a_{17},c,a_{17}\right] & \xmark \\\hline
		42& A_1(A_1A_1)'A_3D_4^c & \mathbb{Z}_2^3 & \begin{smallmatrix} 0002s \\ 1010v \\ 1100s \\\end{smallmatrix} & \begin{smallmatrix} 1100s \\\end{smallmatrix} & \left[c\right] & \xmark \\\hline
		43& C_1C_9 & 1 & & \begin{smallmatrix} 11 \\\end{smallmatrix} & \left[s,s\right] & \xmark \\\hline
	\end{tabular}\caption{Maximal symmetry enhancements in the \BIIb{} theory in $D = 8$.}\label{tab:BIIb8D}\end{table}\begin{table}[H]\renewcommand{\arraystretch}{\tabscale}\centering\begin{tabular}{|c|@{}>{$}c<{$}@{}|>{$}c<{$}|@{}>{$}c<{$}@{}|@{}>{$}c<{$}@{}|@{}>{$}c<{$}@{}|@{}>{}c<{}@{}|}\hline \# & L & H & \{k\} & \spinorsC  & \scalars & \,TF\, \\ \hline\hline 
		44& C_2C_8 & 1 & & \begin{smallmatrix} 11 \\\end{smallmatrix} & \left[c,c\right] & \xmark \\\hline
		45& A_1C_1C_8 & 1 & & \begin{smallmatrix} 011 \\\end{smallmatrix} & \left[c\right] & \xmark \\\hline
		46& (A_1A_1)'C_8 & \mathbb{Z}_2 & \begin{smallmatrix} 001 \\\end{smallmatrix} & & - & \xmark \\\hline
		47& (A_1C_2)'C_7 & 1 & & \begin{smallmatrix} 011 \\\end{smallmatrix} & \left[c\right] & \xmark \\\hline
		48& C_4'C_6 & 1 & & \begin{smallmatrix} 11 \\\end{smallmatrix} & \left[c,c\right] & \xmark \\\hline
		49& A_1C_3C_6 & 1 & & \begin{smallmatrix} 011 \\\end{smallmatrix} & \left[c\right] & \xmark \\\hline
		50& A_2C_2C_6 & \mathbb{Z}_2 & \begin{smallmatrix} 010 \\\end{smallmatrix} & \begin{smallmatrix} 011 \\\end{smallmatrix} & \left[s,c,s\right] & \xmark \\\hline
		51& A_1(A_1C_2)'C_6 & \mathbb{Z}_2 & \begin{smallmatrix} 0010 \\\end{smallmatrix} & \begin{smallmatrix} 0011 \\\end{smallmatrix} & \left[c,c\right] & \xmark \\\hline
		52& A_1A_2C_1C_6 & 1 & & \begin{smallmatrix} 0011 \\\end{smallmatrix} & - & \xmark \\\hline
		53& A_1A_3'C_6 & \mathbb{Z}_2 & \begin{smallmatrix} 101 \\\end{smallmatrix} & & - & \xmark \\\hline
		54& (A_1A_1)'A_2C_6 & \mathbb{Z}_2 & \begin{smallmatrix} 0101 \\\end{smallmatrix} & & - & \xmark \\\hline
		55& C_5C_5 & 1 & & \begin{smallmatrix} 11 \\\end{smallmatrix} & \left[s,s\right] & \xmark \\\hline
		56& A_1C_4'C_5 & 1 & & \begin{smallmatrix} 011 \\\end{smallmatrix} & \left[c\right] & \xmark \\\hline
		57& A_2(A_1C_2)'C_5 & 1 & & \begin{smallmatrix} 0011 \\\end{smallmatrix} & - & \xmark \\\hline
		58& A_4C_1C_5 & 1 & & \begin{smallmatrix} 011 \\\end{smallmatrix} & - & \xmark \\\hline
		59& A_2C_4'C_4' & \mathbb{Z}_2 & \begin{smallmatrix} 011 \\\end{smallmatrix} & \begin{smallmatrix} 011 \\\end{smallmatrix} & \left[c,c,c\right] & \cmark \\\hline
		60& A_1A_1C_4'C_4' & \mathbb{Z}_2 & \begin{smallmatrix} 0011 \\\end{smallmatrix} & \begin{smallmatrix} 0011 \\\end{smallmatrix} & \left[c,c\right] & \cmark \\\hline
		61& A_1A_2C_3C_4' & 1 & & \begin{smallmatrix} 0011 \\\end{smallmatrix} & - & \xmark \\\hline
		62& A_4C_2C_4' & 1 & & \begin{smallmatrix} 011 \\\end{smallmatrix} & - & \xmark \\\hline
		63& A_1A_3C_2C_4 & \mathbb{Z}_2 & \begin{smallmatrix} 1010 \\\end{smallmatrix} & \begin{smallmatrix} 0011 \\\end{smallmatrix} & \left[c\right] & \xmark \\\hline
		64& A_2A_2C_2C_4' & 1 & & \begin{smallmatrix} 0011 \\\end{smallmatrix} & - & \xmark \\\hline
		65& A_3(A_1C_2)'C_4' & \mathbb{Z}_2 & \begin{smallmatrix} 0110 \\\end{smallmatrix} & \begin{smallmatrix} 0011 \\\end{smallmatrix} & \left[c\right] & \xmark \\\hline
		66& A_1A_2(A_1C_2)'C_4' & \mathbb{Z}_2 & \begin{smallmatrix} 00110 \\\end{smallmatrix} & \begin{smallmatrix} 00011 \\\end{smallmatrix} & \left[a_{18},c,a_{18}\right] & \xmark \\\hline
		67& A_1A_2(A_1C_2)'C_4 & \mathbb{Z}_2 & \begin{smallmatrix} 10010 \\\end{smallmatrix} & \begin{smallmatrix} 00011 \\\end{smallmatrix} & \left[a_{18},c,a_{18}\right] & \xmark \\\hline
		68& A_5C_1C_4' & 1 & & \begin{smallmatrix} 011 \\\end{smallmatrix} & \left[c\right] & \xmark \\\hline
		69& A_1A_4C_1C_4' & 1 & & \begin{smallmatrix} 0011 \\\end{smallmatrix} & \left[c\right] & \xmark \\\hline
		70& (A_1A_1)'A_4C_4' & \mathbb{Z}_2 & \begin{smallmatrix} 1101 \\\end{smallmatrix} & & - & \xmark \\\hline
		71& A_1A_2A_3'C_4' & \mathbb{Z}_2 & \begin{smallmatrix} 0021 \\\end{smallmatrix} & & - & \xmark \\\hline
		72& A_1(A_1A_1)'A_3C_4 & \mathbb{Z}_2^2 & \begin{smallmatrix} 00021 \\ 10101 \\\end{smallmatrix} & \begin{smallmatrix} 10120 \\\end{smallmatrix} & \left[c\right] & \xmark \\\hline
		73& (A_1A_1)'A_2A_2C_4' & \mathbb{Z}_2 & \begin{smallmatrix} 11001 \\\end{smallmatrix} & & - & \xmark \\\hline
		74& A_2A_2C_3C_3 & 1 & & \begin{smallmatrix} 0011 \\\end{smallmatrix} & - & \xmark \\\hline
		75& A_5C_2C_3 & 1 & & \begin{smallmatrix} 011 \\\end{smallmatrix} & \left[c\right] & \xmark \\\hline
		76& A_4(A_1C_2)'C_3 & 1 & & \begin{smallmatrix} 0011 \\\end{smallmatrix} & \left[c\right] & \xmark \\\hline
		77& A_1A_3(A_1C_2)'C_3 & \mathbb{Z}_2 & \begin{smallmatrix} 12100 \\\end{smallmatrix} & \begin{smallmatrix} 00011 \\ 12100 \\\end{smallmatrix} & \left[a_{19},a_{19}\right] & \xmark \\\hline
		78& A_1A_5C_1C_3 & \mathbb{Z}_2 & \begin{smallmatrix} 0310 \\\end{smallmatrix} & \begin{smallmatrix} 0011 \\ 1300 \\\end{smallmatrix} & - & \xmark \\\hline
		79& (A_1A_1)'A_5C_3 & \mathbb{Z}_2 & \begin{smallmatrix} 0130 \\\end{smallmatrix} & \begin{smallmatrix} 0130 \\\end{smallmatrix} & \left[c\right] & \xmark \\\hline
		80& A_2A_2A_3'C_3 & 1 & & & - & \xmark \\\hline
		81& A_3A_3C_2C_2 & \mathbb{Z}_2^2 & \begin{smallmatrix} 0211 \\ 2011 \\\end{smallmatrix} & \begin{smallmatrix} 0011 \\ 2200 \\\end{smallmatrix} & \left[s,s\right] & \xmark \\\hline
		82& A_5(A_1C_2)'C_2 & \mathbb{Z}_2 & \begin{smallmatrix} 3100 \\\end{smallmatrix} & \begin{smallmatrix} 0011 \\ 3100 \\\end{smallmatrix} & - & \xmark \\\hline
		83& A_2A_3(A_1C_2)'C_2 & \mathbb{Z}_2 & \begin{smallmatrix} 02011 \\\end{smallmatrix} & \begin{smallmatrix} 00011 \\\end{smallmatrix} & - & \xmark \\\hline
		84& A_7C_1C_2 & \mathbb{Z}_2 & \begin{smallmatrix} 400 \\\end{smallmatrix} & \begin{smallmatrix} 400 \\ 011 \\\end{smallmatrix} & \left[s\right] & \xmark \\\hline
		85& A_2A_5C_1C_2 & 1 & & \begin{smallmatrix} 0011 \\\end{smallmatrix} & - & \xmark \\\hline
		86& A_3'A_5C_2 & \mathbb{Z}_2 & \begin{smallmatrix} 031 \\\end{smallmatrix} & & - & \xmark \\\hline
	\end{tabular}\caption{Maximal symmetry enhancements in the \BIIb{} theory in $D = 8$ (continued).}\end{table}\begin{table}[H]	\ContinuedFloat\renewcommand{\arraystretch}{\tabscale}\centering\begin{tabular}{|c|@{}>{$}c<{$}@{}|>{$}c<{$}|@{}>{$}c<{$}@{}|@{}>{$}c<{$}@{}|@{}>{$}c<{$}@{}|@{}>{}c<{}@{}|}\hline \# & L & H & \{k\} & \spinorsC  & \scalars & \,TF\,\\\hline\hline
		87& (A_1A_1)'A_3A_3C_2 & \mathbb{Z}_2^2 & \begin{smallmatrix} 01021 \\ 10201 \\\end{smallmatrix} & & - & \xmark \\\hline
		88& A_4(A_1C_2)'(A_1C_2)' & \mathbb{Z}_2 & \begin{smallmatrix} 01111 \\\end{smallmatrix} & \begin{smallmatrix} 00101 \\\end{smallmatrix} & - & \xmark \\\hline
		89& A_1A_3(A_1C_2)'(A_1C_2)' & \mathbb{Z}_2^2 & \begin{smallmatrix} 020101 \\ 001111 \\\end{smallmatrix} & \begin{smallmatrix} 000101 \\ 021010 \\\end{smallmatrix} & \left[c\right] & \xmark \\\hline
		90& A_2A_2(A_1C_2)'(A_1C_2)' & \mathbb{Z}_2 & \begin{smallmatrix} 001111 \\\end{smallmatrix} & \begin{smallmatrix} 000101 \\\end{smallmatrix} & - & \xmark \\\hline
		91& A_6C_1(A_1C_2)' & 1 & & \begin{smallmatrix} 0101 \\\end{smallmatrix} & - & \xmark \\\hline
		92& A_1A_5C_1(A_1C_2)' & \mathbb{Z}_2 & \begin{smallmatrix} 13000 \\\end{smallmatrix} & \begin{smallmatrix} 00101 \\ 03010 \\\end{smallmatrix} & \left[c\right] & \xmark \\\hline
		93& A_1A_5C_1(A_1C_2)' & \mathbb{Z}_2 & \begin{smallmatrix} 03100 \\\end{smallmatrix} & \begin{smallmatrix} 00101 \\ 03010 \\\end{smallmatrix} & \left[c\right] & \xmark \\\hline
		94& A_2A_4C_1(A_1C_2)' & 1 & & \begin{smallmatrix} 00101 \\\end{smallmatrix} & - & \xmark \\\hline
		95& (A_1A_1)'A_5(A_1C_2)' & \mathbb{Z}_2^2 & \begin{smallmatrix} 00301 \\ 11011 \\\end{smallmatrix} & & - & \xmark \\\hline
		96& A_3'A_4(A_1C_2)' & \mathbb{Z}_2 & \begin{smallmatrix} 2011 \\\end{smallmatrix} & & - & \xmark \\\hline
		97& A_1A_3A_3'(A_1C_2)' & \mathbb{Z}_2^2 & \begin{smallmatrix} 00211 \\ 12001 \\\end{smallmatrix} & & - & \xmark \\\hline
		98& (A_1A_1)'A_2A_3(A_1C_2)' & \mathbb{Z}_2^2 & \begin{smallmatrix} 100201 \\ 010210 \\\end{smallmatrix} & \begin{smallmatrix} 010210 \\\end{smallmatrix} & - & \xmark \\\hline
		99& A_8C_1C_1 & 1 & & \begin{smallmatrix} 011 \\\end{smallmatrix} & \left[s,a_{20}\right] & \xmark \\\hline
		100& A_1A_7C_1C_1 & \mathbb{Z}_4 & \begin{smallmatrix} 1201 \\\end{smallmatrix} & \begin{smallmatrix} 0400 \\ 0011 \\\end{smallmatrix} & \left[c,s\right] & \cmark \\\hline
		101& A_4A_4C_1C_1 & 1 & & \begin{smallmatrix} 0011 \\\end{smallmatrix} & e_{4} & \xmark \\\hline
		102& (A_1A_1)'A_7C_1 & \mathbb{Z}_2 & \begin{smallmatrix} 0141 \\\end{smallmatrix} & & - & \xmark \\\hline
		103& A_1A_3'A_5C_1 & \mathbb{Z}_2 & \begin{smallmatrix} 0231 \\\end{smallmatrix} & & - & \xmark \\\hline
		104& (A_1A_1)'A_2A_5C_1 & \mathbb{Z}_2 & \begin{smallmatrix} 01030 \\\end{smallmatrix} & \begin{smallmatrix} 10030 \\\end{smallmatrix} & e_{5} & \xmark \\\hline
		105& (A_1A_1)'A_3'A_5 & \mathbb{Z}_2^2 & \begin{smallmatrix} 0123 \\ 1003 \\\end{smallmatrix} & & - & \xmark \\\hline
		106& A_2A_2A_3'A_3' & \mathbb{Z}_2 & \begin{smallmatrix} 0022 \\\end{smallmatrix} & & - & \xmark \\\hline
		107& (A_1A_1)'(A_1A_1)'A_3A_3 & \mathbb{Z}_2^3 & \begin{smallmatrix} 000022 \\ 010120 \\ 101002 \\\end{smallmatrix} & \begin{smallmatrix} 010102 \\ 101020 \\\end{smallmatrix} & \left[a_{21},a_{21}\right] & \xmark \\\hline
	\end{tabular}\caption{Maximal symmetry enhancements in the \BIIb{} theory in $D = 8$ (continued).}\end{table}

\begin{table}[H]\renewcommand{\arraystretch}{\tabscale}
	\centering\begin{tabular}{|@{}>{\scriptsize}c<{}@{}|@{}>{$}c<{$}@{}|@{}>{$}c<{$}@{}|@{}>{$}c<{$}@{}|@{}>{$}c<{$}@{}|@{}>{$}c<{$}@{}|@{}>{$}c<{$}@{}|}\hline \# & L & H & \{k\}  & \spinorsSC& \scalars & L'\\ \hline\hline 
		1&C_2^tE_8 & 1 & - & - & \left[v,v\right] & 2 A_1 \\\hline 
		2&C_{10} & 1 & - & - & \left[v,v\right] & 2 A_1 \\\hline 
		3&A_1^tC_1E_8 & 1 & - & - & \left[v\right] & A_1 \\\hline 
		4&C_3E_7 & 1 & - & - & \left[v\right] & A_1 \\\hline
		5&(A_1C_2^t)'E_7 & \mathbb{Z}_2 & \begin{smallmatrix} 011 \\\end{smallmatrix} & - & \left[v,\tilde{v}\right] & A_1 \\\hline
		6&(A_1^tC_2)'E_7 & \mathbb{Z}_2 & \begin{smallmatrix} 011 \\\end{smallmatrix} & - & \left[v,\tilde{v}\right] & A_1 \\\hline
		7&A_1^tA_2E_7 & \mathbb{Z}_2 & \begin{smallmatrix} 101 \\\end{smallmatrix} & - & \left[v\right] & A_1 \\\hline 
		8&A_1^tC_3E_6 & 1 & - & - & \left[v\right] & A_1 \\\hline 
		9&(A_1A_1^t)'D_8' & \mathbb{Z}_2^2 & \begin{smallmatrix} 00c \\ 11s \\\end{smallmatrix} & - & \left[v,\tilde{v}\right] & A_1 \\\hline 
		10&C_4'D_6' & \mathbb{Z}_2 & \begin{smallmatrix} 1v \\\end{smallmatrix} & - & \left[v,\tilde{v}\right] & A_1 \\\hline 
		11&A_2C_2^tD_6 & \mathbb{Z}_2 & \begin{smallmatrix} 01c \\\end{smallmatrix} & - & \left[v\right] & A_1 \\\hline 
		12&A_1^tA_3D_6 & \mathbb{Z}_2^2 & \begin{smallmatrix} 02v \\ 10s \\\end{smallmatrix} & - & \left[v\right] & A_1 \\\hline 
		13&A_1^tC_4D_5 & \mathbb{Z}_2 & \begin{smallmatrix} 012 \\\end{smallmatrix} & - & \left[v\right] & A_1 \\\hline 
		14&A_1^tC_9 & 1 & - & - & \left[v\right] & A_1 \\\hline 
		15&A_2C_8 & \mathbb{Z}_2 & \begin{smallmatrix} 01 \\\end{smallmatrix} & - & \left[v\right] & A_1 \\\hline
		16&(A_1A_1^t)'C_8 & \mathbb{Z}_2 & \begin{smallmatrix} 001 \\\end{smallmatrix} & - & \left[v,\tilde{v}\right] & A_1 \\\hline  
		17&A_5C_5 & 1 & - & - & \left[v\right] & A_1 \\\hline 
		18&A_1^tA_4C_5 & 1 & - & - & \left[v\right] & A_1 \\\hline
		19&A_8C_2^t & 1 & - & - & \left[v,a_{20}\right] & A_1 \\\hline
		20&A_2A_6C_2^t & 1 & - & - & \left[v\right] & A_1 \\\hline 
		21&A_3A_5C_2^t & \mathbb{Z}_2 & \begin{smallmatrix} 031 \\\end{smallmatrix} & - & \left[v\right] & A_1 \\\hline 
		22&A_1^tA_9 & \mathbb{Z}_2 & \begin{smallmatrix} 15 \\\end{smallmatrix} & - & \left[v\right] & A_1 \\\hline 
		23&A_1^tA_2A_7 & \mathbb{Z}_2 & \begin{smallmatrix} 004 \\\end{smallmatrix} & - & \left[v\right] & A_1 \\\hline 
		24&A_1A_1^tA_3A_5 & \mathbb{Z}_2^2 & \begin{smallmatrix} 0123 \\ 1003 \\\end{smallmatrix} & - & \left[v\right] & A_1 \\\hline
		
		25&C_1C_2^tE_7 & 1 & - & \begin{smallmatrix} 110 \\\end{smallmatrix} & \left[\tilde{v},\tilde{v}\right] & - \\\hline	
		26&A_1C_1C_1E_7 & \mathbb{Z}_2 & \begin{smallmatrix} 0011 \\\end{smallmatrix} & \begin{smallmatrix} 0110 \\ 1001 \\\end{smallmatrix} & \left[s,s\right] & - \\\hline
		27&A_2C_1E_7 & 1 & - & - & - & - \\\hline
		28&(A_1A_1^t)'C_1E_7 & \mathbb{Z}_2 & \begin{smallmatrix} 0011 \\\end{smallmatrix} & \begin{smallmatrix} 0101 \\\end{smallmatrix} & \left[\tilde{v}\right] & - \\\hline
		29&C_1C_3E_6 & 1 & - & \begin{smallmatrix} 110 \\\end{smallmatrix} & - & - \\\hline 
		30&C_2C_2^tE_6 & 1 & - & \begin{smallmatrix} 110 \\\end{smallmatrix} & - & - \\\hline
		31&C_1(A_1C_2^t)'E_6 & 1 & - & \begin{smallmatrix} 1010 \\\end{smallmatrix} & \left[\tilde{v}\right] & - \\\hline 
		32&A_1^tA_2C_1E_6 & 1 & - & - & e_{6} & - \\\hline 
		33&C_2D_8' & \mathbb{Z}_2 & \begin{smallmatrix} 0s \\\end{smallmatrix} & \begin{smallmatrix} 0s \\\end{smallmatrix} & \left[\tilde{v},\tilde{v}\right] & - \\\hline 
		34&A_1C_1D_8' & \mathbb{Z}_2 & \begin{smallmatrix} 00s \\\end{smallmatrix} & \begin{smallmatrix} 00s \\\end{smallmatrix} & \left[\tilde{v}\right] & - \\\hline
		35&(A_1^tC_2)'D_7' & \mathbb{Z}_2 & \begin{smallmatrix} 112 \\\end{smallmatrix} & - & \left[\tilde{v}\right] & - \\\hline 
		36&A_1A_1^tC_1D_7 & \mathbb{Z}_2 & \begin{smallmatrix} 0112 \\\end{smallmatrix} & \begin{smallmatrix} 1102 \\\end{smallmatrix} & \left[a_{22},a_{22}\right] & - \\\hline 
		37&A_1^tC_3D_6' & \mathbb{Z}_2 & \begin{smallmatrix} 10s \\\end{smallmatrix} & \begin{smallmatrix} 10s \\\end{smallmatrix} & \left[\tilde{v}\right] & - \\\hline 
		38&C_2C_2^tD_6 & \mathbb{Z}_2 & \begin{smallmatrix} 11v \\\end{smallmatrix} & \begin{smallmatrix} 110 \\\end{smallmatrix} & \left[\tilde{v},\tilde{v}\right] & - \\\hline 
		39&A_3C_1D_6 & \mathbb{Z}_2 & \begin{smallmatrix} 20v \\\end{smallmatrix} & \begin{smallmatrix} 20v \\\end{smallmatrix} & \left[a_{23},a_{23}\right] & - \\\hline 
		40&A_1A_1^tC_2D_6 & \mathbb{Z}_2^2 & \begin{smallmatrix} 001c \\ 010s \\\end{smallmatrix} & \begin{smallmatrix} 010s \\\end{smallmatrix} & \left[v'_{1},v''_{1}\right] & - \\\hline 
		41&C_1(A_1C_2)'D_6 & \mathbb{Z}_2 & \begin{smallmatrix} 100s \\\end{smallmatrix} & \begin{smallmatrix} 1010 \\ 010s \\\end{smallmatrix} & \left[\tilde{v}\right] & - \\\hline 
		42&A_1^tA_2C_1D_6 & \mathbb{Z}_2 & \begin{smallmatrix} 001c \\\end{smallmatrix} & \begin{smallmatrix} 100c \\\end{smallmatrix} & - & - \\\hline 
		43&(A_1A_1^t)'A_2D_6' & \mathbb{Z}_2^2 & \begin{smallmatrix} 100s \\ 010c \\\end{smallmatrix} & \begin{smallmatrix} 010c \\\end{smallmatrix} & \left[a_{24},\tilde{v},a_{24}\right] & - \\\hline 
		\end{tabular}\hspace{0.3em}\begin{tabular}{|@{}>{\scriptsize}c<{}@{}|@{}>{$}c<{$}@{}|@{}>{$}c<{$}@{}|@{}>{$}c<{$}@{}|@{}>{$}c<{$}@{}|@{}>{$}c<{$}@{}|@{}>{$}c<{$}@{}|}\hline \# & L & H & k  & \spinorsSC& \scalars & L'\\ \hline\hline   
		44&A_1A_1A_1A_1^tD_6 & \mathbb{Z}_2^3 & \begin{smallmatrix} 0001s \\ 0100c \\ 1010v \\\end{smallmatrix} & \begin{smallmatrix} 0101v \\\end{smallmatrix} & \left[v'_{2},v''_{2}\right] & - \\\hline 
		45&C_2^tC_3D_5 & 1 & - & \begin{smallmatrix} 110 \\\end{smallmatrix} & \left[a_{25},a_{25}\right] & - \\\hline
		46&(A_1C_2^t)'C_2D_5 & \mathbb{Z}_2 & \begin{smallmatrix} 0112 \\\end{smallmatrix} & \begin{smallmatrix} 0110 \\\end{smallmatrix} & \left[\tilde{v}\right] & - \\\hline 
		47&A_1^tA_2C_2D_5 & \mathbb{Z}_2 & \begin{smallmatrix} 1012 \\\end{smallmatrix} & - & - & - \\\hline 
		48&(A_1A_1)'(A_1^tC_2)'D_5 & \mathbb{Z}_2^2 & \begin{smallmatrix} 01102 \\ 10012 \\\end{smallmatrix} & \begin{smallmatrix} 01102 \\\end{smallmatrix} & \left[\tilde{v}\right] & - \\\hline 
		49&A_1A_1^tA_3D_5 & \mathbb{Z}_2^2 & \begin{smallmatrix} 0022 \\ 1102 \\\end{smallmatrix} & \begin{smallmatrix} 1120 \\\end{smallmatrix} & \left[a_{26},a_{26}\right] & - \\\hline 
		50&A_1A_1D_4D_4 & \mathbb{Z}_2^3 & \begin{smallmatrix} 00cs \\ 00sc \\ 110v \\\end{smallmatrix} & \begin{smallmatrix} 00vv \\\end{smallmatrix} & \left[v'_{3},v''_{3}\right] & - \\\hline
		51&A_1A_1^tC_4D_4 & \mathbb{Z}_2^2 & \begin{smallmatrix} 001s \\ 110s \\\end{smallmatrix} & \begin{smallmatrix} 110s \\\end{smallmatrix} & \left[v'_{4},v''_{4}\right] & - \\\hline 
		52&A_1A_3C_2D_4^c & \mathbb{Z}_2^2 & \begin{smallmatrix} 020v \\ 101v \\\end{smallmatrix} & \begin{smallmatrix} 020v \\\end{smallmatrix} & \left[\tilde{v}\right] & - \\\hline 
		53&A_1A_3C_2D_4^s & \mathbb{Z}_2^2 & \begin{smallmatrix} 020v \\ 101v \\\end{smallmatrix} & \begin{smallmatrix} 020v \\\end{smallmatrix} & \left[\tilde{v}\right] & - \\\hline 
		54&A_1C_2A_1C_2D_4 & \mathbb{Z}_2^2 & \begin{smallmatrix} 0101s \\ 1010s \\\end{smallmatrix} & \begin{smallmatrix} 01010 \\ 1010s \\\end{smallmatrix} & \left[v'_{5},v''_{5}\right] & - \\\hline
		55&A_1A_2(A_1^tC_2)'D_4^c & \mathbb{Z}_2^2 & \begin{smallmatrix} 0011c \\ 1001v \\\end{smallmatrix} & \begin{smallmatrix} 1010s \\\end{smallmatrix} & \left[a_{17},\tilde{v},a_{17}\right] & - \\\hline 
		56&A_1(A_1A_1^t)'A_3D_4^s & \mathbb{Z}_2^3 & \begin{smallmatrix} 0002c \\ 0110s \\ 1100v \\\end{smallmatrix} & \begin{smallmatrix} 1010c \\\end{smallmatrix} & \left[\tilde{v}\right] & - \\\hline 
		57&C_2^tC_8 & 1 & - & \begin{smallmatrix} 11 \\\end{smallmatrix} & \left[\tilde{v},\tilde{v}\right] & - \\\hline 
		58&A_1C_1C_8 & 1 & - & \begin{smallmatrix} 011 \\\end{smallmatrix} & \left[\tilde{v}\right] & - \\\hline 
		59&C_3C_7 & 1 & - & \begin{smallmatrix} 11 \\\end{smallmatrix} & \left[s,s\right] & - \\\hline 
		60&(A_1C_2^t)C_7 & 1 & - & \begin{smallmatrix} 011 \\\end{smallmatrix} & \left[\tilde{v}\right] & - \\\hline
		61&A_1^tA_2C_7 & 1 & - & - & - & - \\\hline 
		62&C_4'C_6 & 1 & - & \begin{smallmatrix} 11 \\\end{smallmatrix} & \left[\tilde{v},\tilde{v}\right] & - \\\hline 
		63&A_1C_3C_6 & 1 & - & \begin{smallmatrix} 011 \\\end{smallmatrix} & \left[\tilde{v}\right] & - \\\hline 
		64&A_2C_2^tC_6 & \mathbb{Z}_2 & \begin{smallmatrix} 010 \\\end{smallmatrix} & \begin{smallmatrix} 011 \\\end{smallmatrix} & \left[\tilde{v},\tilde{v},\tilde{v}\right] & - \\\hline
		65&A_1A_1C_2^tC_6 & \mathbb{Z}_2 & \begin{smallmatrix} 0010 \\\end{smallmatrix} & \begin{smallmatrix} 0011 \\\end{smallmatrix} & \left[v'_{6},v''_{6}\right] & - \\\hline 
		66&A_4C_6 & 1 & - & - & - & - \\\hline 
		67&A_1^tA_3'C_6 & \mathbb{Z}_2 & \begin{smallmatrix} 101 \\\end{smallmatrix} & - & \left[\tilde{v}\right] & - \\\hline 
		68&A_2A_2C_6 & 1 & - & - & - & - \\\hline 
		69&(A_1A_1^t)'A_2C_6 & \mathbb{Z}_2 & \begin{smallmatrix} 0101 \\\end{smallmatrix} & - & \left[\tilde{v}\right] & - \\\hline 
		70&A_1C_4'C_5 & 1 & - & \begin{smallmatrix} 011 \\\end{smallmatrix} & \left[\tilde{v}\right] & - \\\hline 
		71&A_1A_2C_2^tC_5 & 1 & - & \begin{smallmatrix} 0011 \\\end{smallmatrix} & - & - \\\hline 
		72&A_2A_2C_1C_5 & 1 & - & \begin{smallmatrix} 0011 \\\end{smallmatrix} & - & - \\\hline 
		73&A_1A_1^tA_3C_5 & \mathbb{Z}_2 & \begin{smallmatrix} 1120 \\\end{smallmatrix} & \begin{smallmatrix} 1120 \\\end{smallmatrix} & \left[a_{27},a_{27}\right] & - \\\hline 
		74&A_2C_4'C_4' & \mathbb{Z}_2 & \begin{smallmatrix} 011 \\\end{smallmatrix} & \begin{smallmatrix} 011 \\\end{smallmatrix} & \left[s,\tilde{v},s\right] & - \\\hline 
		75&A_1A_1C_4C_4 & \mathbb{Z}_2 & \begin{smallmatrix} 0011 \\\end{smallmatrix} & \begin{smallmatrix} 0011 \\\end{smallmatrix} & \left[v'_{7},v''_{7}\right] & - \\\hline 
		76&A_1A_2C_3C_4 & 1 & - & \begin{smallmatrix} 0011 \\\end{smallmatrix} & - & - \\\hline 
		77&A_4C_2^tC_4 & 1 & - & \begin{smallmatrix} 011 \\\end{smallmatrix} & - & - \\\hline 
		78&A_1A_3C_2^tC_4 & \mathbb{Z}_2 & \begin{smallmatrix} 1010 \\\end{smallmatrix} & \begin{smallmatrix} 0011 \\\end{smallmatrix} & \left[\tilde{v}\right] & - \\\hline 
		79&A_2A_2C_2^tC_4 & 1 & - & \begin{smallmatrix} 0011 \\\end{smallmatrix} & - & - \\\hline 
		80&A_1A_2(A_1C_2)'C_4' & \mathbb{Z}_2 & \begin{smallmatrix} 00110 \\\end{smallmatrix} & \begin{smallmatrix} 00011 \\\end{smallmatrix} & \left[a_{18},\tilde{v},a_{18}\right] & - \\\hline 
		81&A_1A_2(A_1C_2^t)'C_4 & \mathbb{Z}_2 & \begin{smallmatrix} 10010 \\\end{smallmatrix} & \begin{smallmatrix} 00011 \\\end{smallmatrix} & \left[a_{18},\tilde{v},a_{18}\right] & - \\\hline 
		82&A_5C_1C_4' & 1 & - & \begin{smallmatrix} 011 \\\end{smallmatrix} & \left[\tilde{v}\right] & - \\\hline 
		83&A_1A_4C_1C_4' & 1 & - & \begin{smallmatrix} 0011 \\\end{smallmatrix} & \left[\tilde{v}\right] & - \\\hline 
		84&A_1A_1A_3C_1C_4 & \mathbb{Z}_2 & \begin{smallmatrix} 01210 \\\end{smallmatrix} & \begin{smallmatrix} 00011 \\ 11200 \\\end{smallmatrix} & \left[a_{28},a_{28}\right] & - \\\hline 
		85&A_1^tA_5C_4 & \mathbb{Z}_2 & \begin{smallmatrix} 130 \\\end{smallmatrix} & \begin{smallmatrix} 130 \\\end{smallmatrix} & - & - \\\hline
	\end{tabular}\caption{Maximal symmetry enhancements in the \BIIa{} theory in $D = 8$.}\end{table}\begin{table}[H]\ContinuedFloat\renewcommand{\arraystretch}{\tabscale}\centering\begin{tabular}{|@{}>{\scriptsize}c<{}@{}|@{}>{$}c<{$}@{}|@{}>{$}c<{$}@{}|@{}>{$}c<{$}@{}|@{}>{$}c<{$}@{}|@{}>{$}c<{$}@{}|@{}>{\scriptsize$}c<{$}@{}|}\hline \#&L & H & \{k\}  & \spinorsSC& \scalars & L'\\ \hline\hline 
		86&A_1A_1^tA_4C_4 & \mathbb{Z}_2 & \begin{smallmatrix} 1101 \\\end{smallmatrix} & - & - & - \\\hline 	
		87&A_3A_3C_4 & \mathbb{Z}_2^2 & \begin{smallmatrix} 021 \\ 201 \\\end{smallmatrix} & \begin{smallmatrix} 220 \\\end{smallmatrix} & \left[s,s\right] & - \\\hline 
		88&A_1^tA_2A_3C_4 & \mathbb{Z}_2 & \begin{smallmatrix} 0021 \\\end{smallmatrix} & - & - & - \\\hline 
		89&A_1(A_1A_1^t)'A_3C_4 & \mathbb{Z}_2^2 & \begin{smallmatrix} 00021 \\ 10101 \\\end{smallmatrix} & \begin{smallmatrix} 10120 \\\end{smallmatrix} & \left[\tilde{v}\right] & - \\\hline		
		90&A_1A_1^tA_2A_2C_4 & \mathbb{Z}_2 & \begin{smallmatrix} 11001 \\\end{smallmatrix} & - & - & - \\\hline 
		91&A_4C_3C_3 & 1 & - & \begin{smallmatrix} 011 \\\end{smallmatrix} & - & - \\\hline 
		92&A_5C_2^tC_3 & 1 & - & \begin{smallmatrix} 011 \\\end{smallmatrix} & \left[\tilde{v}\right] & - \\\hline 
		93&A_4(A_1C_2^t)'C_3 & 1 & - & \begin{smallmatrix} 0011 \\\end{smallmatrix} & \left[\tilde{v}\right] & - \\\hline 
		94&A_1A_5C_1C_3 & \mathbb{Z}_2 & \begin{smallmatrix} 0310 \\\end{smallmatrix} & \begin{smallmatrix} 0011 \\ 1300 \\\end{smallmatrix} & - & - \\\hline 
		95&A_7C_3 & \mathbb{Z}_2 & \begin{smallmatrix} 40 \\\end{smallmatrix} & \begin{smallmatrix} 40 \\\end{smallmatrix} & \left[s\right] & - \\\hline 
		96&A_1^tA_6C_3 & 1 & - & - & - & - \\\hline 
		97&A_2A_5C_3 & 1 & - & - & e_7 & - \\\hline 
		98&(A_1A_1^t)'A_5C_3 & \mathbb{Z}_2 & \begin{smallmatrix} 0130 \\\end{smallmatrix} & \begin{smallmatrix} 0130 \\\end{smallmatrix} & \left[\tilde{v}\right] & - \\\hline  
		99&A_1^tA_2A_4C_3 & 1 & - & - & - & - \\\hline 
		100&A_1A_1A_4C_2C_2^t & \mathbb{Z}_2 & \begin{smallmatrix} 11011 \\\end{smallmatrix} & \begin{smallmatrix} 00011 \\\end{smallmatrix} & - & - \\\hline
		101&A_1A_2A_3C_2C_2^t & \mathbb{Z}_2 & \begin{smallmatrix} 00211 \\\end{smallmatrix} & \begin{smallmatrix} 00011 \\\end{smallmatrix} & - & - \\\hline 
		102&A_1A_1A_2A_2C_2C_2^t & \mathbb{Z}_2 & \begin{smallmatrix} 110011 \\\end{smallmatrix} & \begin{smallmatrix} 000011 \\\end{smallmatrix} & - & - \\\hline 
		103&A_1A_6C_1C_2^t & 1 & - & \begin{smallmatrix} 0011 \\\end{smallmatrix} & - & - \\\hline 
		104&A_2A_5C_1C_2^t & 1 & - & \begin{smallmatrix} 0011 \\\end{smallmatrix} & e_8 & - \\\hline 
		105&A_1A_2A_4C_1C_2^t & 1 & - & \begin{smallmatrix} 00011 \\\end{smallmatrix} & - & - \\\hline 
		106&A_2A_2A_3C_1C_2^t & 1 & - & \begin{smallmatrix} 00011 \\\end{smallmatrix} & - & - \\\hline
			107&A_1A_1A_1^tA_5C_2 & \mathbb{Z}_2^2 & \begin{smallmatrix} 00031 \\ 11101 \\\end{smallmatrix} & - & - & - \\\hline
\end{tabular}\hspace{0.1em}\begin{tabular}{|@{}>{\scriptsize}c<{}@{}|@{}>{$}c<{$}@{}|@{}>{$}c<{$}@{}|@{}>{$}c<{$}@{}|@{}>{$}c<{$}@{}|@{}>{$}c<{$}@{}|@{}>{\scriptsize$}c<{$}@{}|}\hline \# & L & H & \{k\}  & \spinorsSC& \scalars & L'\\ \hline\hline  
 
108&A_4A_4C_2^t & 1 & - & - & e_9 & - \\\hline 
			109&A_1A_1^tA_3A_3C_2 & \mathbb{Z}_2^2 & \begin{smallmatrix} 10021 \\ 01201 \\\end{smallmatrix} & - & \left[a_{29},a_{29}\right] & - \\\hline 
		110&A_1A_1^tA_1A_2A_3C_2 & \mathbb{Z}_2^2 & \begin{smallmatrix} 011020 \\ 100021 \\\end{smallmatrix} & \begin{smallmatrix} 011020 \\\end{smallmatrix} & - & - \\\hline
		111&A_1A_3(A_1C_2)'(A_1C_2)' & \mathbb{Z}_2^2 & \begin{smallmatrix} 020101 \\ 001111 \\\end{smallmatrix} & \begin{smallmatrix} 000101 \\ 021010 \\\end{smallmatrix} & \left[\tilde{v}\right] & - \\\hline  
		112&A_1A_5C_1(A_1C_2)' & \mathbb{Z}_2 & \begin{smallmatrix} 13000 \\\end{smallmatrix} & \begin{smallmatrix} 00101 \\ 03010 \\\end{smallmatrix} & \left[\tilde{v}\right] & - \\\hline  
		113&A_3'A_4(A_1^tC_2)' & \mathbb{Z}_2 & \begin{smallmatrix} 2011 \\\end{smallmatrix} & - & \left[\tilde{v}\right] & - \\\hline 
		114&A_1A_7C_1C_1 & \mathbb{Z}_4 & \begin{smallmatrix} 1201 \\\end{smallmatrix} & \begin{smallmatrix} 0400 \\ 0011 \\\end{smallmatrix} & \left[s,\tilde{v}\right] & - \\\hline 
		115&A_2A_6C_1C_1 & 1 & - & \begin{smallmatrix} 0011 \\\end{smallmatrix} & \left[s\right] & - \\\hline 
		116&A_2A_3A_3C_1C_1 & \mathbb{Z}_4 & \begin{smallmatrix} 01111 \\\end{smallmatrix} & \begin{smallmatrix} 02200 \\ 00011 \\\end{smallmatrix} & \left[a_{30},s,a_{31}\right] & - \\\hline
		117&A_9C_1 & 1 & - & - & - & - \\\hline 
		118&A_1^tA_8C_1 & 1 & - & - & \left[a_{20}\right] & - \\\hline 
		119&A_2A_7C_1 & \mathbb{Z}_2 & \begin{smallmatrix} 040 \\\end{smallmatrix} & \begin{smallmatrix} 040 \\\end{smallmatrix} & - & - \\\hline 
		120&A_1A_1^tA_7C_1 & \mathbb{Z}_2 & \begin{smallmatrix} 1041 \\\end{smallmatrix} & - & \left[a_{12}\right] & - \\\hline 
		121&A_1^tA_2A_6C_1 & 1 & - & - & - & - \\\hline 
		122&A_4A_5C_1 & 1 & - & - & - & - \\\hline 
		123&A_1^tA_3A_5C_1 & \mathbb{Z}_2 & \begin{smallmatrix} 0031 \\\end{smallmatrix} & \begin{smallmatrix} 1030 \\\end{smallmatrix} & - & - \\\hline 
		124&A_1A_1^tA_2A_5C_1 & \mathbb{Z}_2 & \begin{smallmatrix} 00031 \\\end{smallmatrix} & \begin{smallmatrix} 01030 \\\end{smallmatrix} & e_{10} & - \\\hline 
		125&A_1A_1^tA_3A_4C_1 & \mathbb{Z}_2 & \begin{smallmatrix} 01201 \\\end{smallmatrix} & \begin{smallmatrix} 11200 \\\end{smallmatrix} & - & - \\\hline
		126&A_1^tA_2A_2A_4C_1 & 1 & - & - & - & - \\\hline
		127&A_5A_5 & \mathbb{Z}_2 & \begin{smallmatrix} 33 \\\end{smallmatrix} & - & - & - \\\hline 
		128&A_1^tA_4A_5 & \mathbb{Z}_2 & \begin{smallmatrix} 103 \\\end{smallmatrix} & \begin{smallmatrix} 103 \\\end{smallmatrix} & - & - \\\hline 
		\multicolumn{7}{c}{}
	\end{tabular}\caption{Maximal symmetry enhancements in the \BIIa{} theory in $D = 8$ (continued).}\end{table}

\begin{table}[H]\renewcommand{\arraystretch}{\tabscale}\centering\begin{tabular}{|c|>{$}c<{$}|>{$}c<{$}|>{$}c<{$}|>{$}c<{$}|>{$}c<{$}|>{$}c<{$}|>{$}c<{$}|>{}c<{}|}\hline \# & L & H & \{k\}  & \spinorsS & \spinorsC & \scalars & L' & TF \\ \hline\hline 
		1&A_1A_2E_7 & \mathbb{Z}_2 & \begin{smallmatrix} 101 \\\end{smallmatrix} & - & - & \left[v\right] & A_1 & \xmark \\\hline 
		2&A_4E_6 & 1 & - & - & - & \left[v\right] & A_1 & \xmark \\\hline 
		3&D_5'D_5' & \mathbb{Z}_2 & \begin{smallmatrix} 22 \\\end{smallmatrix} & - & - & \left[v,\tilde{v}\right] & A_1 & \cmark \\\hline 
		4&(A_1A_1)'A_2D_6' & \mathbb{Z}_2^2 & \begin{smallmatrix} 010c \\ 100s \\\end{smallmatrix} & \begin{smallmatrix} 100s \\\end{smallmatrix} & \begin{smallmatrix} 010c \\\end{smallmatrix} & \left[c,\tilde{v},s\right] & - & \xmark \\\hline 
		5&A_1A_1A_3D_5 & \mathbb{Z}_2^2 & \begin{smallmatrix} 0022 \\ 1102 \\\end{smallmatrix} & \begin{smallmatrix} 1120 \\\end{smallmatrix} & \begin{smallmatrix} 0022 \\\end{smallmatrix} & \left[s,s\right] & - & \xmark \\\hline 
		6&A_1A_2A_7 & \mathbb{Z}_2 & \begin{smallmatrix} 004 \\\end{smallmatrix} & - & - & \left[\begin{smallmatrix} 004 \\\end{smallmatrix}\right] & - & \xmark \\\hline
		7&A_1A_2A_7 & \mathbb{Z}_2 & \begin{smallmatrix} 004 \\\end{smallmatrix} & \begin{smallmatrix} 004 \\\end{smallmatrix} & - & - & - & \xmark \\\hline 
		8&A_2A_2A_6 & 1 & - & - & - & - & - & \xmark \\\hline 
		9&A_1A_4A_5 & \mathbb{Z}_2 & \begin{smallmatrix} 103 \\\end{smallmatrix} & \begin{smallmatrix} 103 \\\end{smallmatrix} & - & \left[s\right] & - & \xmark \\\hline 
		10&A_1A_3A_3A_3 & \mathbb{Z}_2^2 & \begin{smallmatrix} 0022 \\ 0202 \\\end{smallmatrix} & \begin{smallmatrix} 0220 \\\end{smallmatrix} & \begin{smallmatrix} 0202 \\\end{smallmatrix} & \left[\begin{smallmatrix} 0022 \\\end{smallmatrix}\right] & - & \xmark \\\hline 
		11&A_2A_2A_3A_3 & \mathbb{Z}_2 & \begin{smallmatrix} 0022 \\\end{smallmatrix} & - & - & - & - & \xmark \\\hline 
		12&A_1A_1A_1A_1A_3A_3 & \mathbb{Z}_2^3 & \begin{smallmatrix} 000022 \\ 010102 \\ 101002 \\\end{smallmatrix} & \begin{smallmatrix} 010120 \\ 101002 \\\end{smallmatrix} & \begin{smallmatrix} 010102 \\ 101020 \\\end{smallmatrix} & \left[v'_8,v''_8\right] & - & \xmark \\\hline 
	\end{tabular}\caption{Maximal symmetry enhancements in the \BI{} theory in $D = 8$.}\label{tab:BI8D}\end{table}

\begin{table}[H]\renewcommand{\arraystretch}{\tabscaleirrep}\centering\begin{tabular}{|>{$}c<{$}||@{}>{$}c<{$}@{}|>{$}c<{$}|}\hline  & \text{ rep } & L \\ \hline\hline
		a_1 & \begin{smallmatrix}120\end{smallmatrix} & C_1D_5X_4 \\ \hline
		a_2 & \begin{smallmatrix}1010\end{smallmatrix} & A_1A_2B_3F_4 \\ \hline
		a_3 & \begin{smallmatrix}110\end{smallmatrix} & C_1C_1X_8 \\ \hline
		a_4 & \begin{smallmatrix}00c\\110\end{smallmatrix} & C_1C_1D_8 \\ \hline
		a_5 & \begin{smallmatrix}11\end{smallmatrix} & C_1C_9 \\ \hline
		a_6 & \begin{smallmatrix}011\end{smallmatrix} & A_2C_2C_6 \\ \hline
		a_7 & \begin{smallmatrix}11\end{smallmatrix} & C_5C_5 \\ \hline
		a_8 & \begin{smallmatrix}0201\\1010\end{smallmatrix} & A_1A_2B_3C_4 \\ \hline
		a_9 & \begin{smallmatrix}0011\\2200\end{smallmatrix} & A_3A_3C_2C_2 \\ \hline
		a_{10} & \begin{smallmatrix}400\\011\end{smallmatrix} & A_7C_1C_2 \\ \hline
		a_{11} & \begin{smallmatrix}0400\\0011\end{smallmatrix} & A_1A_7C_1C_1 \\ \hline
	\end{tabular}
	\begin{tabular}{|>{$}c<{$}||@{}>{$}c<{$}@{}|>{$}c<{$}|}\hline  & \text{rep} &L \\ \hline\hline
		a_{12} & \begin{smallmatrix}40\end{smallmatrix} & A_7 X_3 \\ \hline	
		a_{13} & \begin{smallmatrix}1003\end{smallmatrix} & A_1A_2^2A_2^2A_5 \\ \hline
		a_{14} & \begin{smallmatrix}010s\end{smallmatrix} & A_1A_1A_2D_6 \\ \hline
		a_{15} & \begin{smallmatrix}1c0\end{smallmatrix} & C_1D_4D_5 \\  \hline
		a_{16} & \begin{smallmatrix}012\end{smallmatrix} & C_1C_4D_5 \\ \hline
		a_{17} & \begin{smallmatrix}1s0\end{smallmatrix} & C_2D_4 X_4 \\\hline
		a_{18} & \begin{smallmatrix}11001\end{smallmatrix} & A_1A_1C_2A_2C_4 \\ \hline
		a_{19} & \begin{smallmatrix}00120\\11001\end{smallmatrix} & A_1A_1C_2A_3C_3 \\\hline
		a_{20} & \begin{smallmatrix}30\end{smallmatrix} & A_8X_2 \\ \hline
		a_{21} & \begin{smallmatrix}010120\\101002\end{smallmatrix} & (A_1A_1)(A_1A_1)A_3A_3 \\\hline
		a_{22} & \begin{smallmatrix}0012\end{smallmatrix} & A_1A_1^tC_1D_7 \\\hline
	\end{tabular}				\begin{tabular}{|>{$}c<{$}||@{}>{$}c<{$}@{}|>{$}c<{$}|}\hline  & \text{rep} & \text{L} \\ \hline\hline
		a_{23} & \begin{smallmatrix}210\end{smallmatrix} & A_3C_1D_6 \\\hline
		a_{24}   & \begin{smallmatrix}0012\end{smallmatrix}  & (A_1A_1^t)A_2D_6  \\\hline
		a_{25} & \begin{smallmatrix}012\end{smallmatrix} & C_2^tC_3D_5 \\\hline
		a_{26} & \begin{smallmatrix}0022\end{smallmatrix} & A_1A_1^tA_3D_5 \\\hline
		a_{27} & \begin{smallmatrix}0021\end{smallmatrix} & A_1A_1^tA_3C_5 \\\hline
		a_{28} & \begin{smallmatrix}00210 \\11001 \end{smallmatrix} & A_1A_1A_3C_1C_4 \\\hline
		a_{29} & \begin{smallmatrix}00220\end{smallmatrix}& A_1A_1^tA_3A_3C_2 \\\hline
		a_{30} & \begin{smallmatrix}00210\\02001\end{smallmatrix} & A_2A_3A_3C_1C_1 \\\hline
		a_{31} & \begin{smallmatrix}02010\\00201\end{smallmatrix} & A_2A_3A_3C_1C_1 \\\hline
		\multicolumn{3}{c}{}\\
			\multicolumn{3}{c}{}
	\end{tabular}\caption{Accidental representations for the massless scalars in $D = 8$.}\label{tab:acc}\end{table}

\begin{table}[H]\renewcommand{\arraystretch}{\tabscaleirrep}\centering\begin{tabular}{|>{$}c<{$}|>{$}c<{$}|>{$}c<{$}|}\hline  & \text{irrep} & L \\ \hline\hline 
		e_1 & \left(1,3,1,27\right)_{\times 4} & A_1A_2C_1E_6 \\
		e_2 & \left(3,15,1,1\right)_{\times 4} & A_2A_5C_1C_2 \\
		e_3 & \left(3,3,1,1\right)_{\times 4} & A_2A_2^2A_3B_3 \\
		e_{4} & \left(\underline{5,10},1,1\right)_{\times 4} & A_4A_4C_1C_1 \\
		e_{5} & (2,1,3,6,1)_{\times 4} & A_1A_1A_2A_5C_1 \\
		e_{6} & \left(1,3,3,1\right)_{\times 4} & A_1A_2A_2^2C_5 \\
		e_{7} & \left(3,15,1\right)_{\times 4} & A_2A_5C_3 \\
		e_{9} & \left(\underline{5,10},1\right)_{\times 4}& A_4A_4C_2 \\
		e_{8} & \left(3,15,1,1\right)_{\times 4} & A_2A_5C_1C_2 \\
		e_{10} & \left(1,1,3,15,1\right)_{\times 4} & A_1A_1A_2A_5C_1 \\
		\hline\end{tabular}\caption{Exceptional representations for the massless scalars in $D = 8$. The subscript $\times 4$ means to take four copies of the representation related through automorphisms of the Dynkin diagram of the algebra. For example: $(3,3,1,1)_{\times 4} = (3,3,1,1)+(\bar 3,3,1,1) + (3,\bar 3, 1,1) + (\bar 3, \bar 3, 1,1)$. The $2$ of $A_1$ is an expectator. In all cases except $e_4$ and $e_9$, the right-moving $U(1)$ charges ($p_R$'s) are given by two norm 1 vectors $u,u'$ with $u \cdot u' = 1/3$, and their negatives. In the cases $e_4,e_9$ there are in total eight norm 1 vectors forming two $2A_1$ systems, one rotated with respect to the other at an angle $\theta = \arccos (4/5)$.}\label{tab:exc}\end{table}

\begin{table}[H]\renewcommand{\arraystretch}{\tabscaleirrep}\centering\begin{tabular}{|>{$}c<{$}|>{$}c<{$}|>{$}c<{$}|}\hline  & \text{irrep} & L \\ \hline\hline 
		v'_{1} & (1,1,1,77)+(1,3,5,1)+(3,1,1,1) & \multirow{2}{*}{$A_1A_1^tC_2D_6$} \\
		v''_{1} & (1,1,1,66)+(1,1,5,1)+(3,3,1,1) &  \\ \hline
		v'_{2} & (1,1,1,1,66)+(3,3,1,1,1)+(1,1,3,3,1) &\multirow{2}{*}{$ A_1A_1A_1A_1^tD_6$} \\
		v''_{2} & (1,1,1,1,66)+(3,1,1,3,1)+(1,3,3,1,1) &  \\ \hline
		v'_{3} & \left(1,1,1,35_c\right)+\left(1,1,35_v,1\right)+(\underline{1,3},1,1) & \multirow{2}{*}{$A_1A_1D_4D_4$} \\
		v''_{3} & \left(1,1,35_c,1\right)+\left(1,1,1,35_v\right)+(\underline{1,3},1,1) &  \\\hline 
		v'_{4} & \left(1,1,1,35_s\right)+(1,1,27,1)+(\underline{1,3},1,1) & \multirow{2}{*}{$A_1A_1C_4D_4$} \\
		v''_{4} & \left(1,1,1,35_v\right)+(1,1,27,1)+(\underline{1,3},1,1) &\\ \hline
		v'_{5} & (1,1,1,1,28)+(1,3,1,5,1)+(3,1,5,1,1) & \multirow{2}{*}{$A_1A_1C_2C_2D_4$} \\
		v''_{5} & (1,1,1,1,28)+(1,3,5,1,1)+(3,1,1,5,1) & \\\hline
		v'_{6} & (1,1,1,65)+(1,3,5,1)+(3,1,1,1) & \multirow{2}{*}{$A_1A_1C_2^tC_6$} \\
		v''_{6} & (1,1,1,65)+(3,1,5,1)+(1,3,1,1) &  \\ \hline
		v'_{7} & (1,1,\underline{1,42})+(\underline{1,3},1,1) & \multirow{2}{*}{$A_1A_1C_4C_4$} \\
		v''_{7} & (1,1,\underline{1,27})+(\underline{1,3},1,1) &  \\ \hline
		v'_8 & (1,1,1,1,\underline{1,15})+(3,3,1,1,1,1)+(1,1,3,3,1,1) & \multirow{2}{*}{$ A_1A_1  A_1A_1 A_3A_3$} \\
		v''_8 & (1,1,1,1,\underline{1,15})+(3,1,1,3,1,1)+(1,3,3,1,1,1) &   \\
		\hline\end{tabular}\caption{Representations of type $\tilde v$ appearing in pairs.}\label{tab:altv}\end{table}

\bibliographystyle{JHEP}
\bibliography{nonsusy}
\end{document}